\documentclass[acmsmall,10pt]{acmart}
\renewcommand\footnotetextcopyrightpermission[1]{}
\usepackage{subcaption} 

\usepackage[utf8]{inputenc} 
\usepackage[T1]{fontenc}    
\usepackage{hyperref}       
\usepackage{url}            
\usepackage{amsfonts}       
\usepackage{nicefrac}       
\usepackage{microtype}      
\usepackage{algpseudocode}
\usepackage{amsmath}
\usepackage{alltt}
\usepackage{syntax}
\usepackage{bussproofs}
\usepackage{mathpartir}
\usepackage{float}
\usepackage{color}
\usepackage{subcaption}
\usepackage{stmaryrd}
\usepackage{balance}
\usepackage{amsthm}
\usepackage{tikz}
\usepackage{rotating}
\usepackage[framemethod=tikz]{mdframed}
\usepackage{listings}
\usepackage{multirow}
\usepackage{tabularx}
\usepackage{multicol}
\usepackage{wrapfig}
\usepackage[binary-units=true, scientific-notation=true]{siunitx}

\usetikzlibrary{shapes.misc,shapes.geometric,arrows,positioning}

\newenvironment{wraplst*}
               {\hspace*{1.4em}\begin{minipage}{0.953\linewidth}}
               {\end{minipage}}

\newenvironment{wraplst}
               {\hspace*{1.8em}\begin{minipage}{0.94\linewidth}}
               {\end{minipage}}

\newenvironment{wrappedwraplst*}
               {\begin{flushright}\begin{minipage}{0.91\linewidth}}
               {\end{minipage}\end{flushright}}

\lstdefinelanguage{JavaScript}{
  keywords={break, case, catch, continue, debugger, default, delete, do, else, finally, for, function, if, in, instanceof, new, return, switch, this, throw, try, typeof, var, void, while, with},
  morecomment=[l]{//},
  morecomment=[s]{/*}{*/},
  morestring=[b]',
  morestring=[b]",
  sensitive=true
}

\setcopyright{none}             

\newcommand{\tool}{Shuffle}
\newcommand{\cons}{\phi}
\newcommand{\denotation}[1]{\llbracket #1 \rrbracket}

\definecolor{WowColor}{rgb}{.75,0,.75}
\definecolor{SubtleColor}{rgb}{1,0,0}

\newcounter{margincounter}

\newcommand\independent{\protect\mathpalette{\protect\independenT}{\perp}}
\def\independenT#1#2{\mathrel{\rlap{$#1#2$}\mkern2mu{#1#2}}}

\begin{document}

\newtheorem{thm}{Theorem}


\title[]{Verifying Handcoded Probabilistic Inference Procedures}         
\suppressfloats


\authorsaddresses{}
\author{Eric Atkinson}
\affiliation{
  \institution{MIT}            
}

\author{Cambridge Yang}
\affiliation{
  \institution{MIT}            
}

\author{Michael Carbin}
\orcid{nnnn-nnnn-nnnn-nnnn}             
\affiliation{
  \institution{MIT}           
}



\begin{abstract}
Researchers have recently proposed several systems that ease the process of
performing Bayesian probabilistic inference. These include systems for
automatic inference algorithm synthesis as well as stronger abstractions for
manual algorithm development. However, existing systems whose performance
relies on the developer manually constructing a part of the inference algorithm
have limited support for reasoning about the correctness of the resulting
algorithm. 

In this paper, we present Shuffle, a programming language for manually
developing inference procedures that 1) enforces the basic rules of probability
theory, 2) enforces the statistical dependencies of the algorithm's
corresponding probabilistic model, and 3) generates an optimized
implementation. We have used Shuffle to develop inference algorithms for
several standard probabilistic models. Our results demonstrate that Shuffle
enables a developer to deliver correct and performant implementations of these
algorithms.

\end{abstract}
\maketitle
\thispagestyle{plain}


\newcommand{\term}{CP}
\newcommand{\combine}{\mathrm{\texttt{::}}}
\newcommand{\progcmd}[1]{\;\mathrm{\tt #1}\;}
\newcommand{\sr}[0]{\mathrm{sr}}
\newcommand{\plaintext}[1]{\; \mathrm{#1} \;}
\renewcommand{\syntleft}{$}          
\renewcommand{\syntright}{$}         

\newcommand{\defs}[4]{{\texttt{def} \; {#1} \; \texttt{(} {#2} \texttt{)} \texttt{:} \; {#3} \; \texttt{=} \; {#4}}}
\newcommand{\defm}[4]{{\texttt{def} \; {#1} \; \texttt{(} {#2} \texttt{)} \texttt{:} \; {#3} \; \texttt{=} \; {#4}}}
\newcommand{\defsi}[5]{{\texttt{def} \; {#1} \; {#2} \; \texttt{(} {#3} \texttt{)} \texttt{:} \; {#4} \; \texttt{=} \; {#5}}}
\newcommand{\prodd}[4]{{\texttt{prod} \; {#1} \; \texttt{in} \; {#2} \; \texttt{where} \; #3 \; \texttt{:} \; {#4}}}
\newcommand{\lift}[1]{{\texttt{lift \{} \; {#1} \; \texttt{\}}}}
\newcommand{\elift}[1]{{\texttt{elift\{} \; {#1} \; \texttt{\}}}}
\newcommand{\map}[2]{{\texttt{map} \; {#1} \; \texttt{by} \; {#2}}}
\newcommand{\factor}[2]{{\texttt{factor} \; {#1} \; \texttt{by} \; {#2}}}
\newcommand{\fix}[1]{{\texttt{fix} \; {#1}}}
\newcommand{\join}[3]{{\texttt{join}\; {#1} \; \texttt{in} \; {#2} \; \texttt{:} \; {#3}}}
\newcommand{\indcoerce}[2]{{\texttt{(ind } \; {#1} \; \texttt{)} \; {#2}}}
\newcommand{\seq}[2]{{#1} \; \texttt{;} \; {#2}}
\newcommand{\var}[2]{{#1} \texttt{[} {#2} \texttt{]}} 
\newcommand{\varset}[4]{{#1} \texttt{\{} {#2} \; \texttt{in} \; {#3} \texttt{:} \; {#4} \texttt{\}}}
\newcommand{\samp}[2]{{#1} \; \texttt{:=} \; \texttt{sample} \; {#2}}
\newcommand{\callnone}[1]{{{#1} \; \texttt{()}}}
\newcommand{\ite}[3]{\texttt{if} \; {#1} \; \{ \; {#2} \; \} \; \{ \; {#3} \; \} }
\newcommand{\callone}[2]{{{#1} \texttt{(} #2 \texttt{)}}}
\newcommand{\calltwo}[3]{{{#1} \texttt{(} #2 \texttt{,} #3 \texttt{)}}}
\newcommand{\comma}{\mathrm{\texttt{,}}}
\newcommand{\lb}{n_1}
\newcommand{\ub}{n_2}
\newcommand{\qv}{q}
\newcommand{\qvspace}{\mathcal{Q}}
\newcommand{\range}[2]{\texttt{[}{#1}\texttt{,}{#2}\texttt{]}}
\newcommand{\argchar}{a}
\newcommand{\prog}{\beta}
\newcommand{\booltype}{\mathrm{\tt Bool}}

\newcommand{\assumplog}{\mathcal{L}}

\newcommand{\basictype}[4]{{
    #1\mathrm{\texttt{(}} {#2} \; | \; {#3} \mathrm{\texttt{,}} {#4} \mathrm{\texttt{)}}
}}
\newcommand{\basictypesplit}[4]{
    #1\mathrm{\texttt{(}} {#2} |$ $ {#3} \mathrm{\texttt{,}} {#4} \mathrm{\texttt{)}}
}
\newcommand{\basictypenocons}[3]{{
	#1 \mathrm{\texttt{(}} {#2} | {#3} \mathrm{\texttt{)}}
}}
\newcommand{\basictypenoconssplit}[3]{
	#1 \mathrm{\texttt{(}}$ $ {#2}|$ ${#3} \mathrm{\texttt{)}}
}

\newcommand{\dtype}[9][\vdash]{\ensuremath{{{#2}, \, {#3}, \, {#4} \; {#1} \; {#5} \; \textnormal{\texttt{:}} \; \basictype{\textnormal{\texttt{density}}}{#7}{#8}{#9}}}}
\newcommand{\dtfn}[9][\vdash]{\ensuremath{{{#2}, {#3}, {#4} \; {#1} \; {#5} \; \textnormal{\texttt{:}} \; {#6},\basictype{\textnormal{\texttt{density}}}{#7}{#8}{#9}}}}
\newcommand{\dtdef}[4][\vdash]{\ensuremath{\dtype[#1]{\mathcal{M}}{\Gamma}{\assumplog}{#2}{\hat{\qv}}{#3}{#4}{\cons}}}

\newcommand{\stype}[9][\vdash]{\ensuremath{{{#2}, {#3}, {#4} \; {#1} \; {#5} \; \textnormal{\texttt{:}} \; \basictype{\textnormal{\texttt{sampler}}}{#7}{#8}{#9}}}}
\newcommand{\stfn}[9][\vdash]{\ensuremath{{{#2}, {#3}, {#4} \; {#1} \; {#5} \; \textnormal{\texttt{:}} \; {#6},\basictype{\textnormal{\texttt{sampler}}}{#7}{#8}{#9}}}}
\newcommand{\stdef}[4][\vdash]{\ensuremath{\stype[#1]{\mathcal{M}}{\Gamma}{\assumplog}{#2}{\hat{\qv}}{#3}{#4}{\cons}}}

\newcommand{\ktype}[9][\vdash]{\ensuremath{{{#2}, {#3}, {#4} \; {#1} \; {#5} \; \textnormal{\texttt{:}} \; \basictype{\textnormal{\texttt{kernel}}}{#7}{#8}{#9}}}}
\newcommand{\ktfn}[9][\vdash]{\ensuremath{{{#2}, {#3}, {#4} \; {#1} \; {#5} \; \textnormal{\texttt{:}} \; {#6},\basictype{\textnormal{\texttt{kernel}}}{#7}{#8}{#9}}}}
\newcommand{\ktdef}[4][\vdash]{\ensuremath{\ktype[#1]{\mathcal{M}}{\Gamma}{\assumplog}{#2}{\hat{\qv}}{#3}{#4}{\cons}}}

\newcommand{\etype}[9][\vdash]{\ensuremath{{{#2}, {#3}, {#4} \; {#1} \; {#5} \; \textnormal{\texttt{:}} \; \basictype{\textnormal{\texttt{estimator}}}{#7}{#8}{#9}}}}
\newcommand{\etfn}[9][\vdash]{\ensuremath{{{#2}, {#3}, {#4} \; {#1} \; {#5} \; \textnormal{\texttt{:}} \; {#6},\basictype{\textnormal{\texttt{estimator}}}{#7}{#8}{#9}}}}
\newcommand{\etdef}[4][\vdash]{\ensuremath{\etype[#1]{\mathcal{M}}{\Gamma}{\assumplog}{#2}{\hat{\qv}}{#3}{#4}{\cons}}}

\newcommand{\propnolog}[4][\vDash]{\ensuremath{{#2}, {#3} \; {#1} \; {#4}}}
\newcommand{\typegeneric}[6][\vdash]{\ensuremath{{#2}, {#3}, {#4} \; {#1} \; {#5} \; \mathrm{\texttt{:}} \; #6}}
\newcommand{\typegenericind}[5]{\typegeneric[\vdash_I]{#1}{#2}{#3}{#4}{#5}}

\newcommand{\forallt}[2]{\forall {#1}. \; {#2}}

\newcommand{\tmap}[7]{${#1}, {#2}, {#3} \vdash \mathrm{QMAP}({#4},{#5},{#6},{#7})$}

\newcommand{\validjudge}[5]{{${#1}, {#2}, {#3} \vdash \; \mathrm{Valid{#4}} ({#5})$}}

\newcommand{\wellformed}[2]{\ensuremath{\mathrm{WF}({#1},{#2})}}

\newcommand{\dmul}{
\ensuremath{
    \inferrule[dmul]{
        \typegeneric{\mathcal{M}}{\Gamma}{\mathcal{L}}{d_1}{\basictype{\texttt{density}}{A}{B \comma C}{\cons}}\\\\
        \typegeneric{\mathcal{M}}{\Gamma}{\mathcal{L}}{d_2}{\basictype{\texttt{density}}{B}{C}{\cons}}\\\\
        \propnolog{\mathcal{M}}{\Gamma}{\mathrm{ValidInfer}(A \comma B | C, \cons)}
    } {
        \typegeneric{\mathcal{M}}{\Gamma}{\mathcal{L}}{d_1 \opfmt{*} d_2}{\basictype{\texttt{density}}{A \comma B}{C}{\cons}}
    }
}
}

\newcommand{\ddiv}{
\ensuremath{
    \inferrule[ddiv]{
        \dtype{\mathcal{M}}{\Gamma}{\assumplog}{d_1}{{\qv}}{A \comma B}{C}{\cons}\\\\
        \dtype{\mathcal{M}}{\Gamma}{\assumplog}{d_2}{{\qv}}{B}{C}{\cons}\\\\
        \propnolog{\mathcal{M}}{\Gamma}{\mathrm{ValidInfer}(A | B \comma C, \cons)}
    } {
        \dtdef{d_1 \opfmt{/} d_2}{A}{B \comma C}
    }
}
}

\newcommand{\ddivtwo}{
\ensuremath{
    \inferrule[ddiv2]{
        \dtdef{d_1}{A \comma B}{C}\\\\
        \dtdef{d_2}{A}{B \comma C}
    } {
        \dtdef{d_1 \opfmt{/} d_2}{B}{C}
    }
}
}

\newcommand{\dint}{
\ensuremath{
    \inferrule[dint] {
        \dtdef{d}{A \comma B}{C}
    } {
        \dtdef{\intd{d}{B}}{A}{C}
    }
}
}

\newcommand{\slift}{
\ensuremath{
    \inferrule[slift]{
        \dtdef{d}{\var{v}{a}}{B}
    } {
        \stdef{\samp{\var{v}{a}}{d}}{\var{v}{a}}{B}
    }
}
}

\newcommand{\sbind}{
\ensuremath{
    \inferrule[sbind]{
        \typegeneric{\mathcal{M}}{\Gamma}{\mathcal{L}}{s_1}{\basictype{\texttt{sampler}}{B}{C}{\cons}}\\\\
        \typegeneric{\mathcal{M}}{\Gamma}{\mathcal{L}}{s_2}{\basictype{\texttt{sampler}}{A}{B \comma C}{\cons}}\\\\
        \propnolog{\mathcal{M}}{\Gamma}{\mathrm{ValidInfer}(A \comma B|C,\cons)}
    } {
        \typegeneric{\mathcal{M}}{\Gamma}{\mathcal{L}}{s_1 \opfmt{;} s_2}{\basictype{\texttt{sampler}}{A \comma B}{C}{\cons}}
    }
}
}

\newcommand{\klift}{
\ensuremath{
    \inferrule[klift]{
        \stdef{s}{A}{B}\\
        \pi_2(\mathcal{M})(v) = (\delta_1, \delta_2)\\
        \mathcal{M}, \assumplog \vDash \mathrm{ReachesAll}(s)
    } {
        \ktype{\mathcal{M}}{\Gamma}{\assumplog}{\lift{s}}{{\qv}}{A}{B}{\cons}
    }
}
}

\newcommand{\kliftcombine}{
\ensuremath{
    \inferrule[klift-combine]{
        \ktdef{\lift{k_1} \opfmt{;} \lift{k_2}}{A}{B}
    } {
        \ktdef{\lift{k_1 \opfmt{;} k_2}}{A}{B}
    }
}
}

\newcommand{\kcombine}{
\ensuremath{
    \inferrule[kcombine]{
        \typegeneric{\mathcal{M}}{\Gamma}{\mathcal{L}}{k_1}{\basictype{\texttt{kernel}}{A}{B \comma C}{\cons}}\\\\
        \typegeneric{\mathcal{M}}{\Gamma}{\mathcal{L}}{k_2}{\basictype{\texttt{kernel}}{B}{A \comma C}{\cons}}\\\\
        \propnolog{\mathcal{M}}{\Gamma}{\mathrm{ValidInfer}(A \comma B | C , \cons)}
    } {
        \typegeneric{\mathcal{M}}{\Gamma}{\mathcal{L}}{k_1 \opfmt{;} k_2}{\basictype{\texttt{kernel}}{A \comma B}{C}{\cons_1 \opfmt{\&\&} \cons_2}}
    }
}
}

\newcommand{\kfix}{
\ensuremath{
    \inferrule[kfix]{
       \ktdef{k}{A}{B} 
    } {
        \stdef{\fix{k}}{A}{B}
    }
}
}

\newcommand{\defr}{
\ensuremath{
    \inferrule[def]{
        \typegeneric{\mathcal{M}}{\Gamma[\qv \mapsto \delta] \;}{\assumplog}{\prog}{\basictype{t_b}{A}{B}{\cons}}\\
        \propnolog{\mathcal{M}}{\Gamma}{\mathrm{ValidInfer}(A | B, \cons)}
    } {
        \typegeneric{\mathcal{M}}{\Gamma}{\assumplog}{\defs{x}{\qv \opfmt{in} \delta}{t}{\prog}}{\Gamma[x \mapsto (\qv, \delta,\basictype{t_b}{A}{B}{\cons \opfmt{\&\&} \qv \in \delta})]}
    }
}
}

\newcommand{\defrec}{
\ensuremath{
    \inferrule[def-rec]{
        \typegeneric{\mathcal{M}}{\Gamma[ \qv \mapsto \delta ][x \mapsto (\qv', \delta, \basictype{t_b}{A[{\qv'}/{\qv}]}{B[{\qv'}/{\qv}]}{(\cons[{\qv'}/{\qv}]) \opfmt{\&\&} {\qv'} < {\qv}})]}{\assumplog}{\prog}{\basictype{t_b}{A}{B}{\cons}}\\
        \propnolog{\mathcal{M}}{\Gamma}{\mathrm{BaseCase}(\qv,\delta,A)}\\
        \propnolog{\mathcal{M}}{\Gamma}{\mathrm{ValidInfer}(A | B , \cons)}
    } {
        \typegeneric{\mathcal{M}}{\Gamma}{\assumplog}{\defsi{\texttt{rec}}{x}{\qv \opfmt{in} \delta}{\basictype{t_b}{A}{B}{\cons}}{\prog}}{\Gamma[x \mapsto (\qv,\delta, \basictype{t_b}{A}{B}{\cons \opfmt{\&\&} \qv \in \delta})]}
    }
}
}

\newcommand{\defind}{
\ensuremath{
    \inferrule[def-ind]{
        \typegenericind{\mathcal{M}}{\Gamma[\qv \mapsto \delta]}{\assumplog}{\prog}{\basictype{t_b}{A}{B}{\cons}}\\
        \propnolog{\mathcal{M}}{\Gamma}{\mathrm{ValidInfer}(A | B , \cons)}
    } {
        \typegeneric{\mathcal{M}}{\Gamma}{\assumplog}{\defsi{\texttt{independent}}{x}{\qv \opfmt{in} \delta}{t}{\prog}}{\Gamma[x \mapsto (\qv,\delta, \basictype{t_b}{A}{B}{\cons \opfmt{\&\&} \qv \in \delta})]}
    }
}
}

\newcommand{\shfcond}{
\ensuremath{
    \inferrule[if] {
        \typegeneric{\mathcal{M}}{\Gamma}{\assumplog}{\prog_t}{\basictype{t_b}{A_t}{B_t}{\cons_t}}\\
        \typegeneric{\mathcal{M}}{\Gamma}{\assumplog}{\prog_f}{\basictype{t_b}{A_f}{B_f}{\cons_f}}\\
        \typegeneric{\mathcal{M}}{\Gamma}{\assumplog}{\cons_i}{\booltype}
    } {
        \typegeneric{\mathcal{M}}{\Gamma}{\assumplog}{\ite{\cons_i}{\prog_t}{\prog_f}}{\basictype{t_b}{(\cons_i, A_t, A_f) \;}{\; (\cons_i, B_t, B_f)}{(\cons_i \opfmt{\&\&} \cons_t) \opfmt{||} (\neg\cons_i \opfmt{\&\&} \cons_f)}}
    }
}
}

\newcommand{\env}{
\ensuremath{
    \inferrule[inv] {
        \typegeneric{\mathcal{M}}{\Gamma}{\assumplog}{x}{{\qv}, {\delta}, \basictype{t_b}{A}{B}{\cons}}\\
        \propnolog{\mathcal{M}}{\Gamma}{\mathrm{ValidInfer}(A[{\argchar}/{q}] | B[{\argchar}/{q}], \cons[{\argchar}/{q}])}
    } {
        \typegeneric{\mathcal{M}}{\Gamma}{\assumplog}{\callone{x}{{\argchar}}}{\basictype{t_b}{A}{B}{\cons}[{\argchar}/{q}]}
    }
}
}

\newcommand{\invrec}{
\ensuremath{
    \inferrule[inv-rec] {
        \typegeneric{\mathcal{M}}{\Gamma}{\assumplog}{x}{{\qv}, {\delta}, \basictype{t_b}{A}{B}{\cons},\texttt{rec}}\\\\
        \propnolog{\mathcal{M}}{\Gamma}{\mathrm{ValidInfer}(A[{\argchar}/{q}] | B[{\argchar}/{q}], \cons[{\argchar}/{q}])}
    } {
        \typegeneric{\mathcal{M}}{\Gamma}{\assumplog}{\callone{x}{{\argchar}}}{\basictype{t_b}{A}{B}{\cons}[{\argchar}/{q}]}
    }
}
}

\newcommand{\envrec}{
\ensuremath{
    \inferrule[env-rec] {
        \typegeneric{\mathcal{M}}{\Gamma}{\assumplog}{\prog}{t}\\\\
        \prog' \neq \prog
    } {
        \typegeneric{\mathcal{M}}{\Gamma :: [\prog' : t']}{\assumplog}{\prog}{t}
    }
}
}

\newcommand{\envbasex}{
\ensuremath{
    \inferrule[env-x] { } {
        \typegeneric{\mathcal{M}}{\Gamma}{\assumplog}{x}{\Gamma(x)}
    }
}
}

\newcommand{\envbaseq}{
\ensuremath{
    \inferrule[env-q] { } {
        \typegeneric{\mathcal{M}}{\Gamma}{\assumplog}{\qv}{\Gamma(\qv)}
    }
}
}

\newcommand{\envnat}{
\ensuremath{
    \inferrule[env-nat] {
        n \in \denotation{\delta}
    } {
        \typegeneric{\mathcal{M}}{\Gamma}{\assumplog}{n}{\delta}
    }
}
}

\newcommand{\envvar}{
\ensuremath{
    \inferrule[env-var] {
        \pi_2(\mathcal{M})(v) = (\delta_1,\delta_2)\\\\
        \typegeneric{\mathcal{M}}{\Gamma}{\assumplog}{a}{\delta_1}
    } {
        \typegeneric{\mathcal{M}}{\Gamma}{\assumplog}{\var{v}{a}}{\delta_2}
    }
}
}

\newcommand{\envlst}{
\ensuremath{
    \inferrule[env-lst] {
        \typegeneric{\mathcal{M}}{\Gamma}{\assumplog}{a}{\delta}\\\\
        \typegeneric{\mathcal{M}}{\Gamma}{\assumplog}{{a}}{{\delta}}
    } {
        \typegeneric{\mathcal{M}}{\Gamma}{\assumplog}{a \comma {a}}{\delta, {\delta}}
    }
}
}

\newcommand{\envempty}{
\ensuremath{
    \inferrule[env-empty-lst] { } {
        \typegeneric{\mathcal{M}}{\Gamma}{\assumplog}{\cdot}{\cdot}
    }
}
}

\newcommand{\consprec}{
\ensuremath{
    \inferrule[constraint] {
        \typegeneric{\mathcal{M}}{\Gamma}{\assumplog}{a}{\delta}\\\\
        \typegeneric{\mathcal{M}}{\Gamma}{\assumplog}{a}{\delta}
    } {
        \typegeneric{\mathcal{M}}{\Gamma}{\assumplog}{a \prec a}{\booltype}
    }
}
}

\newcommand{\consneg}{
\ensuremath{
    \inferrule[constraint-neg] {
        \typegeneric{\mathcal{M}}{\Gamma}{\assumplog}{\cons}{\booltype}
    } {
        \typegeneric{\mathcal{M}}{\Gamma}{\assumplog}{\neg \cons}{\booltype}
    }
}
}

\newcommand{\conscon}{
\ensuremath{
    \inferrule[constraint-and] {
        \typegeneric{\mathcal{M}}{\Gamma}{\assumplog}{\cons_1}{\booltype}\\\\
        \typegeneric{\mathcal{M}}{\Gamma}{\assumplog}{\cons_2}{\booltype}
    } {
        \typegeneric{\mathcal{M}}{\Gamma}{\assumplog}{\cons_1 \opfmt{\&\&} \cons_2}{\booltype}
    }
}
}

\newcommand{\consdis}{
\ensuremath{
    \inferrule[constraint-or] {
        \typegeneric{\mathcal{M}}{\Gamma}{\assumplog}{\cons_1}{\booltype}\\\\
        \typegeneric{\mathcal{M}}{\Gamma}{\assumplog}{\cons_2}{\booltype}
    } {
        \typegeneric{\mathcal{M}}{\Gamma}{\assumplog}{\cons_1 \opfmt{||} \cons_2}{\booltype}
    }
}
}

\newcommand{\model}{
\ensuremath{
    \inferrule[model] {
        (x,d,(\qv,\delta),A,B,\cons) \in \pi_3(\mathcal{M})
    } {
        \typegeneric{\mathcal{M}}{\Gamma}{\assumplog}{x}{(\qv,\delta,\basictype{\textnormal{\texttt{density}}}{A}{B}{\cons \opfmt{\&\&} \qv \in \delta})}
    }
}
}

\newcommand{\progcompose}{
\ensuremath{
    \inferrule[prog-compose]{
        \typegeneric{\mathcal{M}}{\Gamma_0}{\assumplog}{p_1}{\Gamma_1}\\\\
        \typegeneric{\mathcal{M}}{\Gamma_1}{\assumplog}{p_1}{\Gamma_2}
    } {
        \typegeneric{\mathcal{M}}{\Gamma_0}{\assumplog}{p_1 \opfmt{;} p_2}{\Gamma_2}
    }
}
}

\newcommand{\coerce}{
\ensuremath{
    \inferrule[c] {
        \typegeneric{\mathcal{M}}{\Gamma}{\assumplog}{\prog}{t_1}\\
        \mathcal{M} \vdash t_1 \rightarrow t_2
    } {
        \typegeneric{\mathcal{M}}{\Gamma}{\assumplog}{\prog}{t_2}
    }
}
}

\newcommand{\coerceind}{
\ensuremath{
    \inferrule[ind] {
        \typegeneric{\mathcal{M}}{\Gamma}{\assumplog}{\prog}{t_1}\\
        \mathcal{M}, \assumplog \vdash t_1 \rightarrow_I t_2
    } {
        \typegenericind{\mathcal{M}}{\Gamma}{\assumplog}{\prog}{t_2}
    }
}
}

\newcommand{\coerceindinline}{
\ensuremath{
    \inferrule[coerce] {
        \typegeneric{\mathcal{M}}{\Gamma}{\assumplog}{\prog}{\basictype{t_b}{A}{B}{\cons}}\\\\
        \mathcal{M}, \assumplog \vdash \basictype{t_b}{A}{B}{\cons} \rightarrow_I \basictype{t_b}{A}{B \comma C}{\cons}\\\\
        \propnolog{\mathcal{M}}{\Gamma}{\mathrm{ValidInfer}(A | B \comma C , \cons)}
    } {
        \typegenericind{\mathcal{M}}{\Gamma}{\assumplog}{\indcoerce{C}{\prog}}{\basictype{t_b}{A}{B \comma C}{\cons}
}
    }
}
}

\newcommand{\coerceindtype}{
\ensuremath{
    \inferrule[cind] {
        \typegenericind{\mathcal{M}}{\Gamma}{\assumplog}{\prog}{t}
    } {
        \typegeneric{\mathcal{M}}{\Gamma}{\assumplog}{\prog}{t}
    }
}
}

\newcommand{\rec}{
\ensuremath{
    \inferrule[rec] {
        \typegeneric{\mathcal{M}}{\Gamma }{\assumplog}{\prog}{\basictype{t_b}{A}{B}{\cons}}\\
        \forall \sigma.\; \denotation{\qv_0}(\sigma) < \min(\denotation{\delta_0}) \Rightarrow \denotation{A}(\sigma) = \emptyset\\
        \propnolog{\mathcal{M}}{\Gamma}{\mathrm{ValidInfer}(A | B , \cons)}
    } {
        \typegeneric[\vdash_{r}]{\mathcal{M}}{\Gamma}{\assumplog}{\prog}{\basictype{t_b}{A}{B}{\cons}}
    }
}
}

\newcommand{\crec}{
\ensuremath{
    \inferrule[crec] {
        \typegeneric[\vdash_{r}]{\mathcal{M}}{\Gamma}{\assumplog}{\prog}{t}
    } {
        \typegeneric{\mathcal{M}}{\Gamma}{\assumplog}{\prog}{t}
    }
}
}

\newcommand{\logtc}{
\(
    \inferrule[l1] {
        \Gamma \vDash \Gamma_m\\
        \mathcal{M}' = (\mathcal{D},\pi_2(\mathcal{M}),\pi_3(\mathcal{M}))\\
        \mathcal{M}', \Gamma_m \vDash \cons_2 \Rightarrow A \equiv C\\
        \mathcal{M}', \Gamma_m \vDash \cons_2 \Rightarrow B \equiv D\\
        \mathcal{M}', \Gamma_m \vDash \cons_2 \Rightarrow \cons_1
    } {
        \mathcal{M} \vdash \basictype{t_b}{A}{B}{\cons_1} \rightarrow \basictype{t_b}{C}{D}{\cons_2}
    }
\)
}

\newcommand{\logtcindep}{
\(
    \inferrule[l2] {
        \Gamma \vDash \Gamma_m\\
        \mathcal{M}' = (\mathcal{D},\pi_2(\mathcal{M}),\pi_3(\mathcal{M}))\\
        \mathcal{M}, \assumplog \vDash \cons \Rightarrow A \independent C \; | \; B\\
        \mathcal{M'},\Gamma_m \vDash \cons \Rightarrow (A \cap C = \emptyset)
    } {
        \mathcal{M}, \assumplog \vdash \basictype{t_b}{A}{B}{\cons} \rightarrow_I \basictype{t_b}{A}{B \comma C}{\cons}
    }
\)
}

\newcommand{\estsyntax}{
    \textit{E} &  \rightarrow \; x \texttt{(} \textit{A} \texttt{)} \; \mid \; \elift{S} \;\\ & \; \mid \; \factor{E}{D} \; \mid \; \ite{\cons}{E}{E} \\ & \; \mid \; \texttt{(1.0,return)}
}

\newcommand{\estsyntaxtext}{
An estimator, $E$, defines a probability distribution as a weighted sampler. A
{\tool} user can construct an estimator out of a sampler and use a density to
reweight the samples.
}

\newcommand{\estsemantics}{
\begin{figure}
\begin{mdframed}
    \begin{mathpar}
        \denotation{\progcmd{factor} e \opfmt{by} d}(\mathcal{M},\sigma,\sr) =
        \plaintext{let} (w,\sigma') = \denotation{e}(\mathcal{M},\sigma,\sr) \plaintext{in} (w * \denotation{d}(\mathcal{M},\sigma'), \sigma')

        \denotation{\ite{\cons}{e_t}{e_f}}(\mathcal{M},\sigma,\sr) =
        \begin{cases}
            \denotation{e_t}(\mathcal{M},\sigma,\sr) & \hphantom{\neg}\denotation{\cons}(\mathcal{M},\sigma) \\
            \denotation{e_f}(\mathcal{M},\sigma,\sr) & \neg\denotation{\cons}(\mathcal{M},\sigma) 
        \end{cases}

        \denotation{\elift{s}}(\mathcal{M},\sigma,\sr) = 
        (1, \denotation{s}(\mathcal{M},\sigma,\sr))
        
        \denotation{\texttt{(1.0,return)}}(\mathcal{M},\sigma,\sr) = (1,\sigma)\\

    \denotation{\callone{x}{\argchar}}(\mathcal{M}, \sigma[x \mapsto (\qv, \prog)],\sr)  =   \denotation{\prog}(\mathcal{M}, \sigma[\qv \mapsto \denotation{\argchar}(\sigma)],\sr)\\
    \denotation{\callone{x}{\argchar}}(\mathcal{M}, \sigma,\sr)  =  \bot_\sigma \; \textrm{where} \; x \not\in \textrm{dom}(\sigma)\\
\end{mathpar}
\end{mdframed}
\caption{Denotational semantics for estimators}
\label{fig:estsemantics}
\end{figure}

\paragraph{\bf Lift.} A developer can lift a sampler to an estimator. The resulting estimator always returns the value $1$ as the weight of a sampler.

\vspace{-.15cm}
\paragraph{\bf Factor.} The $\progcmd{factor} e \opfmt{by} d$ modifies the weight of the estimator $e$.   

\vspace{-.15cm}
\paragraph{\bf Conditionals.} The syntax $\ite{\cons}{e_t}{e_f}$ returns the
value of the estimator $e_t$ if the constraint $\cons$ is true, and that of$e_f$ 
if the constraint is false.

\vspace{-.15cm}
\paragraph{\bf Invocation.} The syntax $\callone{x}{\hat{\argchar}}$ invokes an
estimator named $x$ that exists in the environment. The call evaluates the
estimator in an environment where the quantified variables are rebound to their
parameters $\hat{\argchar}$. 
}

\newcommand{\eststatement}{An estimator is a function that produces a sample
and corresponding weight such that the expectation of a positive function $f$
under the estimator is correct. An estimator must also preserve the model
relation on the sample portion of its output.

\begin{definition}[Estimator]
    if \etdef[\vDash]{e}{A}{B} then for all $\sigma$ such that $\sigma \vDash \Gamma, \mathcal{M}, \assumplog$, and any $f \in \Sigma_{rv} \rightarrow \mathbb{R}^+$, 
\begin{alignat*}{1}
    &\denotation{\cons}(\mathcal{M},\sigma) \Rightarrow \int_\sr \frac{\pi_1(\denotation{e}(\mathcal{M},\sigma,\sr)) * f(\pi_2(\denotation{e}(\mathcal{M},\sigma,\sr)))}{\int_{\sr} \pi_1(\denotation{e}(\mathcal{M},\sigma,\sr))} = \\
    &\int_{\denotation{A}(\mathcal{M},\sigma)} f(\sigma) * \mathcal{J}({A}|{B})(\mathcal{M},\sigma)
\end{alignat*}
and
\[
    \denotation{\cons}(\sigma) \Rightarrow \pi_1(\denotation{e}(\mathcal{M},\sigma,\sr)) \vDash \Gamma, \mathcal{M}, \assumplog
\]
\end{definition}
}

\newcommand{\estlift}{
\ensuremath{
    \inferrule[elift]{
        \stdef{s}{A}{B}\\
    } {
        \etdef{\elift{s}}{A}{B}
    }
}
}
\newcommand{\efactor}{
\ensuremath{
    \inferrule[efact]{
        \etdef{e}{A}{B}\\
        \dtdef{d}{C}{A \comma B}\\
        \mathcal{M}, \Gamma, \assumplog \vDash \mathrm{Valid}(A | B \comma C , \cons)
    } {
        \etdef{\factor{e}{d}}{A}{B \comma C}
    }
}
}

\newcommand{\esttypefigure}{
\begin{figure}
\begin{mdframed}
    \begin{mathpar}
        \estlift\\
        \efactor
    \end{mathpar}
\end{mdframed}
\caption{Type rules for estimators}
\label{fig:estrules}
\end{figure}
}

\newcommand{\esttyperules}{Figure~\ref{fig:estrules} shows the typing rules
for estimators. These give the developer the ability to to conduct {\em
likelihood weighting}. The EFACT rule requires a check that the type is valid
because it might be the case that, for instance, the set of random variables
$C$ depends on the values of random variables in $A$, which would render the
type $\basictype{\texttt{estimator}}{A}{B \comma C}{\cons}$ invalid.}

\newcommand{\estthm}{
\begin{theorem}[Estimator Soundness]\ \\
    if \etfn{\mathcal{M}}{\Gamma}{\assumplog}{e}{\hat{\qv},\hat{\delta}}{A}{B}{\cons}
    then\\
\hspace*{.5cm} \etfn[\vDash]{\mathcal{M}}{\Gamma}{\assumplog}{e}{\hat{\qv},\hat{\delta}}{A}{B}{\cons}
\end{theorem}
}

\newcommand{\opfmt}[1]{\;\mathrm{\texttt{#1}}\;}
\section{Introduction}
\label{sec:inference}

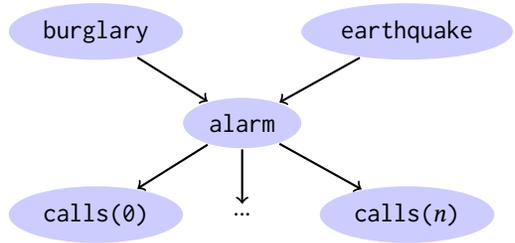
\begin{wrapfigure}[10]{r}{.5\linewidth}
\begin{tikzpicture}[
    varnode/.style={ellipse,fill=blue!20,thick},
    to/.style={->,thick}
]
    \matrix[row sep = 5mm]{
        \node[varnode] (burglary) {\tt burglary}; & &
        \node[varnode] (earthquake) {\tt earthquake};\\
        & \node[varnode] (alarm) {\tt alarm}; & \\
        \node[varnode] (callsone) {\tt calls(0)}; &
        \node[draw=none,fill=none] (dots) {...}; &
        \node[varnode] (callsn) {\tt calls($n$)};\\
    };

    \draw[to] (burglary) -- (alarm);
    \draw[to] (earthquake) -- (alarm);
    \draw[to] (alarm) -- (callsone);
    \draw[to] (alarm) -- (dots);
    \draw[to] (alarm) -- (callsn);
\end{tikzpicture}
\caption{The Burglary Bayesian Network}
\label{fig:burglary-network}
\end{wrapfigure}

Bayesian Probabilistic Inference provides a well-studied formalism for modeling
and reasoning about uncertain computations. As an example,
Figure~\ref{fig:burglary-network} presents a diagram of the classic Burglary
model~\cite{aima}. This model captures the probabilistic response of a house's
security alarm when it may be triggered by either an active burglary or an
earthquake. Furthermore, if either event triggers an alarm, a number of people
may call the authorities to report the alarm.

\paragraph{\bf Modeling.} The first step in developing a probabilistic model is
to model the state of the world with {\em random variables}.  Each node in the
diagram corresponds to a boolean random variable. Each variable is random
because it has a {\em probability distribution} associated with its values.
For example, the classic presentation of this model assumes a priori that a
burglary has a $0.1\%$ change of occurring and an earthquake has a $0.2\%$
chance of occurring. Further, each edge in the graph specifies a conditional
dependence. For example, if both a burglary and an earthquake occur, then the
alarm has a $95\%$ chance of triggering, and if neither occur, the alarm has a
$0.1\%$ chance of triggering.  Given random variables and a complete
specifications of their probability distributions, the techniques of Bayesian
probabilistic inference enable one to query and compute the answer to the
question, "what is the probability that a burglary happened given knowledge of
who called the authorities?"

Although simply stated here, Bayesian probabilistic inference has been applied
to domains such as perception, state estimation, target tracking, and data
science, where the models capture rich properties of the physical world.

\paragraph{\bf Inference.} The {\em inference} task for a probabilistic model
is to compute the probability distribution of a set of random variables in the
model, potentially {\em conditioned} on the values of the model's other random
variables. For example, the aforementioned query regarding who called the
authorities corresponds to computing the probability distribution
$P(\texttt{burglary}\; | \; \texttt{calls})$.  More formally, the inference
task for this distribution is to produce an {\em inference procedure} $f :
\mathbb{B}^n \rightarrow (\mathbb{B} \rightarrow \mathbb{R})$ where $f$ takes
as input a tuple of values for the $n$ \texttt{call} variables and produces a
function that given a value of \texttt{burglary}, returns the probability of
that value.   Moreover, we can specify the inference problem, generally, as a
traditional synthesis problem:

\begin{center} 
$\mathcal{M} \vDash  \exists f. \forall b_1, \hat{b}_2. f(\hat{b}_2)(b_1) = P(\texttt{burglary} = b_1 \, \mid \, \texttt{calls} = \hat{b}_2)$
\end{center}

The task is therefore to perform {quantifier elimination} and produce an
inference procedure $f$ that satisfies the specification that it computes the
distribution.

\subsection{Approaches to and Systems for Probabilistic Inference}

There is a rich space of systems that seek to tackle the inference problem.
These systems range from synthesis approaches to direct, handcoded
implementations.

\vspace{.10cm}
\paragraph{\bf Synthesis.} We deliberately pose the inference problem as an
synthesis problem because many systems for probabilistic inference take a
synthesis approach. Specifically, for a pre-specified class of models, systems
such as PSI \citep{psi}, Mathematica \citep{mathematica} and Maple
\citep{maple} will automatically generate an inference procedure for a given
model.

Other systems with even looser restrictions on the models will automatically
generate an inference procedure that uses approximate inference techniques
based on the Monte Carlo method, such as Markov-Chain Monte Carlo or Sequential
Monte Carlo . These techniques are approximate in that the
resulting inference procedure approximates the distribution. While these
techniques produce stochastic functions, if we were to view their resulting
inference procedures as deterministic functions, then they solve the following
modified problem:
\begin{center}
$\mathcal{M} \vDash  \exists f. \forall b_1, \hat{b}_2. \underset{n \rightarrow \infty}{\textrm{lim}}f(n, \hat{b}_2)(b_1) = P(\texttt{burglary} = b_1 \, \mid \, \texttt{calls} = \hat{b}_2)$
\end{center}

In words, these techniques produce an inference procedure that, given a
parameter $n$ that controls the amount of work that the procedure performs,
exactly computes the distribution -- as that parameter goes to infinity (i.e.,
as the procedure does more work).

\vspace{.10cm}
\paragraph{\bf Handcoded.} Libraries, such as scipy.stats \citep{scipy} and PyMC \citep{pymc}, support
solving the inference problem by hand. A primary reason why a developer may
choose to handcode an inference algorithm over using a synthesis-based system is
efficiency. Specifically, in the approximate inference procedure case, a
developer may choose a different inference technique that produces a better
estimate of the distribution with the same amount of -- or even less --
computation. For example, a developer could implement a Collapsed Monte Carlo
technique~\citep{gibbscollapsed} -- instead of a potentially less efficient general
Monte Carlo technique -- because the developer knows the dependencies between
the random variables in the model, the exact distributions behind those
dependencies, and can therefore, 1) apply analytical techniques to efficiently
solve part of the inference problem exactly and 2) solve the remainder of the
problem with an approximate technique.  

To support this development approach, these libraries offer basic probabilistic
primitives, such as computing the value of a Gaussian distribution at a point
or producing a value from a Gaussian distribution. A developer can then use
these primitives in their manual implementation of an inference procedure.

\subsection{Probabilistic Inference Programming}

Constructing efficient inference procedures by hand, alternatively {\em
probabilistic inference programming}, requires tackling several programming
idioms in the domain.

\paragraph{\bf Distributions.} The primitive objects in this programming model are
probability distributions. For example, if implementing an analytical approach
for Burglary, a developer will have -- conceptually -- a representation for the
prior distribution for \texttt{burglary} and another for the prior distribution
of \texttt{earthquake}. In a standard programming language one way to represent
these is by two functions, $\texttt{burglaryPrior} : \mathbb{B} \rightarrow
\mathbb{R}$ and $\texttt{earthquakePrior} : \mathbb{B} \rightarrow \mathbb{R}$,
respectively.

\paragraph{\bf Distribution Operations.} The next programming idiom is that
developers will compose new distributions using the representations of other
distributions. For example, if a developer would like to compute the joint
prior probability of both a \texttt{burglary} and -- simultaneously -- an
\texttt{earthquake}, then the developer can rely on the traditional rule from
probability theory that $P(A, B) = P(A) * P(B)$ and multiply together each
variable's prior distributions.
\begin{center}
\begin{verbatim}
def jointBE(b, e) : burglaryPrior(b) * earthquakePrior(e);
\end{verbatim}
\end{center}

\vspace{-.1cm}
\paragraph{\bf Sound Probability Theory.} A proviso to the developer's above
application of probability theory is the this rule is sound only if $A$ and $B$
-- alternatively, \texttt{burglary} and \texttt{earthquake} -- are
statistically independent. Namely, it must be the case that $P(A | B) = P(A)$
({and vice-versa}). In a standard programming language -- with no explicit
representation of the probabilistic model and its dependencies -- such
assertions about the model can only be informally documented as comments.

\vspace{-.1cm}
\paragraph{\bf Code Generation and Optimization.} To effectively program with the abstraction of
probability distributions and their traditional operations, the developer must
carefully translate their implementations of these operations to manage the
realizability of the computation.

For example, developers would ideally like to specify integrations. In general,
these integrations may not be tractable, but for certain models they are. In
such cases, the developer must jointly transform multiple operations into its
analytical solution.  Such transformations result in algebraically simplified
opaque blocks of code that have limited correspondence to the original
high-level operations. Such blocks are akin to the code generated by optimizing
compilers.  

\begin{figure}
\begin{subfigure}{0.45\linewidth}
\begin{wraplst*}
\begin{lstlisting}[
    language=C, escapeinside={(*}{*)},
    keywordstyle=\color{blue},
    escapeinside={(*}{*)}, 
    keywordstyle=\color{blue}, 
    numbers=left,
    frame=single
]
for(i = 0; i < N; i++) {
    s = 0;
    for(j = 0; j < i; j++)
        s += A[j];
    B[i] = s;
}
\end{lstlisting}
\end{wraplst*}
\caption{unoptimized prefix-sum}
\label{fig:prefix-sum-unopt}
\end{subfigure}\quad\quad\quad
\begin{subfigure}{0.45\linewidth}
\begin{wraplst*}
\begin{lstlisting}[
    language=C, escapeinside={(*}{*)},
    keywordstyle=\color{blue},
    escapeinside={(*}{*)}, 
    keywordstyle=\color{blue}, 
    numbers=left,
    frame=single
]
i, s = 0, 0;
for(i = 0; i < N; i++) {
    s += A[i];
    B[i] = s;
}
\end{lstlisting}
\end{wraplst*}
\caption{optimized prefix-sum}
\vspace{-.15cm}
\label{fig:prefix-sum-opt}
\end{subfigure}
\vspace{-0.5cm}
\caption{equivalent prefix-sum programs}
\label{fig:prefix-sum}
\vspace{-0.6cm}
\end{figure}

Another concern is the performance of the inference procedure's implementation.
Sample-based approximate inference techniques often generate code that needs to
be incrementalized~\citep{swift}. As a simplified
example, an inference procedure might need to collect statistics about the
input data and the natural specification of the algorithm results in prefix
sum. The algebraic formulation of prefix sum is written as $\forall i, B[i] =
\sum_{j < i} A[j]$.  Figure~\ref{fig:prefix-sum-unopt} shows this computation
directly translated into imperative code. This code has complexity quadratic to
input's length. In contrast, Figure~\ref{fig:prefix-sum-opt} presents an
equivalent optimized prefix sum that has linear complexity.   In general, these
optimizations are challenging to implement manually and -- like algebraic
simplifications -- they are hard to interpret and maintain.

While probabilistic inference programming presents an opportunity to craft a
programming domain based on the well-developed theory of probability
distributions, delivering practical implementations by hand presents multiple
challenges.

\subsection{Contributions: Shuffle}

In this paper we present Shuffle, a programming language for probabilistic
inference programming that gives developers the abstraction of probability
distributions with also supplementary support to ensure that their inference
procedures are sound, realizable, and optimized.

\vspace{-.12cm}
\paragraph{\bf Modeling} In Shuffle, a developer first writes a probabilistic
model in Shuffle's modeling language. The model's specification includes the
model's random variables, the dependencies between the random variables, and
the probability distributions that characterize these dependencies.

\vspace{-.12cm}
\paragraph{\bf Inference} The developer next writes their inference procedure
using first-class abstractions of probability distributions and
Shuffle-provided operations to compose probability distributions into new
probability distributions.  Shuffle provides the abstraction of probability
distributions where their underlying implementation and corresponding interface
is either 1) a {\em probability density function} -- a function that computes
the probability that a variable takes a value -- 2) a {\em Monte Carlo Sampler}
-- a stochastic function that produces a value of the target variables of a
probability distribution, according to their probability -- or 3) a {\em
Monte-Carlo Markov Chain Transition Kernel} -- stochastic functions that if
repeatedly iterated behave as Monte Carlo Samplers.  

Shuffle also provides a set of strongly-typed operators that are inspired by
the rules of probability theory that enable developers to compose
distributions. An inference procedure written in Shuffle composes together
distributions, with the resulting sequence of compositions corresponding to an
implementation of an inference procedure for a distribution in the model.
Shuffle's distribution abstractions are expressive enough for a developer to
write inference algorithms inference algorithms such as  variable
elimination~\cite{variableelimination}, Gibbs sampling~\cite{gibbssampling}, ,
Metropolis-Hastings~\cite{mha,mhb}, and likelihood
weighting~\cite{likelihoodweighting}.

\vspace{-.12cm}
\paragraph{\bf Type Checking} Given an inference procedure, Shuffle's type
system verifies that each distribution composition is sound with respect to the
rules of probability theory as well as the dependencies expressed in the model.
For example, if $d_1$ is a density for the distribution $\Pr(A\,| \, B,C)$,
where $A$,$B$, and $C$ are sets of random variables, and $d_2$ is a density for
the distribution $\Pr(B|C)$, then Shuffle's type system determines that the
density $d_1 \opfmt{*} d_2$ is a density for the distribution $\Pr(A,B \, |
\,C)$. This fact follows from probability theory. 

\vspace{-.12cm}
\paragraph{\bf Code Generation} Given a type-checked inference procedure,
Shuffle then automatically generates code that implements the procedure.
Shuffle's code generator 1) automatically applies algebraic simplifications to
eliminate integrals when possible, 2) automatically translates the developer's
inference procedure to compute with log probabilities, and 3) automatically
applies incremental optimizations to improve performance.

\vspace{-.12cm}
\paragraph{\bf Case Studies} We evaluate Shuffle by specifying several
probabilistic models and implementing several inference procedures.
  Our evaluation
demonstrates that Shuffle can express several inference procedures for several
models and generates code that is more efficient than Venture~\cite{venture},
another system for probabilistic inference programming. Specifically, we show
speedups of at least 3.1x on these benchmarks.

Altogether, Shuffle enables a developer to build a rich set of inference
procedures with strong guarantees as to the correctness, realizability, and
efficiency of their implementation.

\vspace{-.15cm}

\newcommand{\intd}[2]{{\texttt{int} \; {#1}  \; \texttt{by} \; {#2}}}
\section{Example: Burglary Model}
\label{sec:example}

\begin{figure}
\begin{minipage}{\linewidth}
\centering
\begin{wraplst}
\begin{lstlisting}[
language=C, escapeinside={(*}{*)},
 keywordstyle=\color{blue},
    escapeinside={(*}{*)}, 
    keywordstyle=\color{blue}, 
    morekeywords={def, domain, in, forall,variable, model, normal, uniform, density, Bool, flip}, 
    numbers=left,
    frame=single]
model {
   variable Bool burglary;(*\label{code:burglary}*)
   variable Bool earthquake;(*\label{code:earthquake}*)
   
   def burglaryPrior() : (*\label{code:burglaryPrior}*)density(burglary) = flip(burglary, 0.002);

   def earthquakePrior() : (*\label{code:earthquakePrior}*)density(earthquake) = flip(earthquake, 0.001);
   
   variable Bool alarm;(*\label{code:alarm}*)
 
   def alarmDens() : (*\label{code:alarmDens}*)density(alarm | burglary, earthquake) =
      if (earthquake == 1) {
         if (burglary == 1) { flip(alarm, 0.95) } 
         else { flip(alarm, 0.29) }
      } else {
         if (burglary == 1) { flip(alarm, 0.94) } 
         else { flip(alarm, 0.001) }
      };

   domain People;
   variable Bool[People] calls;(*\label{code:calls}*)

   def callDens(p in People) : (*\label{code:callDens}*)density(calls[p] | alarm) =
      if (alarm == 1) { flip(calls[p], 0.9) } 
      else { flip(calls[p], 0.01) }
}
\end{lstlisting}
\end{wraplst}
\end{minipage}
    \vspace{-.15cm}
    \caption{The Burglary model in {\tool}.}
\label{fig:BurglaryModel}
\vspace{-.35cm}
\end{figure}

To use Shuffle to create an inference procedure, a developer first specifies a
probabilistic model. Figure~\ref{fig:BurglaryModel} presents a Shuffle
specification of the Burglary model presented in Section~\ref{sec:inference}.
The specification gives the model's {\em random variables}, {\em prior
distributions}, and {\em conditional distributions}.

\vspace{-.15cm}\paragraph{\bf Random Variables.} Lines~\ref{code:burglary} and
~\ref{code:earthquake} declare the boolean random variables \texttt{burglary}
and \texttt{earthquake}, the values of which denote whether or not burglary or
earthquake as occurred, respectively.  These variables are {\em random} in that
they have a probability distribution associated with their values.

\vspace{-.15cm} \paragraph{\bf Distributions.}  Each \texttt{def} statement in
the model specifies a name, type, and implementation for a distribution.  For
example, the \texttt{def} statement on Line~\ref{code:burglaryPrior} specifies
the distribution for \texttt{burglary} with the name \texttt{burglaryPrior} and
type \texttt{density(burglary)}. 

The type specification is similar to the traditional probability notation
$P(\texttt{burglary})$ and therefore explicitly links this distribution's name
and its implementation to the \texttt{burglary} random variable. However,
instead of the traditional
$P(\cdot)$ notation, the type has the syntax $\texttt{density}(\cdot)$, which
denotes that the distribution is implemented as a {\em density function}.

Probability densities functions  are a common tool for specifying the
distribution of random variables.  If a random variable $X$ may take on a value
$x : A$ , then then a probability density $f : A \rightarrow R$ at $x$ (i.e.,
$f(x)$) returns a real-valued {\em score} for how likely it is that $X$ takes
on the value $x$. If $A$ is continuous (such as the real numbers), then $f(x)$
is the probability that $X$ lies in an infinitesimal region around $x$,
{divided by the size of that region.} If $A$ is a discrete type (such as the
integers), then $f(x)$ is simply the probability that $X = x$ (and is
frequently instead called a {\em probability mass function}; in this paper we
will not make a distinction). 
 The implementation on Line~\ref{code:burglaryPrior} makes use of
the \texttt{flip($v$,$p$)} primitive, which is a Shuffle primitive that returns
$p$ if its first argument $v$ evaluates to $1$, and $1-p$ otherwise.
Line~\ref{code:earthquakePrior} similarly specifies the distribution for
\texttt{earthquake} with name \texttt{earthquakePrior} and type
\texttt{density(earthquake)}.

\vspace{-.15cm} \paragraph{\bf Conditional Distributions} Line~\ref{code:alarm}
declares the boolean random variable for the occurrence of an alarm
Line~\ref{code:alarmDens} specifies its distribution. As seen in
Section~\ref{sec:inference}, the distribution of \texttt{alarm} depends on, or
is {\em conditioned} on, the values of \texttt{burglary} and
\texttt{earthquake}.  The distribution declaration on Line~\ref{code:alarmDens}
declares that the distribution of alarm is conditioned on the values of of
\texttt{earthquake} and \texttt{burglary}.  Similar to the traditional
probability notation $P(\texttt{alarm} \, | \, \texttt{burglary},
\texttt{earthquake})$, the declaration specifies that implementation has the
type $\texttt{density}(\texttt{alarm} \, | \, \texttt{burglary},
\texttt{earthquake})$, thereby noting both the conditional nature of the
distribution, but also its implementation as a density function. Note that
conditioning extends the capabilities of the density implementation in that the
conditioned variables are made available to the implementation as values that
can be inspected . In this case, \texttt{alarmDens} uses \texttt{if} statements
that test the values of these conditioned variables statements to implement its
distribution specification.

\vspace{-.15cm}
\paragraph{\bf Random Variable Sets.} The final random variable declaration on
Line~\ref{code:calls} specifies a {\em random variable set}, \texttt{calls}, where each
random variable in the set denotes whether an individual person calls the
authorities.  Variable sets enable Shuffle programs to specify first-order
models where, in the model, no a priori bound on the number of people need to
be specified.  Variable sets map an {\em index domain} to a {\em target set}.
For example the index domain of \texttt{calls} is \texttt{People}, and its
target set is the domain \texttt{Bool}.  Domains are, in general, named subsets
of the natural numbers, and the target set may be either a domain or the real
numbers.  Enforcing a finite index domain enables Shuffle's type system to
reason about variable set membership, while allowing the target set to be the
real numbers enables developers to construct models that have real-valued
random variables.

\vspace{-.15cm} \paragraph{\bf Quantified Distributions.}  The distribution
declaration on Line~\ref{code:callDens} declares that the distribution of
\texttt{calls[p]}, where \texttt{p} is a quantifier over \texttt{People}, is
conditioned on \texttt{alarm}. Quantifiers enable quantification over domains
and can appear in the type (as they do in this declaration) and also can be
referred to in the distribution's implementation.  This distribution therefore
defines the distribution for each variable in \texttt{calls} in turn.

\paragraph{\bf Summary.}

After specifying the model, the developer next specifies an inference procedure
that implements a distribution of interest.  In the burglary example, the
distribution of interest is $P(\texttt{burglary}\; | \; \texttt{calls})$. A Shuffle
developer can compute this distribution using either an {\em exact} or {\em
approximate} inference procedure. An exact inference procedure computes a
density for these distributions therefore the resulting code that implements
the distribution would have the type $\texttt{density}(\texttt{burglary}\; | \;
\texttt{calls})$.  An approximate inference procedure computes other
representations of these distributions using Shuffle's sampler and kernel
distribution abstractions.

\subsection{Exact Inference}

\begin{figure}
\begin{minipage}{\linewidth}
\begin{wraplst}
\begin{lstlisting}[
    keywordstyle=\color{blue},
    escapeinside={(*}{*)}, 
    keywordstyle=\color{blue}, 
    morekeywords={int, by, independent, ind, rec, export, max, def, domain, in, forall,variable, model, normal, uniform, density, macro}, 
    numbers=left,
    frame=single]
def alarmMarg() : density(alarm | burglary) =
   int(*\label{code:int}*) alarmDens() * (*\label{code:multiplication}*)((ind burglary) earthquakePrior())(*\label{code:independence}*)
   by earthquake;

def independent callDensI(p in People) : (*\label{code:calldensi}*)
   density(calls[p] | calls{i in People: i < p}, alarm, burglary) =
      callDens(p);

def rec callDensAll(p in People) :(*\label{code:calldensall}*)
   density(calls{i in People: i <= p} | alarm, burglary) =
      callDensI(p) * callDensAll(p-1);

def callsMarg() : density(calls | burglary) =
   int callDensAll(max(People)) * alarmMarg() 
   by alarm;

def macro bayes_rule(v,d) =(*\label{code:macro}*) d / (int d by v);

def burglaryPost() : density(burglary | calls) =
   bayes_rule(burglary, callsMarg() * burglaryPrior());
\end{lstlisting}
\end{wraplst}
\end{minipage}
\vspace{-.15cm}
\caption{Inference program for computing \texttt{density(burglary\;|\;calls)} for the Burglary model.}
\label{fig:burglary-exact}
\vspace{-.35cm}
\end{figure}

Figure~\ref{fig:burglary-exact} presents an exact inference algorithm written
in Shuffle that computes the distribution \texttt{density(burglary\;|\;calls)}
using arithmetic operations on density functions. 

Shuffle's density arithmetic
operations correspond to the rules of probability theory and the type system
enforces that the operands satisfy the statistical properties required for the
operation to be valid.

\vspace{-.15cm} \paragraph{\bf Independence.} In the definition of
\texttt{alarmMarg} in Figure~\ref{fig:burglary-exact}, we use the syntax
\texttt{(ind burglary)} on Line~\ref{code:independence} to coerce the type of
\texttt{earthquakePrior()} using an {\em independence assumption}. In this
case, we coerce the type of \texttt{earthquakePrior()},
\texttt{density(earthquake)}, to the type \texttt{density(earthquake |
burglary)} using the independence assumption \texttt{earthquake $\independent$
burglary}. {While the independence assumption is necessary for Shuffle to
soundly produce this judgment, Shuffle does not verify independence assumptions
internally. Instead, Shuffle generates a log of assumptions for the developer
to manually audit.}

\vspace{-.15cm} 
\paragraph{\bf Density Multiplication.} Shuffle enables a
developer to multiply densities using the syntax $d_1 \opfmt{*} d_2$, where
$d_1$ and $d_2$ are densities.  Line~\ref{code:multiplication} multiplies the
densities \texttt{alarmDens()} and \texttt{((ind burglary) earthquakePrior())},
producing a density of type \texttt{density(alarm, earthquake | burglary)}.
This type results from the rule of probability theory that $\Pr(A,B \, | \, C)
= \Pr(A \, | \, B, C) \cdot \Pr(B \, | \, C)$. Given that the type of
\texttt{alarmDens()} is \texttt{density(alarm | earthquake, burglary)} and the
type of \texttt{((ind burglary) earthquakePrior())} is
\texttt{density(earthquake | burglary)}, this type follows given the
equivalences $A = \texttt{alarm}$, $B = \texttt{earthquake}$, and $C =
\texttt{burglary}$. Shuffle automatically computes and checks this type.

\vspace{-.15cm} \paragraph{\bf Integration.} Shuffle also enables a developer
to {\em integrate} a density. On Line~\ref{code:int}, the developer leverages
integration to {\em eliminate}  the variable \texttt{earthquake}
from the intermediate distribution he or she computed.  Integration implements
the rule of probability $\Pr(B) = \int_A \Pr(A,B)$ using the syntax
$\intd{d}{B}$, where $d$ is a probability density and $B$ is a set of random
variables.  Line~\ref{code:int} integrates the distribution
\texttt{density(alarm, earthquake | burglary)} with respect to the variable
\texttt{earthquake} and therefore eliminates it from the resulting type of the
integration.  Because \texttt{earthquake} is a boolean random variable with
finite possibilities, Shuffle implements this integral by summing the inner
density over all possible values of \texttt{earthquake}, namely $0$ and $1$.

\vspace{-.15cm} \paragraph{\bf Constrained Variable Sets.} Shuffle also enables
a developer to specify types that include constrained relations between
variables. The definition of \texttt{callDensI} on Line~\ref{code:calldensi} is
a quantified density that computes the density of \texttt{call[p]} conditioned,
\texttt{alarm}, and \texttt{burglary} (as before), but also conditioned on all
values \texttt{call[i]} where \texttt{i} is less than \texttt{p}. Constrained
sets enable developers to specify both structured and dynamically changing
dependencies because constraints can depend on either quantifiers (as here) or
even the value of random variables. 

The developer provides an implementation here through an independence
assumption. The \texttt{independent} annotation on the definition operates as
an independence assumption where the set of independent variables is given by
the difference between those in the specified type and that computed by the
type system for the term.  

\vspace{-.15cm} \paragraph{\bf Recursion.} {The definition of
\texttt{callDensAll} on Line~\ref{code:calldensall} demonstrates Shuffle's {\em
recursive procedures}. This definition computes the joint density of all random
variables in the \texttt{calls} collection by computing the product of
\texttt{callDensI(p)} with the recursively defined \texttt{callDensAll(p-1)}.
The effect on the type is an inductive proof. Specifically, assuming that
invocations of {\tt callDensAll} have the annotated type within the body of
\texttt{callDensAll}, Shuffle verifies that the body also has this type.
Shuffle specifies default base cases for all of its objects. In this case,
Shuffle defines that {\tt callDensAll(i-1)}, where \texttt{i} is the smallest
value in the domain \texttt{People}, yields the value $1$. This is because {\tt
callDensAll}'s type, {\tt density(calls\{i in People:\ i <= p\} | alarm,
burglary)}, specifies uncertainty for an empty set of variables when \texttt{i}
falls below the minimum value in {\tt People}. The constant function returning
$1$ is always a correct density for an empty set of random variables.}

\paragraph{\bf Result.}  In the last step of this inference program, the
developer computes the posterior distribution \texttt{density(burglary |
calls)}.  Given the intermediate density calculation steps outlined above, the
developer accomplishes with an application of {\em Bayes' rule} from
probability theory. This rule states that $\Pr(A|B) = \frac{\Pr(A,B)}{\int_B
\Pr(A,B)}$. In Shuffle, a developer can represent this rule with a macro
(Line~\ref{code:macro}) that takes as input a density \texttt{d} and a set of
random variables \texttt{v}. If \texttt{d} has the type
\texttt{density($A, v \,|\, B$)}, where $A$ and $B$ are any sets of random
variables, then the result of \texttt{bayes\_rule(d,v)} has the type
\texttt{density($A \,| \, v, B$)}.

Given these set of definitions, Shuffle automatically translates these
definitions to Python code. In this specific example, Shuffle translates
integrals to summations (because the domains of these variables are discrete).

\subsection{Approximate Inference}

\begin{figure}
\begin{wraplst}
\begin{lstlisting}[ 
    keywordstyle=\color{blue},
    escapeinside={(*}{*)}, 
    keywordstyle=\color{blue}, 
    morekeywords={def, domain, in, independent, ind, int, by, density, sampler, kernel, variable, model, normal, uniform, sample, lift, fix}, 
    numbers=left,
    frame=single]
def independent earthquakeMarg() :
   density(earthquake | alarm, burglary, calls) =
   bayes_rule(earthquake, alarmDens() * (ind burglary) earthquakePrior());

def independent burglaryMarg() :
   density(burglary | earthquake, alarm, calls) =
   bayes_rule(burglary, alarmDens() * (ind earthquake) burglaryPrior());

def alarmMarg() : density(alarm | burglary, earthquake, calls) =
   bayes_rule(alarm, (ind burglary, earthquake) 
                      callDensAll(max(People)) * alarmDens());

def abeKernel() : kernel(alarm, burglary, earthquake | calls) =
   lift {(*\label{code:lift}*)
      alarm := sample alarmMarg();(*\label{code:alarmSample}*)
      burglary := sample burglaryMarg();(*\label{code:burgSample}*)
      earthquake := sample earthquakeMarg()(*\label{code:earthSample}*)
   };(*\label{code:lift-end}*)

def abePost() : sampler(alarm, burglary, earthquake | calls) =
   fix abeKernel;(*\label{code:fix}*)

\end{lstlisting}
\end{wraplst}
\vspace{-.15cm}
\caption{Approximate Inference for the Burglary model.}
\label{fig:burglary-approx}
\vspace{-.35cm}
\end{figure}

Distribution implementation methods that rely purely on computations of density
functions are designed to compute the distribution exactly. An alternative
implementation approach is to approximate the distribution. For example, while
the integrals in Figure~\ref{fig:burglary-exact} are efficiently computable,
other models may not admit tractable integrations and, therefore, may need to
be approximated. Sample-based {\em approximate inference} inference procedures,
such as Gibbs Sampling~\citep{gibbssampling} and Metropolis-Hastings~\citep{mha,mhb},
have been designed to perform inference in such models  . These
inference procedures, called {\em Monte Carlo samplers},  randomly produce
samples from the desired distribution.  The probability that an implementation
produces a given value approximates the probability of that value according to
the specified distribution.  As a result, a client of such a distribution can
use the {\em Monte Carlo method}  to ask questions about a
distribution. Specifically, the developer can phrase a question as a function
$f(x) : X \rightarrow \mathbb{R}$ and then compute the expectation $\int_x f(x)
* P(x)$ approximately by $\sum_i f(x_i)$, where each $x_i$ is a sample from the
distribution.

\paragraph{\bf Samplers.}  In Shuffle, {\em Monte Carlo Samplers}, or simply
{\em samplers}, are imperative procedures that take as input a concrete state
that includes values for all of the model's random variables and produces a
modified state. The probability that a sampler produces a given state
approximates the probability of that state -- i.e., the values of modified
variables -- according to the specified distribution.
Figure~\ref{fig:burglary-approx} presents a Gibbs Sampling implementation for
the posterior distribution $P(\texttt{alarm}, \texttt{burglary},
\texttt{earthquake} \, | \, \texttt{calls})$ as a sampler.   On
Line~\ref{code:alarmSample}, the developer assigns \texttt{alarm} to a random
value drawn according to the distribution \texttt{alarmMarg()}.  Because
\texttt{alarmMarg()} has the type \texttt{density(alarm | burglary, earthquake,
calls)}, the sample statement on that line is a sampler with the type
\texttt{sampler(alarm | burglary, earthquake, calls)}. Specifically, the
statement produces values of \texttt{alarm} conditioned on the values of
\texttt{burglary}, \texttt{earthquake}, and \texttt{calls}. Because the type is
conditioned, the sampler may read the values of \texttt{burglary},
\texttt{earthquake}, and \texttt{calls} while producing its value. Further, a
sampler can only modify the target variables specified in its type. Therefore
this sampler only modifies \texttt{alarm}. Shuffle's typesystem verifies that a
sampler's imperative implementation is consistent with the specification in its
type.

\vspace{-.15cm}
\paragraph{\bf Markov-Chain Transition Kernels.}  The \texttt{lift} block
between Lines~\ref{code:lift}-\ref{code:lift-end} composes together multiple
samplers on this inference procedure's overall path towards producing a
sampler. A key insight into understanding this procedure and its use of
composition is the types of \texttt{alarmMarg}, \texttt{burglaryMarg}, and
\texttt{earthquakeMarg}.  Specifically, the sampler on
Line~\ref{code:alarmSample} produces a sample of \texttt{alarm} conditioned on
the value of \texttt{burglary} (and other variables of the model), the sampler
on Line~\ref{code:burgSample} produces a sample of \texttt{burglary}
conditioned on the new value of \texttt{alarm} (as well as other variables of
the model), and the sampler on Line~\ref{code:earthSample} produces a sample of
\texttt{earthquake} conditioned on the new values of \texttt{alarm} and
\texttt{burglary}.  Because the {target variables} and conditioned variables
of these three samplers intersect, the values of these random variables after
executing all three samplers can be correlated in a way that does not result in
a sampler for the desired distribution:   the joint distribution of $P(\texttt{alarm}, \texttt{burglary},
\texttt{earthquake} \, | \, \texttt{calls})$. 

However, if the developer were to repeatedly apply this block, then -- in the
limit -- that iteration process does produce a correct sample-based
implementation.  A distribution implementation for which repeated iteration
produces a sampler is called a {\em Markov-Chain transition
kernel} .  Therefore, the difference between a sampler and a
kernel is that a sampler directly produces a sample from a distribution whereas
a kernel converges to a sampler under repeated composition with itself.

Shuffle supports transition kernel implementations through its \texttt{kernel}
type. Kernels have a weaker correctness condition which means they have more
flexible composition rules.  Shuffle developers construct kernels using the
\texttt{lift} syntax, as on Line~\ref{code:lift} and the checker computes that
the block lift on Line~\ref{code:lift} has the type \texttt{kernel(alarm,
burglary, earthquake | calls)}.

\vspace{-.15cm} \paragraph{\bf Fixpoints.}  The final step towards producing a
sampler implementation of the posterior distribution is to follow the approach
suggested by the previous paragraph: iterate the kernel implementation until it
reaches a fixpoint.  Line~\ref{code:fix} uses the \texttt{fix} operator, which
denotes the fixpoint of a distribution's kernel implementation. The result this
operator is a sampler for the distribution.

We have designed Shuffle's type system such that all fixpoints can be computed
by repeated iteration and therefore Shuffle generates code for \texttt{fix}
that performs iteration. Shuffle's guarantees hold in the limit and it is the
responsibility of the developer to set the parameters of the generated
implementation that control the number of iterations during execution.
Automatically determining the number of iterations for a Markov-Chain Monte
Carlo algorithm is an undecidable problem.   However,
developers can use a variety of profiling techniques to aid setting these
parameters~\citep{diagnostics}.

\begin{figure}
\centering
\begin{minipage}{.7\textwidth}
\begin{wraplst}
\begin{lstlisting}[keywordstyle=\color{blue},morekeywords={def,True,range, for, if, then, else, in, return},  numbers=left, frame=single]
def bApprox(calls) :
   sum = 0
   for i in range(num_samples):
      (alarm, burglary, earthquake) = abePost(calls)
      sum += (1 if (burglary == True) else 0)
   return sum / num_samples
\end{lstlisting}
\end{wraplst}
\caption{Python code for estimating the probability that a burglary occurs using extracted \texttt{eabPost}.}

\vspace{-.35cm}
\label{fig:burglary-est}
\end{minipage}\;
\begin{minipage}{.28\textwidth}
  \centering
  \vspace{-.1cm}
  \includegraphics[width=0.7\linewidth]{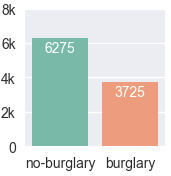}
  \vspace{-.4cm}
  \caption{approximate inference samples}
  \label{fig:burglary-est-plot}

\vspace{-.35cm}
\end{minipage}
\end{figure}

\vspace{-.15cm} 
\paragraph{\bf Summary.} Together, Shuffle's abstractions for densities,
samplers, and kernels enable developers to compose inference procedures with
strongly-typed abstractions and generate efficient code. For example,
Figure~\ref{fig:burglary-est} presents an example of how a developer would use
a sampler generated by Shuffle within a Python program.  Specifically, by
repeatedly calling \texttt{abePost} to produce a stream of samples from the
distribution of $P(\texttt{alarm, burglary, earthquake}\;|\;\texttt{calls})$,
the function \texttt{bApprox}  computes the expectation of the indicator
function, \texttt{burglary == True}, which is therefore an approximation of the
probability that a burglary occurs.  Figure~\ref{fig:burglary-est-plot} shows a
result of approximate inference when 3 out of 10 people call. The plot
illustrates that out of 10k samples we obtained by calling \texttt{abePost},
3725 samples had True for the value of \texttt{burglary}.

\section{Language}

We next present the syntax of models, types, and inference procedures in
Shuffle. For clarify of presentation, we elide a complete presentation of the
syntax of Shuffle's modelling language.  However, the examples in
Section~\ref{sec:example} are representative in that a probabilistic model
defines the model's domain of values, the model's set of random variables, and
the probability densities that relate them. A {\em domain declaration},
specifies a domain $\delta \in \Delta$. A {\em variable declaration}, specifies
a random variable $v \in V$.  A random variable is an array and a domain
$\delta$ specifies its {\em index space}.

\begin{wrapfigure}{l}{.6\textwidth}
\begin{mdframed}
\begin{center}
$n \in \mathbb{N}, \; r \in \mathbb{R}, \; x \in X, \; v \in V, \; \qv \in \qvspace, \; \delta \in \Delta$
\end{center}
\vspace{-.15cm}
\begin{align*}
       T_x  & \rightarrow \; T_b \texttt{(} V_g \mid V_g \texttt{,} \; \cons \texttt{)} \\
       T_b  & \rightarrow \; \texttt{density} \; \mid \; \texttt{sampler} \; \mid \; \texttt{kernel} \; \mid \; \texttt{estimator} \\\\[-.75em]
       V_g & \rightarrow \;  v \texttt{[} A\texttt{]} \; \mid \; \varset{v}{\qv}{\delta}{\cons} \; \mid \; V_g \texttt{,} V_g   \\
        \cons   & \rightarrow \;  A \; \texttt{==}  \; A \; \mid \;  A \; \texttt{<} \; A  \; \mid \; \neg \cons \; \mid \; \cons \; \texttt{\&\&} \; \cons  \\
  A  & \rightarrow \; n \; \mid q \; \mid \;\; \qv \opfmt{-} n \; \mid \; V_s \; \mid \; \texttt{min(} \delta \texttt{)} \; \mid \; \texttt{max(} \delta \texttt{)}
\end{align*}
\end{mdframed}
    \vspace{-.15cm}
\caption{The syntax of {\tool} types}
\label{fig:models}
\end{wrapfigure}

\subsection{\bf Types} 

Figure~\ref{fig:models} presents the syntax Shuffle's
types.  The language of types, $T_x$, denotes that a object is either  a
\texttt{density}, \texttt{sampler}, \texttt{kernel}, or \texttt{estimator} that
computes the probability of a set of random variables conditioned on another
set of random variables, while subject to a constraint on the conditioned
random variables. The random variables within either set may be either a single
random variable from an array of random variables, $v[A]$, or a constrained
subset of the random variables within an array,
$\varset{v}{\qv}{\delta}{\cons}$.  The variable set $v$ is syntactic sugar for
the set $\varset{v}{\qv_0}{\delta}{\texttt{true}}$.

A constraint, $\cons$, that appears in either a type or a random variable
subset notation is a boolean combination of (in)equalities over 1) integers, 2)
quantifier variables from {domains that are isomorphic to the integers}, 3)
a single random variable with a value from a {domain that is isomorphic to
the integers}, and 4) the minimum or maximum elements of a domain.


\begin{wrapfigure}[17]{L}{.6\linewidth}
\begin{mdframed}
\begin{align*}
    P   & \rightarrow \; \Big(\defsi{K_d}{x}{\qv \opfmt{in} \delta}{T_x}{\beta}\Big)^+\\
    K_d & \rightarrow \; (\texttt{independent} \mid \texttt{rec})^? \\\\[-.75em]
   \beta & \rightarrow D \; \mid \; S \; \mid \; K \mid \; E \\\\[-.75em]
   D     & \; \rightarrow \;x \texttt{(} \textit{A} \texttt{)} \; \mid \; D \; \texttt{*} \; D \; \mid \; D \; \texttt{/} \; D \\
         & \;\; \mid \;\; \intd{D}{V_g} \; \mid  \; \ite{\cons}{D}{D} \; \mid \; \texttt{1.0} \\\\[-.75em]
        S   & \rightarrow \; x \texttt{(} \textit{A} \texttt{)} \; \mid \; v \texttt{[} A\texttt{]}   \;\texttt{:=} \; \texttt{sample} \; D  \\
            & \; \mid \;\;\; \texttt{return} \; \mid \; \seq{S}{S} \mid \; \ite{\cons}{S}{S} \; \mid \; \fix{K}\\\\[-.75em]
    K & \rightarrow \; x \texttt{(} \textit{A} \texttt{)} \; \mid \; \lift{S} \\ & \; \mid \; \ite{\cons}{K}{K} \; \mid \; \seq{K}{K} \; \mid \; \texttt{return}
\end{align*}
\end{mdframed}
\caption{The Syntax of {\tool} Inference Procedures}
\label{fig:programs}
\vspace{-.5cm}
\end{wrapfigure}

\subsection{Inference Procedures}
Figure~\ref{fig:programs} presents the syntax of Shuffle's inference procedures. As a simplification for presentation purposes, each definition admits one quantifier, but our implementation permits multiple quantifiers per definition.

\paragraph{\bf Densities.} A probability density, $D$, defines a probability
distribution through density operators. An inference procedure may call an an
atomic density arguments $x \texttt{(} A \texttt{)}$, a multiply two densities,
$D_m \; \texttt{*} \; D_m$,  divide a density by another density, $D_m \;
\texttt{/} \; D_m$, {\em integrate} a density using the syntax $\intd{D}{V_g}$
or conditionally switch between densities. Each of these operators has a
corresponding well-defined semantics in probability theory with respect to its
operands.

\vspace{-.15cm}
\paragraph{\bf Samplers.} A sampler, $S$, defines a probability distribution
through {\em sampler operators}. Sampler operators include sampling a value
from a density, $\samp{v[A]}{D}$, sequentially composing two
samplers together, $S \opfmt{;} S$, and computing the fixed
point of a kernel, $\mathrm{\tt fix} \; K$.

\vspace{-.15cm}
\paragraph{\bf Kernels.} A kernel, $K$, defines a probability distribution in
terms of {\em kernel operators}. Kernel operators include lifting a sampler,
$\lift{S}$, and composing two kernels together with the syntax $K \opfmt{;} K$.

\section{Semantics}
\label{sec:semantics}

\label{sec:model-semantics}

A Shuffle model $\mathcal{M} \in \mathbb{M} = (\Delta \rightarrow (\mathbb{N} \times
\mathbb{N})) \times (V \rightarrow (\Delta \times (\Delta + \{\texttt{R}\}))) \times
H)$ consists of its domain declarations, random variable declarations, and density
definitions, respectively.

\vspace{-.15cm}
\paragraph{\bf Domain Declaration.} We map the set of domain definitions in the
syntactic specification of the model to a function $f : \Delta \rightarrow
(\mathbb{N} \times \mathbb{N})$ that maps domain names to a pair, consisting of
the domain's minimum and maximum elements, respectively.
 We denote the semantics of a domain
$\delta$ by the set of integers between (inclusive) its minimum and maximum
values.
\begin{center}
$\denotation{\delta}(\mathcal{M}) = \{ n \mid \pi_{1}(\pi_{1}(\mathcal{M})(\delta)) \leq n \leq
\pi_{2}(\pi_{1}(\mathcal{M})(\delta)) \}$
\end{center}
We denote the the minimum and maximum of a domain $\delta$ by their definitions in the declaration:
\begin{mathpar}
    \denotation{\mathrm{\texttt{min(}}\delta\mathrm{\texttt{)}}}(\mathcal{M}) =
    \pi_1(\pi_1(\mathcal{M})(\delta))
    
    \denotation{\mathrm{\texttt{max(}}\delta\mathrm{\texttt{)}}}(\mathcal{M}) =
    \pi_2(\pi_1(\mathcal{M})(\delta))
\end{mathpar}

\paragraph{\bf Random Variable Declaration.} We map the set of random variable
declarations in the model to a function $f : V \rightarrow (\Delta \times
(\Delta \cup \mathrm{\tt R}))$ that maps random variable names to a pair
consisting of the name of the random variable's index domain and the name of
its target set, respectively.

\paragraph{\bf Density Definition.} We map a density definition in the syntatic
specification of the model to a tuple $\eta \in H = (X \times D_m \times (Q
\times \Delta) \times V_g \times V_g \times \Phi$ that consists of the syntax
of the density function, a list of pairs of quantifier names and their
associated domain names, a set of random variables denoting the density's
target variables, a set of random variables denoting the density's conditioned
variables, and a predicate $\phi$ denoting the density function's condition of
application.

\subsection{Preliminaries}

\paragraph{\bf Errors.} A Shuffle inference procedure may produce one of two
error values instead of a conventional value: 1) a procedure produces the error
value $\bot_\sigma$ if and only if it requires access to an element of the
environment that is not within the environments domain and 2) a procedure
produces the error value $\bot_0$ if and only if it contains a division by 0.
In the semantics below we use $\bot = \{\bot_\sigma, \bot_0\}$ to refer to the 
domain of errors and elide explicit failure propagation rules.  

\vspace{-.15cm}
\paragraph{\bf Environments.} An environment, $\sigma \in \Sigma = (V \times
\mathbb{N}) + \qvspace + X \rightarrow (\mathbb{R}^+ + \mathbb{N} + (Q,\Delta,\beta))$ is a finite
map from random variables, quantifier variables, and bound distributions to
their respective values. The notation $\sigma(a)$ denotes the value to which
$a$ is mapped by $\sigma$, which can either be 1) a random variable $(v,n)$
where $n$ is a natural number 2) a quantifier variable $\qv$ or 3) a named
distribution $x$. We use the notation $\sigma[a \mapsto b]$ to mean $a$ with
$a$, which could be any of the above, remapped to $b$.

\vspace{-.15cm} 
\paragraph{\bf Variables.} Our formalization relies on several
disjoint variable spaces. A {\em quantifier variable} $\qv \in \qvspace$ is
drawn from the space $\qvspace$; a {\em named distribution} $x \in X$ drawn
from $X$, the space of distribution names; a {\em random variable} $v \in V$ is
drawn from $V$, the space of variable names; and a {\em domain} $\delta \in
\Delta$ is drawn from $\Delta$, the space of domain names.

\vspace{-.15cm}
\paragraph{\bf Variable Sets.} A {\em variable set}  is a comma delimited list
of random variables ($V_g^+$ in Figure~\ref{fig:models}) that we denote by the
symbols $A$, $B$, and $C$.  We specify the semantics of a variable set by the
semantic function $\denotation{A} : \mathbb{M} \times \Sigma \rightarrow
\mathcal{P}(\mathcal{V})$ where $\mathcal{V} = V \times \mathbb{N}$.  The
denotation of a variable set is therefore a set of pairs that each consist of a
random variable name and the corresponding index within that variable. For each
syntactic form, we give variable sets the following denotation:

\begin{itemize}

\item {\bf Set Comprehensions.} For variable sets of the form $A =
    \varset{v}{\qv_0}{\delta_1}{\cons}$, let $\denotation{A}(\mathcal{M},\sigma) = \{ (v, n)
        \; \mid \;  n \in \denotation{\delta}(\mathcal{M}) \wedge \denotation{\cons}(\mathcal{M},\sigma[\qv_0 \mapsto n]) \}$.

\item {\bf Indexed Variables.} The single variable $\var{v}{a}$, in the context
of a variable set is
syntactic sugar for the set $\varset{v}{\qv_0}{\delta}{\qv_0 \opfmt{==} a}$, with
the corresponding denotation given by that for set comprehensions.

\item {\bf Union.} The comma operator $A, B$ unions two {\em
disjoint} variable sets. Namely, the denotation of this operator is 
$\denotation{A, B}(\mathcal{M}, \sigma) = 
     \begin{cases}
         \denotation{A}(\mathcal{M}, \sigma) \cup \denotation{B}(\mathcal{M},\sigma)  & \denotation{A}(\mathcal{M},\sigma) \cap \denotation{B}(\mathcal{M},\sigma) = \emptyset \\
    \bot_\sigma & \mathrm{\textbf{else}}\\
    \end{cases}
$

\end{itemize}

\vspace{-.15cm}
\paragraph{\bf Source of Randomness.}  A {\em source of randomness}, denoted by
``$\sr \in \mathrm{SR} = [0,1]^+$'' is a sequence of uniform distributed values
on the interval $[0,1] \subset \mathbb{R}^+$. Let the notation $\int_\sr
f(\sr)$ denote an integral over each element of this sequence.  We assume the
existence of a function $\mathrm{split} \in \mathrm{SR} \rightarrow \mathrm{SR}
\times \mathrm{SR}$ that produces two identical sources of randomness from one,
such that $\int_{\pi_1(\mathrm{split}(\sr))} f(\pi_1(\mathrm{split}(\sr))) =
\int_{\pi_2(\mathrm{split}(\sr))} f(\pi_2(\mathrm{split}(\sr)))$ for any
positive measurable function $f$.

\subsection{Definitions}

\begin{wrapfigure}[11]{r}{.7\linewidth}
\vspace{-.4cm}
\begin{mdframed}[leftmargin=.1cm,innerleftmargin=.1cm, rightmargin=.1cm,innerrightmargin=.1cm]
\begin{align*}
    \denotation{\defs{x}{\qv \opfmt{in} \delta}{t_x}{\prog_1}}(\mathcal{M}, \sigma) & = & \sigma[x \mapsto (\qv, \prog_1)] \\
    \denotation{\defsi{\texttt{rec}}{x}{\qv \opfmt{in} \delta}{t_x}{\prog_1}}(\mathcal{M}, \sigma) & = & \sigma[x \mapsto (\qv, \mathrm{default}(\prog_1))]\\
    \denotation{p_1 \opfmt{;} p_2}(\mathcal{M}, \sigma) & = & \denotation{p_2}(\mathcal{M}, \denotation{p_1}(\mathcal{M}, \sigma))
\end{align*}
\begin{align*}
    \mathrm{default}(d) & = & \small{\ite{\qv_0 \opfmt{< min(} \delta_0 \; \texttt{)}}{\opfmt{1.0}}{d}} \\
    \mathrm{default}(s) & = & \small{\ite{\qv_0 \opfmt{< min(} \delta_0 \; \texttt{)}}{\opfmt{return}}{s}} \\
    \mathrm{default}(e) & = & \small{\ite{\qv_0 \opfmt{< min(} \delta_0 \; \texttt{)}}{\opfmt{(return,1.0)}}{e}}
\end{align*}
\end{mdframed}
\vspace{-.15cm}
\caption{Semantics of  definitions}
\label{fig:defsemantics}
\end{wrapfigure}
\vspace{-.15cm}

The semantics of the syntax $\defs{x}{q \opfmt{in} \delta}{t_x}{\prog_1}$ is a
new environment with $x$ bound to the procedure $\prog_1$ with the quantifier
$\qv$.  

\vspace{-.15cm}
\paragraph{\bf Recursive Definitions.} A recursive definition denoted
$\defsi{\texttt{rec}}{x}{\qv \opfmt{in} \delta}{t_x}{\prog_1}$ is the same as
that of an ordinary definition, except that recursive invocations refer to a
modified $\prog_1$ that define {\em default base cases} when the argument of
the procedure $\qv$ is out of its domain. The default base case for densities
returns the constant $1$; the default base case for samplers and kernels leaves
the environment unchanged.   Default base
cases enable developers to write programs that type check, as Shuffle's type
system cannot reason about arbitrary base cases.

\subsection{Densities}

Figure~\ref{fig:denssemantics} presents the denotation of a density. The
denotation of a density $d$, denoted by $\denotation{d} \in (\mathbb{M} \times
\Sigma) \rightarrow (\mathbb{R}^+ + \bot)$, is a function from an environment
to a positive real number or an error value. Multiplication, division, and
conditionals have standard semantics.

\begin{wrapfigure}[14]{r}{.7\linewidth}
\begin{mdframed}
\begin{align*}
    \denotation{d_1 \opfmt{*} d_2}(\mathcal{M}, \sigma) & = & \denotation{d_1}(\mathcal{M}, \sigma) * \denotation{d_2}(\mathcal{M}, \sigma)\\
    \denotation{d_1 \opfmt{/} d_2}(\mathcal{M}, \sigma) & = & \begin{cases}
    \frac{\denotation{d_1}(\mathcal{M}, \sigma)}{\denotation{d_2}(\mathcal{M}, \sigma)} & \denotation{d_2}(\mathcal{M}, \sigma) \neq 0\\
    \bot_0 & \mathrm{\textbf{else}}\\
    \end{cases}\\
    \denotation{\intd{d}{V_g}}(\mathcal{M}, \sigma) & = & \int_{\denotation{V_g}(\mathcal{M}, \sigma)} \denotation{d}(\mathcal{M},\sigma) \\
    \denotation{\callone{x}{\argchar}}(\mathcal{M}, \sigma[x \mapsto (\qv, \prog)]) & = &  \denotation{\prog}(\mathcal{M}, \sigma[\qv \mapsto \denotation{\argchar}(\sigma)])\\[.4em]
    \denotation{\callone{x}{\argchar}}(\mathcal{M}, \sigma) & = & \bot_\sigma \; \textrm{where} \; x \not\in \textrm{dom}(\sigma)\\
    \denotation{\ite{\cons}{d_1}{d_2}}(\mathcal{M}, \sigma) &=& \begin{cases}
        \denotation{d_1}(\sigma) &  \hphantom{\neg} \denotation{\cons}(\mathcal{M}, \sigma) \\
        \denotation{d_2}(\sigma) &  \neg \denotation{\cons}(\mathcal{M}, \sigma) \\
    \end{cases}
\end{align*}
\end{mdframed}
\vspace{-.25cm}
\caption{Semantics of densities}
\label{fig:denssemantics}
\end{wrapfigure}

\vspace{-.15cm}
\paragraph{\bf Integration.} An expression $\progcmd{int} d \opfmt{by} V_g$
computes the integral of a probability density, $d$. It computes the integral
of its density parameter over all possible values of the random variables,
$V_g$.

\vspace{-.15cm}
\paragraph{\bf Invocation.} An expression $\callone{x}{\argchar}$ invokes a
density the $x$. The invocation evaluates the density in an environment where
the quantifier variable is rebound to its parameter $\argchar$.

\subsection{Samplers}

Figure~\ref{fig:sampsemantics} presents the denotation of a sampler.  The
denotation of a sampler $s$, denoted by the semantic function $\denotation{s}
\in (\mathbb{M} \times \Sigma \times \mathrm{SR}) \rightarrow (\Sigma + \bot)$,
is a function that takes an environment and a source of randomness, and
produces a new environment or an error value. The new environment will have one
to new values for its target variables that are randomly chosen according to
the sampler's distribution and the value of the source of randomness.

\begin{wrapfigure}[12]{r}{.7\linewidth}
\begin{mdframed}
\begin{align*}
    \denotation{\samp{\var{v}{a}}{d}}(\mathcal{M}, \sigma,\sr) & = &\hspace{-5cm} \sigma[(v,\denotation{a}(\mathcal{M}, \sigma)) \mapsto r] \\[.25em]
    & & \hspace{-3cm} \textrm{where} \; \textrm{InverseTransform}(\mathcal{M}, v, a, d, \sigma, \sr, r ) \\[.25em]
    \denotation{\callone{x}{\argchar}}(\mathcal{M}, \sigma[x \mapsto (\qv, \prog)],\sr) & = & \denotation{\prog}(\mathcal{M}, \sigma[\qv \mapsto \denotation{\argchar}(\sigma)],\sr)\\[.25em]
    \denotation{s_1 \opfmt{;} s_2}(\mathcal{M}, \sigma, \sr) & = & \denotation{s_2}(\mathcal{M}, \denotation{s_1}(\mathcal{M}, \sigma, \sr^2),\sr^1)\\[.25em]
    & & \hspace{-3cm} \textrm{where} \; \sr^1 = \pi_1(\mathrm{split}(\sr)),  \sr^2 = \pi_2(\mathrm{split}(\sr)) 
\end{align*}
\end{mdframed}
\caption{Semantics of samplers (abbreviated)}
\label{fig:sampsemantics}
\end{wrapfigure}

\paragraph{\bf Sampling.} A statement of the form $\samp{\var{v}{a}}{d}$
samples from the density $d$. The sampler updates $\sigma$ so that the mapped
value of $(v,\denotation{a}(\sigma))$ is overwritten with the newly sampled
value. We specify the denotation of the sample command via {\em inverse
transform sampling}. Inverse transform sampling chooses a value $r$ such that
the integral of $d$ on the region $(-\infty,r]$ is equal to a uniform random
value from the source of randomness. In cases where $r$ is discrete, an inverse
transform sampler rounds up so that the uniform random value is smaller than
this integral:  \[{\small
\mathrm{InverseTransform}(\mathcal{M},v,a,d,\sigma,\sr,r) = \arg \min_r
\Big(\int_{x \in (-\infty, r]} \denotation{d}(\sigma[(v,\denotation{a}(\sigma))
\mapsto x])\Big) > \pi_1(\sr) }\]  

\vspace{-.15cm}
\paragraph{\bf Invocation.} An expression $\callone{x}{\argchar}$ invokes the
sampler $x$. The invocation evaluates the sampler in an environment where the
quantifier variable is rebound to its parameter $\argchar$.

\subsection{Kernels}

Figure~\ref{fig:kernsemantics} presents the denotation of a kernel.  The
denotation of a kernel $k$, written $\denotation{k} \in (\mathbb{M} \times
\Sigma \times \mathrm{SR}) \rightarrow (\Sigma + \bot)$, is a function that
takes an environment and a source of randomness, and produces a new environment
or an error value.  The semantics of kernel composition, invocation, and
{conditionals} are the same as that for samplers.

\begin{wrapfigure}[6]{r}{.7\linewidth}
\begin{mdframed}
\[
\begin{array}{lcl}
    \denotation{\lift{s}}(\mathcal{M}, \sigma) & = & \denotation{s}(\mathcal{M}, \sigma, \sr) \\[.7em]
   
    \forall f. \int_\sr f(\denotation{\progcmd{fix} k}(\mathcal{M}, \sigma, \sr)) & = &
    \int_{\sr} f(\denotation{k; \progcmd{fix} k}(\mathcal{M}, \sigma, \sr))\\
  
\end{array}
\]
\end{mdframed}
\caption{Semantics of kernels (abbreviated)}
\label{fig:kernsemantics}
\end{wrapfigure}

\vspace{-.15cm}
\paragraph{\bf Lift.} A developer can lift a sampler to a kernel. The resulting kernel has exactly the same behavior as the original sampler, and is used to represent the same distribution.

\vspace{-.15cm}
\paragraph{\bf Fixed Point.} For a given kernel for a distribution, a developer
can produce a sampler for the same kernel via the \texttt{fix} operator. The
denotational semantics of \texttt{fix} are declarative, as
Figure~\ref{fig:kernsemantics} specifies that the operator must have the
property that the sampled distribution is invariant under composition with the
kernel.  Shuffle type checks its code assuming an exact implementation of
\texttt{fix}, but generates code that approximately implements it by running
the kernel repeatedly in an iterative process. As the number of iterations
grows large, the approximate distribution approaches the true distribution.

\section{Type System}
\label{sec:types}

A typing judgment in {\tool} is a logical proposition of the form
\(\typegeneric{\mathcal{M}}{\Gamma}{\assumplog}{\prog}{t}\) where $\mathcal{M}$
is a {\em model}, $\Gamma$ is a {\em type environment}, $\assumplog$ is an {\em
assumption log}, $\prog$ is a {\tool} inference program, and $t$ is a type.
For example, the type judgment $\dtdef{\prog}{A}{B}$ states that the term
$\prog$ is a density for the conditional distribution \(\Pr(A|B)\), provided
that \(\cons\) is true.  In this section we present {\tool}'s type system,
including the semantics of the model, assumption log, and type environment

\newcommand\smdots{\hbox to 1em{.\hss.\hss.}}

\subsection{Model}
\label{sec:model-semantics}

We next give the model a semantics. Specifically, we define the semantics of
the model by the {\em joint density} of the model's variables. The joint
density computes the probability that the model's variables all together take
on a set of prescribed values.  We define the model's joint probability
density, $\mathcal{J}$, via the model's density function
definitions as follows:
\begin{center}
    $\mathcal{J}(\mathcal{M}, \sigma) = \prod_{(x, d, (q,\delta),A, B, \phi) \in
\pi_{3}(\mathcal{M})} \denotation{d}(\mathcal{M}, q, \phi, \sigma)$
\end{center}
where the notation $ \denotation{d}(\mathcal{M}, q, \phi, \sigma)$
denotes the {\em quantified semantics} of a density function.  The quantified
semantics of a density function denotes the joint density of all the possible
{target} variables that the density defines. We therefore define the
quantified semantics of a density function as the product over all
instantiations of the the density's quantifier variables, which, therefore
determine the set of target variables.  We define the quantified semantics as
follows:
\begin{align*}
\denotation{d}(\mathcal{M}, q, \phi, \sigma) = \underset{n \in \denotation{q}(\mathcal{M})}{\prod} 
        \begin{cases}
        \denotation{d}(\mathcal{M}, \sigma[q \mapsto n]) & \denotation{\phi}(\mathcal{M}, \sigma[q \mapsto n]) \\ 
            1 & \mathrm{\bf else}
        \end{cases}
\end{align*}

{This representation of the model demonstrates that the joint probability
density of the model can be factored into a product of  its constituent density
definitions}.  Building on the definition of the joint density as well as
Bayes' Rule, we define the {\em conditional density} of a set of variables,
$A$, conditioned on the
values of a set of variables $B$,
\vspace{-.25cm} 
 $\mathcal{J}(A \mid  B)(\mathcal{M},\sigma)
= \frac{\int_{\mathcal{V} - (\denotation{A}(\mathcal{M},\sigma) \cup \denotation{B}(\mathcal{M},\sigma))} \mathcal{J}(\mathcal{M},\sigma))}{\int_{\mathcal{V} - \denotation{B}(\mathcal{M}(\mathcal{M},\sigma),\sigma)}
\mathcal{J}}$

\paragraph{\bf Model Type Environment.} A model type environment $\Gamma_m$ is
a finite map from quantified variables to domain names: $\Gamma_m \in Q
\rightarrow \Delta$.

\paragraph{\bf Model Satisfaction} We next define what it means for an
environment to satisfy a model.

\begin{itemize}

\item {\bf Random Variable Satisfaction.} The value of a random variable in
an environment satisfies its domain if the value of the variable at each index
in its domain has a value that is within the variable's codomain.  
\begin{align*} 
 \sigma, \mathcal{M} \vDash v : (\delta_1,\delta_2) & \Rightarrow \forall n_1 \in \denotation{\delta_1}(\mathcal{M}). \; \exists n_2 \in \denotation{\delta_2}(\mathcal{M}). \; \sigma((v,n_1)) = n_2
\end{align*}

\item{\bf Satisfaction.} An environment satisfies a model if its random
variables satisfy their domains. 

\begin{center}
$\sigma \vDash \mathcal{M} \Rightarrow (\forall v \in \mathrm{dom}(\sigma), \delta_1, \delta_2. \; \Big(\pi_{2}(\mathcal{M})(v) = (\delta_1, \delta_2)\Big) \Rightarrow  \sigma, \mathcal{M} \vDash v : (\delta_1,\delta_2))$
\end{center}

\item{\bf Quantifier Variable Satisfaction.} The value of quantifier
variable in an environment satisfies its domain if it is an element of the
domain.
$$
    \sigma, \mathcal{M} \vDash \qv : \delta 
    \Rightarrow \sigma(\qv) \in \denotation{\delta}(\mathcal{M}) 
$$

\item{\bf Environment Satisfaction.} An environment satisfies a
    model and model type environment if 1) it satsifies the model and 2) every
        quantifier variable in the environment is a member of the doman
        prescribed by the type environment.
$$
    \sigma \vDash \mathcal{M}, \Gamma_m \Rightarrow \sigma \vDash \mathcal{M} \wedge
    \forall \qv \in \mathrm{dom}(\sigma). \; \sigma, \mathcal{M} \vDash \qv : \Gamma_m(\qv)
$$

\end{itemize}

\paragraph{\bf Model Validity.} For a Shuffle inference program to be sound
with respect to a model $\mathcal{M}$, the model must be {\em valid}. Shuffle
assumes that each density implementation defined in the model is equal to its
conditional density as defined above in terms of $\mathcal{J}$. A valid model
satisfies a sequence of properties that ensure this is the case.
Appendix~\ref{sec:validity} defines these conditions in more detail.

\begin{thm}[Model Validity]
    \label{thm:validity}
    If $\mathcal{M} \in \mathbb{M}$ is valid, then for any $\eta = (x,d,(\qv,\delta),A,B,\cons) \in \pi_3(\mathcal{M})$
    it must be true that
    $
        \forall \sigma.\; \Big(\sigma \vDash \mathcal{M}, \Gamma_m[\qv \mapsto \delta] 
        \land \denotation{\cons}(\mathcal{M},\sigma)\Big) \Rightarrow 
        \denotation{d}(\mathcal{M}, \sigma) = \mathcal{J}(A|B)(\mathcal{M}, \sigma)
    $
\end{thm}

\subsection{Assumption Log}

\begin{wrapfigure}[6]{r}{.5\linewidth}
\begin{mdframed}
\begin{align*}
\assumplog & \rightarrow & \emptyset \; \mid \; \assumplog :: \alpha \\
\alpha & \rightarrow & (\cons \Rightarrow A \independent B \; \mid \; C) \; \mid \; \mathrm{ReachesAll}(s)
\end{align*}
\end{mdframed}
\vspace{-.15cm}
\caption{Assumption log structure}
\label{fig:alog}
\end{wrapfigure}

Figure~\ref{fig:alog} presents the structure of an assumption log.  An
assumption log, $\assumplog$, is a record of the set of model and inference
program assumptions made by the developer during the construction of their
inference program.  Figure~\ref{fig:alog} presents the structure of an
assumption log.  An assumption log, $\assumplog$, is a record of the set of
model and inference program assumptions made by the developer during the
construction of their inference program.  An entry in an assumption log,
$\alpha$, is a logical proposition the asserts either {\em statistical
independence} or a {\em reachability}.  We denote the semantics of an
assumption log by the semantic function $\denotation{\assumplog} : (\mathbb{M}
\times \Sigma) \rightarrow \mathbb{B}$ with the semantics of a full log given
by $\denotation{\assumplog :: \alpha}(\mathcal{M}, \sigma) =
\denotation{\assumplog}(\mathcal{M}, \sigma) \wedge
\denotation{\alpha}(\mathcal{M}, \sigma)$ and the semantics of individual
entries as follows.

\paragraph{\bf Statistical Independence.} {\tool}'s independence assumptions
assume the conditional statistical independence of two sets of variables in the
model. We define the semantics of the independence notation $\cons \Rightarrow
A \independent B \; | \; C$, meaning that under the constraint $\cons$ the set
of variables $A$ is $B$ independent of $B$ given the values of $C$, using a
standard definition of statistical independence in probability theory
 . Specifically, the joint density, $\mathcal{J}$, gives a
natural specification:
\begin{align*}
\denotation{\cons \Rightarrow A \independent B \; | \; C}(\mathcal{M}, \sigma) = \denotation{\cons}(\mathcal{M}, \sigma) \Rightarrow \Big(\mathcal{J}(A, B \, | \, C)(\mathcal{M}, \sigma) =  \mathcal{J}(A \, | \, C)(\mathcal{M}, \sigma)* \mathcal{J}(B \, | \, C)(\mathcal{M}, \sigma) \Big)
\end{align*}

\paragraph{\bf Reachability.} The predicate $\mathrm{ReachesAll}(s)$ states
that a given sampler $s$ reaches every value in its output space with positive
probability. In other words, for any variable $(v,n) \in \mathcal{V}$, if $s$
modifies $v,n$, then there must be some positive probability of reaching every
value of $n$ in $n$'s domain.
\begin{align*}
\denotation{\mathrm{ReachesAll}(s)}(\mathcal{M}, \sigma) & = & \forall v,n. \; \Big(\exists \sr,r. \; s(\sigma[(v,n) \mapsto r],\sr)(v,n) \neq r \Rightarrow \forall r. \exists \sr. \; s(\sigma,\sr)(v,n) = r\Big)
\end{align*}

\paragraph{\bf Entailment Relation.} We define the entailment relation between
two assumption logs by the notation and semantics, $\mathcal{M}, \assumplog_1
\vDash \assumplog_2 = \forall \sigma. \; \Big(\sigma \vDash \mathcal{M} \land
\denotation{\assumplog_1}(\mathcal{M}, \sigma) \Big) \Rightarrow
\denotation{\assumplog_2}(\mathcal{M}, \sigma)$

\paragraph{\bf Satisfaction Relation.} An environment and model satisfy an
assumption log if the denotation of the assumption log evaluates to
\textit{true} for the environment.
$\sigma, \mathcal{M} \vDash \assumplog  =  (\denotation{\assumplog}(\mathcal{M}, \sigma) = \textit{true})$

\subsection{Type Environment and Typing Context} 

A type environment, $\Gamma \in (Q \cup X) \rightarrow t$ is a finite map from
quantifier variables and distribution variables to types. Note that a type
environment differs from a {\em model} type environment in that a model type
environment only contains bindings for quantifier variables.

\paragraph{\bf Model Environment Entailment.} We use the notation $\Gamma
\vDash \Gamma_m$to mean the type environment $\Gamma \in (Q \cup X) \rightarrow
t$ entails the model type environment $\Gamma_m \in Q \rightarrow \Delta$.
  \[ \Gamma
\vDash \Gamma_m = \forall \qv \in \mathrm{dom}(\Gamma_m). \; \Gamma_m(\qv) =
\Gamma(\qv) \]

\paragraph{\bf Environment Satisfaction.} We use the notation ${\sigma \vDash
\Gamma, \mathcal{M}, \assumplog}$ to denote when an environment $\sigma$ {\em
satisfies} a {{\em type context}}, which consists of an environment
$\Gamma$, a Shuffle model $\mathcal{M}$, and assumption log $\assumplog$.  We
partition the definition of this relation for environment by first specifying
the satisfaction relation for each type of variable that may be present in an
environment.

 \paragraph{\bf Distribution Variable Satisfaction.} The value of a distribution
variable in an environment satifies its type $t$ when the code definition of
the distribution ($\prog$) satifies $t$. 
\begin{center}
$\typegeneric[\vDash]{\sigma, \mathcal{M}}{\Gamma}{\assumplog}{x}{t} \Rightarrow \Big( \exists q, \delta, \prog. \; (\sigma(x) = ((q, \delta),  \prog))  \land \typegeneric[\vDash]{\mathcal{M}}{\Gamma}{\assumplog}{\prog}{t}\Big)$
\end{center}

In the following section we give precise definitions of the satisifcation
relation for each type of code definition (e.g., densities and samplers).

\paragraph{\bf Satisfaction. } Given the above definitions, an
environment satisfies a model, type environment, and assumption log if the
environment has positive joint density (according to its definition via
$\mathcal{J}$) and each of its quantifier variables, random variables, and
distribution variables meet their respective satisfaction relations.
\begin{align*}    
    \sigma \vDash \Gamma, \mathcal{M}, \assumplog = &\exists \Gamma_m. \; \Gamma \vDash \Gamma_m \land \sigma \vDash \mathcal{M}, \Gamma_m \; \land \; \sigma, \mathcal{M} \vDash \assumplog \; \land \\ 
    &\forall x \in \textrm{dom}(\sigma). \; \typegeneric[\vDash]{\sigma, \mathcal{M}}{\Gamma}{\assumplog}{x}{\Gamma(x)} \land \mathcal{J}(\mathcal{M}, \sigma) > 0 \;
\end{align*}


\subsection{Type Semantics}

A type $t$ in {\tool} is an element of the grammar: $T \rightarrow \delta \;
\mid \; (\delta, \; \delta) \; \mid \; \booltype \; \mid \; T_x \; \mid \;
(\qv, \; \delta, \; T_x)$.  In this grammar, $\delta$ is a domain, $\qv$ is a
quantifier variable, $\booltype$ is the boolean type, and  $T_x$ is a
distribution type. This language of types extends the developer-supplied types
$T_x$ as specified in Figure~\ref{fig:models} with types Shuffle derives
internally. 

\begin{definition}[Density] If \dtdef[\vDash]{d}{A}{B}, \\ 
\hspace*{2cm}then $\forall \sigma. \; \Big(\sigma \vDash \Gamma, \mathcal{M}, \assumplog \land \denotation{\cons}(\sigma)  \Big) \Rightarrow \Big(\denotation{d}(\sigma) =
\mathcal{J}(A \, | \, B)(\mathcal{M}, \sigma)\Big)$
\end{definition}
{\noindent} A term $d$ with type $\basictype{\texttt{density}}{A}{B}{\phi}$ is a function
that computes the probability that the variables denoted by $A$ taken on a
given set of values when conditioned on $B$. Given our definition of the
semantics of the full model by its joint density, $\mathcal{J}(\mathcal{M},
\sigma)$, and the subsequent definition of the conditional density,
$\mathcal{J}(A \mid  B)$ (Section~\ref{sec:model-semantics}), we directly use
the definition of the conditional density to define the semantic judgment for
density functions.  Specifically, if a term $d$ satisfies the semantic judgment
for density types, then, for all environments that satisfy the type context as
well as satisfy the type's constraint, $\phi$, then the term computes the
conditional density of $A$ given $B$.

\begin{definition}[Sampler] If \stdef[\vDash]{s}{A}{B}, then\\
 $\forall f \in \Sigma_{\mathrm{rv}} \rightarrow \mathbb{R}^+, \; \sigma. \Big(\sigma \vDash \Gamma, \mathcal{M}, \assumplog \land \denotation{\cons}(\sigma)\Big)$ $\Rightarrow 
\int_\sr f(\denotation{s}(\sigma,\sr)) = 
    \int_{\denotation{A}(\mathcal{M},\sigma)} f(\sigma) * \mathcal{J}(A \,| \, B)(\mathcal{M}, \sigma)
$
\end{definition}
{\noindent}A term with sampler type is a function with it is possible to compute
the expectation of any positive function $f$ under the distribution $P(A \, |
\, B)$.

\begin{definition}[Kernel] If $\ktdef[\vDash]{k}{A}{B}$, then\\[-1em]
\begin{center} 
\hspace*{.75cm}$\forall s. \stdef[\vDash]{s}{A}{B} \Rightarrow$ $\stdef[\vDash]{s \, \texttt{;} \, k}{A}{B}$
\end{center}
\end{definition}
{\noindent}A kernel $k$ is a function such that for any sampler $s$ for a given
distribution and any positive function $f$, the expectation of $f$ under $s$ is
the same as that under the composition of $s$ with $k$.

\subsubsection{Quantified Types} 

Shuffle exports inference procedures that are functions of quantifier
variables. The function is function is correct if instantiations of the
function body are correct. This relationship is defined by the equality:
$\typegeneric[\vDash]{\mathcal{M}}{\Gamma}{\assumplog}{\prog}{(\qv, \delta, t)}
= \typegeneric[\vDash]{\mathcal{M}}{\Gamma}{\assumplog}{\prog}{t} $

\subsection{Type Rules}
\label{subsec:typerules}

\renewcommand{\sr}{\mathrm{sr}}

 Shuffle's type rules make use of the predicate
$\propnolog{\mathcal{M}}{\Gamma}{\mathrm{ValidInfer}(A|B,\cons)}$. This predicate ensures that the type
$\basictype{{t_b}}{A}{B}{\cons}$ is {\em valid}.  For an example of an
invalid type, consider the variable set \texttt{z\{i0 in Dom: z[i0] == j\}}.
Integrating a density with respect this variable set would require changing a
value of \texttt{z}, but this would in and of itself change the set of
integration variables.  When type rules change $A$ and $B$, they must check
that the new type is still valid.  Appendix~\ref{sec:typevalidity} provides
more details on the formal definition of validity and our implementation of the
validity check that uses the Z3 theorem prover~\citep{z3}.

\def\defaultHypSeparation{\hskip 0cm}
\begin{figure}
\small
\begin{mdframed}
\centering
    \begin{mathpar}
    \dmul

    \ddiv\\

    \ddivtwo

    \dint
    \end{mathpar}
\end{mdframed}
\caption{Type rules for probability densities}
\label{fig:densrules}
\vspace{-.25cm}
\end{figure}

\paragraph{\bf Densities.} Figure~\ref{fig:densrules} shows the typing rules
for probability densities. These follow the rules of conditional probability,
and give the developer the ability to multiply, divide, and integrate densities
in a typesafe manner. DMUL takes two densities and multiplies them together
pointwise. It converts one of the conditioned variables in the first density to
a target variable. This assumes that the converted variable is a target
variable in the second density.  The DMUL rule requires a check that the type
is valid because it might be the case that, for instance, the set of random
variables $A$ depends on the values of random variables in $B$, which would
render the type $\basictype{\texttt{density}}{A \comma B}{C}{\cons}$ invalid.
DDIV divides the first density by the second. This has the effect of converting
target variables in the first density to conditioned variables in the new
density. The moved variables must be target variables in the second density.
DDIV2 enables a developer to convert density over two target variable sets to
density over only one of the variable sets. The developer does this by dividing
the density by a density for the eliminated target variable, conditioned on the
uneliminated target variables.   The DINT rule provides another method for
eliminating a target variable set.  

\begin{figure}
\begin{mdframed}
\centering
\small
\begin{mathpar}
        \slift

        \sbind
\end{mathpar}
\end{mdframed}
\caption{Type rules for samplers (abbreviated)}
\label{fig:samplerules}
\vspace{-.4cm}
\end{figure}

\vspace{-.15cm}
\paragraph{\bf Samplers.} Figure~\ref{fig:samplerules} presents the typing
rules for samplers.  The SBIND rule asserts that the type is valid because it
might be the case that, for instance, the set of random variables $A$ depends
on the values of random variables in $B$, which would render the type
$\basictype{\texttt{density}}{A \comma B}{C}{\cons}$ invalid.

\begin{wrapfigure}{r}{.5\textwidth}
\small
\begin{mdframed}
    \centering
    \begin{mathpar}
        \klift\\
        \kcombine\\
        \kfix
    \end{mathpar}
\end{mdframed}
\caption{Type rules for kernels (abbreviated)}
\vspace{-.15cm}
\label{fig:kernrules}
\vspace{-.15cm}
\end{wrapfigure}

\paragraph{\bf Kernels.} Figure~\ref{fig:kernrules} presents the typing rules
for kernels. KLIFT constructs a kernel out of a sampler. The constructed
kernel's behavior is the same as the input sampler's.  However, for the sampler
to be a valid kernel, two conditions must hold.  First, the sampler's output
must be a finite random variable.  
Second, for each value in the target random variable's domain, $s$ must have a
positive probability of producing that value. Together these properties ensure
that every kernel representable in Shuffle admits an approximate implementation
of via repeated iteration~\texttt{fix}~\cite{markovmixing}.  . The type rule enforces the first
constraint with the {second premise} of the rule and the second constraint
with the third premise. The type system does not verify $\mathrm{ReachesAll}$.
Instead, $\mathrm{ReachesAll}$ must be an implication of the provided
assumption log.

KCOMBINE combines two kernels for conditional densities into a kernel for a
{joint density}. Specifically, the two kernels must be conditioned on each
other's target variables. The resulting kernel represents the joint density of
each kernel's output conditioned on any global conditioned variables both
kernels have.  The KCOMBINE rule requires a check that the type is valid
because it might be the case that, for instance, the set of random variables
$A$ depends on the values of random variables in $B$, which would render the
type $\basictype{\texttt{kernel}}{A \comma B}{C}{\cons}$ invalid. KFIX
transforms a kernel for into a sampler for the same distribution using the
\texttt{fix} operator.

\paragraph{\bf Definitions and Conditionals} The rules for definitions and
conditionals have expected definitions. For clarity of presentation, we present
these rules and the remaining rules in Appendix~\ref{sec:otherrules}.

\subsection{Properties}

Shuffle's type system guarantees that densities, samplers, kernels, and
estimators correctly implement the conditional probability distribution implied
by their type. This means that for any type judgment produced by the type
rules, the semantics of the type must be true.

\begin{thm}[Soundness]
    If $\mathcal{M}$ is valid, and \typegeneric{\mathcal{M}}{\Gamma}{\assumplog}{\prog}{t}, 
    then for all $\mathcal{D}$,\\ 
    \typegeneric[\vDash]{(\mathcal{D},\pi_2(\mathcal{M}),\pi_3(\mathcal{M}))}{\Gamma}{\assumplog}{\prog}{t}
\end{thm}

\noindent We present a proof of this theorem in Appendix~\ref{sec:proof}.

\section{The {\tool} System}
\label{sec:system}

{\tool} as a system performs type checking, assumption log generation, and
inference program extraction. A developer therefore receives a concrete
executable inference procedure that has been type checked against the program's
specified types as well as an auditable list of assumptions about the
probabilistic model that must be true for the inference procedure to be
correct.

\subsection{\bf Type Checking}

The {\tool} system implements the type checking rules presented in
Section~\ref{sec:types}. Shuffle uses the Z3 theorem prover~\citep{z3} to check
assertions over sets of quantified and random variables. Shuffle models
quantified variable values, random variable values, and domain bounds as 64-bit
bitvectors, and constraints using a combination of bitvector comparisons and
boolean operations.  Shuffle checks equality and implication relations between
constraints using quantifier-free bitvector theories, and checks type validity
assertions through Z3's quantified bit-vector formulas.

For example, the implementation of \texttt{callDensAll} on
Line~\ref{code:calldensall} in Figure~\ref{fig:burglary-exact} invokes the
density \texttt{callDensAll(p-1)}, which has the target random variable set
\texttt{calls\{i in People: i <= (p-1)\}}. The developer wishes to multiply
this density with \texttt{callDensI}, which has a conditioned random variable
set that contains the set \texttt{calls\{i in People: i < p\}}.  Shuffle uses
Z3 to verify these variable sets are equivalent. Shuffle generates the
following constraint to send to Z3:
\begin{center}
    \begin{wraplst}
    \begin{lstlisting}[
    language=C, escapeinside={(*}{*)},
    keywordstyle=\color{blue},
    escapeinside={(*}{*)}, 
    keywordstyle=\color{blue}, 
    morekeywords={def, rec, domain, in, forall,variable, model, normal, uniform}, 
    numbers=left,
    frame=single]
(People_min >= 0 && People_max >= 0) &&
(People_min <= p && p <= People_max) && 
(People_min <= i && i <= People_max) &&
(i <= p - 1) != (i < p)
    \end{lstlisting}
    \end{wraplst}
\end{center}
Z3 tries to find values for each of the bitvectors \texttt{People\_min},
\texttt{People\_max}, \texttt{p}, and \texttt{i} such that the values satisfy
the constraint. If Z3 returned such an assignment, it would mean that there
exist values for the variables that demonstrate that the two random variable
sets are not equivalent, and Shuffle would terminate with a type error.
However, in this case Z3 determines that this constraint is not satisfiable,
and thus the variable sets are equivalent.


\vspace{-.15cm}
\subsection{Inference Program Extraction.} Shuffle extracts a Python program
for a given type-checked Shuffle program. Shuffle's program extraction is by
and large a straightforward, syntax-directed recursive procedure that produces
a Python program that implements the denotational semantics presented in
Section~\ref{sec:semantics}. Shuffle's extraction procedure differs
operationally from the denotational semantics in that it 1) simplifies integral
expressions it can identify analytical solutions   2) reports an
error when it cannot simplify intractable integral expressions or {sample from a density} 3) uses a
representation for probabilities that ensures numerical stability, and 4)
performs automatic incremental optimizations.  Note that all of these code
generation concerns are issues that -- without Shuffle -- a developer would
otherwise have to perform by hand.
 
\vspace{-.15cm} \paragraph{\bf Simplification.}   Shuffle used standard algebraic
simplification techniques to simplify integrals with known closed-form
solutions. Shuffle currently simplifies conjugate and posterior-predictive
distributions for Gaussian and Dirichlet distributions.

\vspace{-.15cm}
\paragraph{\bf Optimization} {\tool} uses program transformation techniques 
for synthesizing efficient incremental inference procedures. 
Currently, {\tool} uses a variant of the algorithm presented in \citet{arrayagg} to 
incrementalize reduction operations. Generally, the algorithm first identifies 
two loops (generated by \tool's simplication pass) where the inner loop contains an reduction. 
It then attempts to hoist the inner loop out, and incrementally
updates the value of the hoisted loop when necessary. 
{\tool} supports a wider class of incremental optimization than \citet{arrayagg}. It 
identifies non-affine patterns in loop conditions and performs enabling program 
transformation for further potential incrementalization, which we found common among inference procedures.
Note that incrementalization also generalizes Loop Invariant Code Motion, in which case
the hoisted loop does not need incremental updates. 
We discuss an transformation example in Appendix~\ref{sec:optimization-detail}.



\section{Evaluation}
\label{sec:eval}


Shuffle sits within a landscape of probabilistic programming tools and
approaches that range from fully automated systems to systems that encourage
handcoded implementations via a built-in library of primitives and -- in the
extreme -- direct implementations in a standard programming language, such as
C++. Here, we consider Venture as a reference point for the evaluation.  We
also include benchmarks compared with BLOG~\citep{blog} in
Appendix~\ref{sec:morebenchmarks} as complementary to showing Shuffle's
flexibility.

\vspace{-.1cm}
\paragraph{\bf Venture} Venture is a probabilistic programming system that,
like Shuffle, supports programmable inference. Venture's approach makes use of
{\em stochastic procedures}, which are Python objects that encapsulate
complicated density arithmetic. Venture enables developers to manipulate
stochastic procedures with {\em modeling commands} and {\em inference
commands}. Modeling commands compose stochastic procedures together to form a
probabilistic model, and inference commands use invoke a stochastic procedure's
methods to implement inference procedures. Venture developers hand-write the
modeling commands, inference commands, and the contents of stochastic
procedures. This is similar to Shuffle, in which developers hand-write the
models and inference procedures. However, Venture provides no support for
either verifying the correctness of an inference procedure nor {automatically
generating the code of an implementation}. Specifically, in Venture, if a
developer seeks to augment Venture with their own inference procedure, they
must do so manually in Python.

\paragraph{\bf Research Questions} We compare against Venture to evaluate the
following research questions:

\begin{itemize}

\item{\bf Verification Burden.} What is the annotation overhead for the
developer to deliver a typesafe inference procedure, relative to an untyped
procedure? 

\item{\bf Performance.}  Do Shuffle's abstractions increase or decrease
performance relative to implementations of the same algorithms within a
comparable system?
\end{itemize}

\vspace{-.1cm}
\paragraph{\bf Benchmark Models.} To evaluate the first research question,
we implement several different inference procedures for several different
models.  Each of these models has the property that portions of the model can
be solved analytically, but practical algorithms rely on combining exact
analytical methods with approximate, sampling-based methods. 

\begin{itemize}

\item {\bf GMM.} A Gaussian mixture model~\cite{gmm}. A GMM models a sequence
of samples where each sample has a Gaussian distribution centered around a {\em
cluster center}. Based on a dataset size used in prior work~\cite{augurv2},
this model contains 10000 samples and 10 cluster centers.

\item {\bf LDA.} A Latent Dirichlet Allocation model~\cite{lda}. LDA models a
sequence of documents by randomly selecting a topic for each word based on a
document-specific topic distribution. It then randomly selects each word from a
topic-specific word distribution.  Based on dataset sizes used in prior
work~\cite{kos}, this model  uses approximately 466k words, 50 topics, 3430
documents and an vocabulary size of 6.9k. 

\item {\bf DMM.} A Dirichlet Multinomial Mixture model~\cite{dmm}. A DMM
models a sequence of documents by randomly selecting a topic for each document,
and thereafter selecting each word from a topic-specific word distribution.
Based on dataset sizes used in prior work~\cite{twins}, this model contains
approximately 570k words, 4 topics, 278 documents and an vocabulary size of
129. 
\end{itemize}

\vspace{-.1cm}
\paragraph{\bf Benchmark Inference Algorithms.} For each benchmark model, we
implemented several inference algorithms in Shuffle. We compare Shuffle's
support for these algorithms against Venture's for Gibbs
sampling~\cite{gibbssampling} (a Markov-chain Monte Carlo inference algorithm),
Metropolis-Hastings~\cite{mha,mhb} (a Markov-chain Monte Carlo algorithm) and
Likelihood weighting~\citep{likelihoodweighting} (a sequential Monte Carlo
algorithm)

\subsection{Verification Burden}

To quantify Shuffle's verification burden, we compare the length of Shuffle's
inference procedure against the length of Venture's stochastic procedure
implementation.  We also report the number of assertions Shuffle generates in
its assumption log for each benchmark.

\subsubsection{Methodology}

We compared the total number of lines of code -- model plus inference -- in the
Venture and Shuffle implementations of each benchmark model and inference
algorithm. For GMM in Venture, we included the lines of code for all model
definition commands, inference commands, and stochastic procedures we wrote
ourselves. For LDA and DMM, our implementations take advantage of pre-built
stochastic procedures included in Venture. We report two numbers for these
benchmarks: the lines of code we wrote ourselves and the lines of code the
Venture developers used to construct the built-in stochastic procedures
 .

\begin{table*}
    \footnotesize
    \centering
    \begin{tabular}{|c|c|c|c|c|c|c|c|c|c|}
        
    \hline

        \multirow{3}{*}{Model} & \multicolumn{3}{c|}{Gibbs} & \multicolumn{3}{c|}{Metropolis-Hastings} & \multicolumn{3}{c|}{Likelihood Weighting} \\

        \cline{2-10}
    

        & \multicolumn{2}{c|}{LOC} & \multirow{2}{*}{Assump.} & \multicolumn{2}{c|}{LOC} & \multirow{2}{*}{Assump.} & \multicolumn{2}{c|}{LOC} & \multirow{2}{*}{Assump.}\\

        \cline{2-3} \cline{5-6} \cline{8-9}


        & Shuffle & Venture & & Shuffle & Venture & & Shuffle & Venture & \\

        \hline

        GMM & 112 & 104 & 6/1 & 118 & 111 & 7/1 & 86 & 98 & 6/0 \\
        LDA & 179 & 28/91 & 13/1 & 196 & 27/91 & 14/1 & 169 & 25/91 & 12/0 \\
        DMM & 180 & 33/91 & 11/1 & 190 & 40/91 & 12/1 & 190 & 40/91 & 11/0 \\

        \hline
    \end{tabular}

\caption{Evaluation of Shuffle's verification burden. We use the notation $a/b$ in the Venture line count to mean that $a$ is the number of lines of code written by the inference developer and $b$ is the number of lines of code in the stochastic procedures that are built in to Venture. We use the notation $a/b$ in the ``Assump.'' column to mean that $a$ is the number of independence assertions and $b$ is the number of reachability assertions.}
    \label{tab:resultsloc}
\vspace{-.5cm}
\end{table*}

\subsubsection{Results}
Table~\ref{tab:resultsloc} presents the size in lines of code for both {\tool}
and Venture implementations of the benchmarks.   The GMM
implementations in Shuffle and Venture require similar amounts of code, but LDA
and DMM require significantly more code in Shuffle than in Venture, even
accounting for the code required to implement the stochastic procedure in
Venture. This is primarily due to the fact that Venture enables a developer to
reuse a stochastic procedure in multiple places, where as Shuffle does not
support reuse and therefore a Shuffle developers need to perform density
arithmetic from scratch for each density with a unique type. This motivates a
future direction for Shuffle that supports polymorphism or higher-order
features that enable reuse.

\subsection{Performance}

To evaluate Shuffle's performance, we compare the performance of Shuffle's
generated code versus the runtime of the Venture-implemented inference
procedures for our benchmarks. A key aim of this evaluation is to demonstratate
that our abstractions, which more closely model standard abstractions in
probability than standard Python code, do not introduce performance overhead.

\begin{table*}
    \footnotesize
    \centering

\begin{tabular}{|c|c|c|c|c|c|c|c|c|c|}
    \hline

        \multirow{2}{*}{Model} & \multicolumn{3}{c|}{Gibbs} & \multicolumn{3}{c|}{Metropolis-Hastings} & \multicolumn{3}{c|}{Likelihood Weighting} \\
    
    \cline{2-10} 

              & Shuffle & Venture & Speedup & Shuffle & Venture & Speedup & Shuffle & Venture & Speedup \\

        \hline
        GMM & \SI{1.3}{\ms} & 33\si{\ms} & ${\bf 25x}$ & \SI{1.7e-1}{\ms} & 5.3 \si{\ms} & ${\bf 31x}$ & \SI{7.5e-1}{\s} & 16.2 \si{\s} & ${\bf 21x}$ \\
        LDA & \SI{6.1e-3}{\s} & 10.4 \si{\s} & ${\bf 1700x}$ & \SI{1.1e-1}{\ms} & 270 \si{\ms} & ${\bf 2400x}$ & 45.3 \si{\s} & 800 \si{\s} & ${\bf 17x}$ \\
        DMM & 1.1 \si{\s} & 72 \si{\s} & ${\bf 65x}$ & \SI{6.8e-1}{\s} & 2.1 \si{\s} & ${\bf 3.1x}$ & 30 \si{\s} & 580 \si{\s} & ${\bf 19x}$ \\
        \hline
    \end{tabular}
    \caption{Performance of Shuffle compared to Venture. Speedups are speedup of Shuffle over Venture.}
    \label{tab:resultsperf}

\begin{tabular}{|c|c|c|c|c|c|c|}
	\hline
	\multirow{2}{*}{Model} & \multicolumn{2}{c|}{Gibbs} & \multicolumn{2}{c|}{Metropolis-Hastings} & \multicolumn{2}{c|}{Likelihood Weighting} \\
	\cline{2-7} 
	& Unoptimized & Speedup  & Unoptimized & Speedup & Unoptimized & Speedup \\
	\hline
	GMM & \SI{1.8e2}{\ms} & ${\bf 138x}$ & \SI{6e1}{\ms} & ${\bf 353x}$ & \SI{23}{\s} & ${\bf 30x}$ \\
	LDA & \textbf{timeout} & ${\bf -}$ & \textbf{timeout} & ${\bf -}$ & \textbf{timeout} & ${\bf -}$ \\
	DMM & \SI{2.1e2}{\s} & ${\bf 191x}$  & \SI{3.5e1}{\s} & ${\bf 51x}$  & \textbf{timeout} & ${\bf -}$ \\
	\hline
\end{tabular}
\caption{Performance of Shuffle without optimization. Speedups are speedup of optimized over unoptimized}
\label{tab:resultsperfunopt}
\vspace{-.75cm}
\end{table*}

\subsubsection{Methodology}


For each benchmark, we measured wall clock time over 20 runs. 
We time out benchmarks that did not finish in 30 hours. The Gibbs and
Metropolis-Hastings tests measure the time for one Gibbs update to a single
random variable. The Likelihood Weighting test measures the time to perform
inference on 1 particle. We also measure each benchmark's peak memory
consumption, as reported by \texttt{time -v} command.  All benchmarks are
performed on m5.12xlarge Amazon EC2 instance, which has 192 \si{\giga\byte}
memory. All inference procedures are single threaded, and use only 1 CPU core. 
 
Each benchmark uses synthetically generated observed data from the prior
distribution of the statistical model. Because the comparison is between
computational efficiencies of identical algorithms, we do not anticipate that
changes in the data values alone will have an impact on performance. The size
of the datasets are similar in scale to real-world examples of GMM
~\citep{augurv2},   LDA ~\citep{kos},
DMM~\citep{twins} models.

\subsubsection{Results} Figure~\ref{tab:resultsperf} present the results of our
performance evaluation. The results show that Shuffle's abstractions do not
introduce overhead and that the resulting inference procedures are at least as
fast as Venture. In fact, Shuffle's inference procedures are significantly
faster than Venture on all of our benchmarks, with a minimum speedup of 3.1x.
We note that this significant speedup is due to Venture's overhead on manipulating 
its internal data structures for tracking the incremental updates, which in
some cases leads to asympotitc degrade in performance; for DMM-MH, the current implementation
of Shuffle fails to perform one potential incremental update, 
which explains the relatively smaller speedup of 3.1x. 

We also measure Shuffle's performance without incremental optimization.  The
results in Figure~\ref{tab:resultsperfunopt} demonstrate incremental
optimizations are critical to performance, resulting in at least a 30x speedup
over an unoptimized Shuffle program.

Appendix~\ref{sec:perf-discussion} presents more results on Shuffle's
performance, including standard deviations for performance experiments, peak
memory consumption comparisons, and a more detailed discussion of Shuffle's
performance improvement over Venture.

\section{Related Work}

\paragraph{\bf Automated Inference. } Church~\cite{church} and
WebPPL~\cite{webppl} enable a user to specify Turing-complete stochastic
programs as models, but restrict inference algorithms to all-purpose algorithms
such as Metropolis-Hastings \cite{mha,mhb}. JAGS~\cite{jags} provides a
notation for expressing graphical models and automatically performs sampling
for a fixed set of distributions. JAGS therefore provides automated support for
a subset of {\tool}'s rules. For example, JAGS can automatically generate a
collapsed sampler for GMM. However, it can do so only if the model is specified
with a monolithic GMM primitive. This stands in contrast to {\tool}, which, via
its compositional nature, enables a user to prove the correctness of collapsed
sampling for a wide class of models. 

\vspace{-.15cm}
\paragraph{\bf Manual Unverified Inference.} Other systems,
such as Venture~\cite{venture} and PyMC~\cite{pymc} enable a user to augment
the system's inference procedure with arbitrary code. However, when the user
augments the inference algorithm with arbitrary code, there is no guarantee
that the resulting inference algorithm is correct. In contrast, the code that a
user generates with {\tool} is in accordance with the {\tool}'s proof rules and
therefore enjoys {\tool}'s correctness guarantees.  Park et al. developed a
language that includes samplers and other objects as first-class primitives~\cite{Park:2005}.
The type of a term in their language communicates the base type of the object
(e.g., a sampler).  While, Shuffle shares its base operations with their
language, Shuffle's novel contribution is to extend the types to describe the
conditional distribution that the object represents.

\vspace{-.15cm} \paragraph{\bf Compiled Inference.} AugurV2~\cite{augurv2}
provides a language of coarse-grained operators to build inference procedures
out of, like {\tool}. AugurV2 supports a richer set of kernels than Shuffle.
AugurV2 also provides more support for parallelism and alternative compilation
targets.  However, AugurV2 does not provide correctness guarantees as strong as
{\tool}'s. In particular, AugurV2's kernels are not guaranteed to converge
iteratively to the target distribution. AugurV2 also does not have density
operations to support collapsed Gibbs samplers. Thus AugurV2 does not support
any of the benchmarks from Section~\ref{sec:eval}, although it does support
other inference procedures for the GMM and LDA models.  \citet{shuffle}
motivate the need for correct-by-construction probabilistic inference. Shuffle
is the first complete system that provides a programming language that moves
probabilistic inference towards that goal.

\vspace{-.15cm}
\paragraph{\bf Program Transformation.} Hakaru~\cite{hakaru} and related
systems~\cite{transformations} enable developers to perform probabilistic
inference by applying transformations to a program that specifies the
underlying probabilistic model. The resulting transformed program implements an
executable inference procedure for the query of interest.  Shuffle's approach
is complementary in that it advocates ground-up composition of inference
algorithms from base primitives. In addition, Shuffle's type system enables
developers to explicitly specify the behavior of an inference procedure, as
opposed to transformation systems that implicitly specify this behavior
outside of the language.

The PSI solver~\cite{psisolver} transforms probabilistic models into densities
representing inference procedures. PSI can find densities for a larger class of
models than Shuffle, but doesn not support samplers or kernels.



\balance

\bibliography{paper}


\begin{thebibliography}{33}


\ifx \showCODEN    \undefined \def \showCODEN     #1{\unskip}     \fi
\ifx \showDOI      \undefined \def \showDOI       #1{#1}\fi
\ifx \showISBNx    \undefined \def \showISBNx     #1{\unskip}     \fi
\ifx \showISBNxiii \undefined \def \showISBNxiii  #1{\unskip}     \fi
\ifx \showISSN     \undefined \def \showISSN      #1{\unskip}     \fi
\ifx \showLCCN     \undefined \def \showLCCN      #1{\unskip}     \fi
\ifx \shownote     \undefined \def \shownote      #1{#1}          \fi
\ifx \showarticletitle \undefined \def \showarticletitle #1{#1}   \fi
\ifx \showURL      \undefined \def \showURL       {\relax}        \fi
\providecommand\bibfield[2]{#2}
\providecommand\bibinfo[2]{#2}
\providecommand\natexlab[1]{#1}
\providecommand\showeprint[2][]{arXiv:#2}

\bibitem[\protect\citeauthoryear{Atkinson and Carbin}{Atkinson and
  Carbin}{2016}]%
        {shuffle}
\bibfield{author}{\bibinfo{person}{Eric Atkinson} {and}
  \bibinfo{person}{Michael Carbin}.} \bibinfo{year}{2016}\natexlab{}.
\newblock \showarticletitle{Towards Correct-by-Construction Probabilistic
  Inference}. In \bibinfo{booktitle}{\emph{LearningSys}}.
\newblock


\bibitem[\protect\citeauthoryear{Blei, Ng, and Jordan}{Blei
  et~al\mbox{.}}{2003}]%
        {lda}
\bibfield{author}{\bibinfo{person}{D.~M. Blei}, \bibinfo{person}{A.~Y. Ng},
  {and} \bibinfo{person}{M.~I. Jordan}.} \bibinfo{year}{2003}\natexlab{}.
\newblock \showarticletitle{Latent Dirichlet Allocation}. In
  \bibinfo{booktitle}{\emph{J. Mach. Learn. Res.}}, Vol.~\bibinfo{volume}{3}.
\newblock


\bibitem[\protect\citeauthoryear{Cowles and Carlin}{Cowles and Carlin}{1996}]%
        {diagnostics}
\bibfield{author}{\bibinfo{person}{Mary~Kathry Cowles} {and}
  \bibinfo{person}{Bradley~P. Carlin}.} \bibinfo{year}{1996}\natexlab{}.
\newblock \showarticletitle{Markov Chain Monte Carlo Convergence Diagnostics: A
  Comparative Review}. In \bibinfo{booktitle}{\emph{JASA}},
  Vol.~\bibinfo{volume}{91}.
\newblock


\bibitem[\protect\citeauthoryear{Daniel~Huang}{Daniel~Huang}{2017}]%
        {augurv2}
\bibfield{author}{\bibinfo{person}{Greg~Morisett Daniel~Huang,
  Jean-Baptiste~Tristan}.} \bibinfo{year}{2017}\natexlab{}.
\newblock \showarticletitle{Compiling Markov Chain Monte Carlo Algorithms for
  Probabilistic Modeling}. In \bibinfo{booktitle}{\emph{PLDI}}.
\newblock


\bibitem[\protect\citeauthoryear{David A.~Levin}{David A.~Levin}{2008}]%
        {markovmixing}
\bibfield{author}{\bibinfo{person}{Elizabeth M.~Wilmer David A.~Levin,
  Yuval~Peres}.} \bibinfo{year}{2008}\natexlab{}.
\newblock \showarticletitle{Markov Chains and Mixing Times}.
\newblock


\bibitem[\protect\citeauthoryear{Fung and Chang}{Fung and Chang}{1989}]%
        {likelihoodweighting}
\bibfield{author}{\bibinfo{person}{Robert~M. Fung} {and}
  \bibinfo{person}{Kuo-Chu Chang}.} \bibinfo{year}{1989}\natexlab{}.
\newblock \showarticletitle{Weighing and Integrating Evidence for Stochastic
  Simulation on Bayesian Networks}. In \bibinfo{booktitle}{\emph{UAI}}.
\newblock


\bibitem[\protect\citeauthoryear{Gehr, Misailovic, and Vechev}{Gehr
  et~al\mbox{.}}{2016a}]%
        {psi}
\bibfield{author}{\bibinfo{person}{Timon Gehr}, \bibinfo{person}{Sasa
  Misailovic}, {and} \bibinfo{person}{Martin Vechev}.}
  \bibinfo{year}{2016}\natexlab{a}.
\newblock \showarticletitle{PSI: Exact Symbolic Inference for Probabilistic
  Programs}. In \bibinfo{booktitle}{\emph{Computer Aided Verification}},
  \bibfield{editor}{\bibinfo{person}{Swarat Chaudhuri} {and}
  \bibinfo{person}{Azadeh Farzan}} (Eds.). \bibinfo{publisher}{Springer
  International Publishing}, \bibinfo{address}{Cham}, \bibinfo{pages}{62--83}.
\newblock
\showISBNx{978-3-319-41528-4}


\bibitem[\protect\citeauthoryear{Gehr, Misailovic, and Vechev}{Gehr
  et~al\mbox{.}}{2016b}]%
        {psisolver}
\bibfield{author}{\bibinfo{person}{Timon Gehr}, \bibinfo{person}{Sasa
  Misailovic}, {and} \bibinfo{person}{Martin Vechev}.}
  \bibinfo{year}{2016}\natexlab{b}.
\newblock \showarticletitle{{P}{S}{I}: {E}xact symbolic inference for
  probabilistic programs}. In \bibinfo{booktitle}{\emph{CAV}}.
\newblock


\bibitem[\protect\citeauthoryear{Geman and Geman}{Geman and Geman}{1984}]%
        {gibbssampling}
\bibfield{author}{\bibinfo{person}{Stuart Geman} {and} \bibinfo{person}{Donald
  Geman}.} \bibinfo{year}{1984}\natexlab{}.
\newblock \showarticletitle{Stochastic Relaxation, Gibbs Distributions, and the
  Bayesian Restoration of images}. In \bibinfo{booktitle}{\emph{IEEE
  Transactions on Pattern Analysis and Machine Intelligence}}.
\newblock


\bibitem[\protect\citeauthoryear{Goodman, Mansinghka, Roy, Bonawitz, and
  Tenenbaum}{Goodman et~al\mbox{.}}{2008}]%
        {church}
\bibfield{author}{\bibinfo{person}{Noah~D. Goodman}, \bibinfo{person}{Vikash~K.
  Mansinghka}, \bibinfo{person}{Daniel~M. Roy}, \bibinfo{person}{Keith
  Bonawitz}, {and} \bibinfo{person}{Joshua~B. Tenenbaum}.}
  \bibinfo{year}{2008}\natexlab{}.
\newblock \showarticletitle{Church: A language for generative models}. In
  \bibinfo{booktitle}{\emph{UAI}}.
\newblock


\bibitem[\protect\citeauthoryear{Goodman and Stuhlm\"{u}ller}{Goodman and
  Stuhlm\"{u}ller}{2014}]%
        {webppl}
\bibfield{author}{\bibinfo{person}{Noah~D Goodman} {and}
  \bibinfo{person}{Andreas Stuhlm\"{u}ller}.} \bibinfo{year}{2014}\natexlab{}.
\newblock \showarticletitle{{The Design and Implementation of Probabilistic
  Programming Languages}}.
\newblock
\newblock
\shownote{Accessed: 2016-10-7.}


\bibitem[\protect\citeauthoryear{Hastings}{Hastings}{1970}]%
        {mhb}
\bibfield{author}{\bibinfo{person}{W.~K. Hastings}.}
  \bibinfo{year}{1970}\natexlab{}.
\newblock \showarticletitle{Monte Carlo Sampling Methods Using Markov Chains
  and Their Applications}. In \bibinfo{booktitle}{\emph{Biometrika}},
  Vol.~\bibinfo{volume}{57}.
\newblock


\bibitem[\protect\citeauthoryear{Holmes, Harris, and Quince}{Holmes
  et~al\mbox{.}}{2012}]%
        {dmm}
\bibfield{author}{\bibinfo{person}{Ian Holmes}, \bibinfo{person}{Keith Harris},
  {and} \bibinfo{person}{Christopher Quince}.} \bibinfo{year}{2012}\natexlab{}.
\newblock \showarticletitle{Dirichlet Multinomial Mixtures: Generative Models
  for Microbial Metagenomics}. In \bibinfo{booktitle}{\emph{PloS one}}.
\newblock


\bibitem[\protect\citeauthoryear{Jones, Oliphant, Peterson,
  et~al\mbox{.}}{Jones et~al\mbox{.}}{01  }]%
        {scipy}
\bibfield{author}{\bibinfo{person}{Eric Jones}, \bibinfo{person}{Travis
  Oliphant}, \bibinfo{person}{Pearu Peterson}, {et~al\mbox{.}}}
  \bibinfo{year}{2001--}\natexlab{}.
\newblock \bibinfo{title}{{SciPy}: Open source scientific tools for {Python}}.
\newblock   (\bibinfo{year}{2001--}).
\newblock
\urldef\tempurl%
\url{http://www.scipy.org/}
\showURL{%
\tempurl}
\newblock
\shownote{[Online; accessed <today>].}


\bibitem[\protect\citeauthoryear{Leonardo De~Moura}{Leonardo De~Moura}{2008}]%
        {z3}
\bibfield{author}{\bibinfo{person}{Nikolaj~Bj{\o}rner Leonardo De~Moura}.}
  \bibinfo{year}{2008}\natexlab{}.
\newblock \showarticletitle{Z3: An Efficient SMT Solver}. In
  \bibinfo{booktitle}{\emph{TACAS / ETAPS}}.
\newblock


\bibitem[\protect\citeauthoryear{Liu}{Liu}{1994}]%
        {gibbscollapsed}
\bibfield{author}{\bibinfo{person}{Jun~S. Liu}.}
  \bibinfo{year}{1994}\natexlab{}.
\newblock \showarticletitle{The Collapsed Gibbs Sampler in Bayesian
  Computations with Applications to a Gene Regulation Problem}. In
  \bibinfo{booktitle}{\emph{Journal of the American Statistical Association}},
  Vol.~\bibinfo{volume}{89}.
\newblock


\bibitem[\protect\citeauthoryear{Liu, Stoller, Li, and Rothamel}{Liu
  et~al\mbox{.}}{2005}]%
        {arrayagg}
\bibfield{author}{\bibinfo{person}{Yanhong~A. Liu}, \bibinfo{person}{Scot~D.
  Stoller}, \bibinfo{person}{Ning Li}, {and} \bibinfo{person}{Tom Rothamel}.}
  \bibinfo{year}{2005}\natexlab{}.
\newblock \showarticletitle{Optimizing Aggregate Array Computations in Loops}.
  In \bibinfo{booktitle}{\emph{TOPLAS}}, Vol.~\bibinfo{volume}{27}.
\newblock
Issue 1.


\bibitem[\protect\citeauthoryear{{Mansinghka}, {Selsam}, and
  {Perov}}{{Mansinghka} et~al\mbox{.}}{2014}]%
        {venture}
\bibfield{author}{\bibinfo{person}{V. {Mansinghka}}, \bibinfo{person}{D.
  {Selsam}}, {and} \bibinfo{person}{Y. {Perov}}.}
  \bibinfo{year}{2014}\natexlab{}.
\newblock \showarticletitle{{Venture: a higher-order probabilistic programming
  platform with programmable inference}}. In \bibinfo{booktitle}{\emph{ArXiv
  e-prints}}.
\newblock


\bibitem[\protect\citeauthoryear{{Maple Inc.}}{{Maple Inc.}}{2018}]%
        {maple}
\bibfield{author}{\bibinfo{person}{{Maple Inc.}}}
  \bibinfo{year}{2018}\natexlab{}.
\newblock \bibinfo{title}{Maple}.
\newblock   (\bibinfo{year}{2018}).
\newblock
\urldef\tempurl%
\url{www.maplesoft.com/products/maple/}
\showURL{%
\tempurl}


\bibitem[\protect\citeauthoryear{{Metropolis}, {Rosenbluth}, {Rosenbluth},
  {Teller}, and {Teller}}{{Metropolis} et~al\mbox{.}}{1953}]%
        {mha}
\bibfield{author}{\bibinfo{person}{N. {Metropolis}}, \bibinfo{person}{A.~W.
  {Rosenbluth}}, \bibinfo{person}{M.~N. {Rosenbluth}}, \bibinfo{person}{A.~H.
  {Teller}}, {and} \bibinfo{person}{E. {Teller}}.}
  \bibinfo{year}{1953}\natexlab{}.
\newblock \showarticletitle{{Equation of State Calculations by Fast Computing
  Machines}}. In \bibinfo{booktitle}{\emph{Journal of Chemical Physics}},
  Vol.~\bibinfo{volume}{21}.
\newblock


\bibitem[\protect\citeauthoryear{Milch}{Milch}{2006}]%
        {blog}
\bibfield{author}{\bibinfo{person}{Brian~Christopher Milch}.}
  \bibinfo{year}{2006}\natexlab{}.
\newblock \emph{\bibinfo{title}{Probabilistic Models with Unknown Objects}}.
\newblock \bibinfo{thesistype}{Ph.D. Dissertation}. \bibinfo{address}{Berkeley,
  CA, USA}.
\newblock Advisor(s) Russell, Stuart J.
\newblock
\newblock
\shownote{AAI3253991.}


\bibitem[\protect\citeauthoryear{Murphy}{Murphy}{2012}]%
        {gmm}
\bibfield{author}{\bibinfo{person}{Kevin~P. Murphy}.}
  \bibinfo{year}{2012}\natexlab{}.
\newblock \bibinfo{booktitle}{\emph{Machine Learning: A Probabilistic
  Perspective}}.
\newblock \bibinfo{publisher}{MIT Press}, \bibinfo{address}{Cambridge,
  Massachusets}.
\newblock


\bibitem[\protect\citeauthoryear{Narayanan, Carette, Romano, Shan, and
  Zinkov}{Narayanan et~al\mbox{.}}{2016}]%
        {hakaru}
\bibfield{author}{\bibinfo{person}{Praveen Narayanan}, \bibinfo{person}{Jacques
  Carette}, \bibinfo{person}{Wren Romano}, \bibinfo{person}{Chung{-}chieh
  Shan}, {and} \bibinfo{person}{Robert Zinkov}.}
  \bibinfo{year}{2016}\natexlab{}.
\newblock \showarticletitle{Probabilistic Inference by Program Transformation
  in Hakaru (System Description)}. In \bibinfo{booktitle}{\emph{FLOPS}}.
\newblock


\bibitem[\protect\citeauthoryear{Newman}{Newman}{2008}]%
        {kos}
\bibfield{author}{\bibinfo{person}{David Newman}.}
  \bibinfo{year}{2008}\natexlab{}.
\newblock \showarticletitle{Bag of Words Dataset}. In
  \bibinfo{booktitle}{\emph{UCI Machine Learning Respository}}.
\newblock


\bibitem[\protect\citeauthoryear{Park, Pfenning, and Thrun}{Park
  et~al\mbox{.}}{2005}]%
        {Park:2005}
\bibfield{author}{\bibinfo{person}{Sungwoo Park}, \bibinfo{person}{Frank
  Pfenning}, {and} \bibinfo{person}{Sebastian Thrun}.}
  \bibinfo{year}{2005}\natexlab{}.
\newblock \showarticletitle{A Probabilistic Language Based Upon Sampling
  Functions}. In \bibinfo{booktitle}{\emph{Proceedings of the 32Nd ACM
  SIGPLAN-SIGACT Symposium on Principles of Programming Languages}}
  \emph{(\bibinfo{series}{POPL '05})}. \bibinfo{publisher}{ACM},
  \bibinfo{address}{New York, NY, USA}, \bibinfo{pages}{171--182}.
\newblock
\showISBNx{1-58113-830-X}
\urldef\tempurl%
\url{https://doi.org/10.1145/1040305.1040320}
\showDOI{\tempurl}


\bibitem[\protect\citeauthoryear{Patil, Huard, and Fonnesbeck}{Patil
  et~al\mbox{.}}{2010}]%
        {pymc}
\bibfield{author}{\bibinfo{person}{Anand Patil}, \bibinfo{person}{David Huard},
  {and} \bibinfo{person}{Christopher Fonnesbeck}.}
  \bibinfo{year}{2010}\natexlab{}.
\newblock \showarticletitle{PyMC: Bayesian Stochastic Modelling in Python}.
\newblock   \bibinfo{volume}{35} (\bibinfo{year}{2010}).
\newblock


\bibitem[\protect\citeauthoryear{Plummer}{Plummer}{2015}]%
        {jags}
\bibfield{author}{\bibinfo{person}{Martyn Plummer}.}
  \bibinfo{year}{2015}\natexlab{}.
\newblock \bibinfo{booktitle}{\emph{JAGS Version 4.0.0 user manual}}.
\newblock \bibinfo{publisher}{Addison-Wesley}, \bibinfo{address}{Reading,
  Massachusetts}.
\newblock


\bibitem[\protect\citeauthoryear{Russell and Norvig}{Russell and
  Norvig}{2011}]%
        {aima}
\bibfield{author}{\bibinfo{person}{Stuart Russell} {and} \bibinfo{person}{Peter
  Norvig}.} \bibinfo{year}{2011}\natexlab{}.
\newblock \bibinfo{booktitle}{\emph{Artificial Intelligence: A Modern
  Approach}}.
\newblock \bibinfo{publisher}{Prentice Hall}, \bibinfo{address}{Upper Saddle
  River, NJ}.
\newblock


\bibitem[\protect\citeauthoryear{\'{S}cibior, Kammar, V\'{a}k\'{a}r, Staton,
  Yang, Cai, Ostermann, Moss, Heunen, and Ghahramani}{\'{S}cibior
  et~al\mbox{.}}{2018}]%
        {transformations}
\bibfield{author}{\bibinfo{person}{Adam \'{S}cibior}, \bibinfo{person}{Ohad
  Kammar}, \bibinfo{person}{Matthijs V\'{a}k\'{a}r}, \bibinfo{person}{Sam
  Staton}, \bibinfo{person}{Hongseok Yang}, \bibinfo{person}{Yufei Cai},
  \bibinfo{person}{Klaus Ostermann}, \bibinfo{person}{Sean~K. Moss},
  \bibinfo{person}{Chris Heunen}, {and} \bibinfo{person}{Zoubin Ghahramani}.}
  \bibinfo{year}{2018}\natexlab{}.
\newblock \showarticletitle{Denotational Validation of Higher-order Bayesian
  Inference}. In \bibinfo{booktitle}{\emph{POPL}}.
\newblock


\bibitem[\protect\citeauthoryear{Turnbaugh, Hamady, Yatsunenko, Cantarel,
  Duncan, Ley, Sogin, Jones, A~Roe, Affourtit, Egholm, Henrissat, C~Heath,
  Knight, and I~Gordon}{Turnbaugh et~al\mbox{.}}{2008}]%
        {twins}
\bibfield{author}{\bibinfo{person}{Peter Turnbaugh}, \bibinfo{person}{Micah
  Hamady}, \bibinfo{person}{Tanya Yatsunenko}, \bibinfo{person}{Brandi
  Cantarel}, \bibinfo{person}{Alexis Duncan}, \bibinfo{person}{Ruth Ley},
  \bibinfo{person}{Mitchell Sogin}, \bibinfo{person}{Joe Jones},
  \bibinfo{person}{Bruce A~Roe}, \bibinfo{person}{Jason Affourtit},
  \bibinfo{person}{Michael Egholm}, \bibinfo{person}{Bernard Henrissat},
  \bibinfo{person}{Andrew C~Heath}, \bibinfo{person}{Rob Knight}, {and}
  \bibinfo{person}{Jeffrey I~Gordon}.} \bibinfo{year}{2008}\natexlab{}.
\newblock \showarticletitle{A core gut microbiome in obese and lean twins}.
\newblock   \bibinfo{volume}{457} (\bibinfo{date}{12} \bibinfo{year}{2008}),
  \bibinfo{pages}{480--4}.
\newblock


\bibitem[\protect\citeauthoryear{{Wolfram Research, Inc.}}{{Wolfram Research,
  Inc.}}{2018}]%
        {mathematica}
\bibfield{author}{\bibinfo{person}{{Wolfram Research, Inc.}}}
  \bibinfo{year}{2018}\natexlab{}.
\newblock \bibinfo{title}{Mathematica}.
\newblock   (\bibinfo{year}{2018}).
\newblock
\urldef\tempurl%
\url{https://www.wolfram.com/mathematica/}
\showURL{%
\tempurl}


\bibitem[\protect\citeauthoryear{Wu, Li, Russel, and Bodik}{Wu
  et~al\mbox{.}}{2016}]%
        {swift}
\bibfield{author}{\bibinfo{person}{Yi Wu}, \bibinfo{person}{Lei Li},
  \bibinfo{person}{Stuart Russel}, {and} \bibinfo{person}{Rastislav Bodik}.}
  \bibinfo{year}{2016}\natexlab{}.
\newblock In \bibinfo{booktitle}{\emph{IJCAI}}.
\newblock


\bibitem[\protect\citeauthoryear{Zhang and Poole}{Zhang and Poole}{1994}]%
        {variableelimination}
\bibfield{author}{\bibinfo{person}{Nevin~Lianwen Zhang} {and}
  \bibinfo{person}{David Poole}.} \bibinfo{year}{1994}\natexlab{}.
\newblock \showarticletitle{A simple approach to Bayesian network
  computations}. In \bibinfo{booktitle}{\emph{Canadian Conference on Artificial
  Intelligence}}.
\newblock


\end{thebibliography}

\newpage

\appendix
\section{Validity Conditions for Models and Types}
\label{sec:validity}
\subsection{Model Validity}


\paragraph{\bf Variable Set Validity.} The variable sets $A$ and $B$ and
constraint $\cons$ in a type $\basictype{t_b}{A}{B}{\cons})$ can freely depend
on the values of random variables and therefore some specifications of $A$,
$B$, or $\cons$ may be undefined .  A quantified triple of $A$,
$B$, and $\cons$ are {\em valid}, written $\mathcal{M}, \Gamma_m \vDash
\mathrm{Valid}(A | B,\cons)$ if the following properties hold.

\begin{itemize}
    \item{\bf Error Free.} Evaluations of $\phi$, $B$, and $A$ must not error. 
\[
    \forall \sigma. \; \Big(\sigma \vDash \mathcal{M}, \Gamma_m\Big) \Rightarrow
    \Big(\denotation{\phi}(\mathcal{M},\sigma) \neq \bot_\sigma \land
    \denotation{A}(\mathcal{M},\sigma) \neq \bot_\sigma \land
    \denotation{B}(\mathcal{M},\sigma) \neq \bot_\sigma \Big)
\]

\item{\bf Stratified Conditioned Variables.} The conditioned set of random
variables $B$ in a type must have a total order $<$ on the set of variable
names $V$ such that if $v_i < v_j$, then the subset of $v_i$ contained in $B$
does not depend on the subset of $v_j$ contained in $B$. This ensures that the
integral over $B$ in $\mathcal{J}(A|B)$ is well-defined. For example, consider
the variable set \texttt{z\{i0 in Dom: z[i0] == j\}}. Integrating a function
over this variable set would require changing a value of \texttt{z}, but this
would change the set of integration variables. By contrast if a total order
exists as defined above where $v_i < v_j$, integrating over $B$ requires
integrating over $v_j$ for every possible value of $v_i$, and then integrating
the resulting function by $v_i$.
\begin{align*}
    \exists i \in V \rightarrow \mathbb{N}.& \;
    \forall \sigma, v_1, v_2. \; 
    \Big(i(v_1) < i(v_2) \land \sigma \vDash \mathcal{M}, \Gamma_m\Big) \Rightarrow\\
    &\Big(\forall (v_*,n_*),n_\mathrm{new}, \sigma'. \\
    &\Big[(v_*,n_*) \not \in \denotation{B \cap v_2}(\mathcal{M},\sigma) \land
    \sigma' = \sigma[(v_*,n_* \mapsto n_\mathrm{new}] \land
    \sigma' \vDash \mathcal{M}\Big] \Rightarrow \\
    &\Big[\denotation{B \cap v_1}(\mathcal{M},\sigma) = 
    \denotation{B \cap v_1}(\mathcal{M},\sigma')\Big] \Big)
\end{align*}
        where $\denotation{B \cap v_i}(\mathcal{M},\sigma) = \{(v_i,n) | (v_i, n)
        \in \denotation{B}(\mathcal{M},\sigma)\}$ is the variable set $B$
        projected onto the variable name $v_i$.

\item{\bf Computable Constraint.} The constraint $\cons$ must be
computable given the only the values of the conditioned variables $B$.
Therefore any variable that the value of $\cons$ depends on must be contained
in $B$. This is necessary to ensure integration within an inference procedure
is correct. For example, the constraint \texttt{z[i] == max(Dom)} would make
Shuffle's DINT type rule invalid for the program \texttt{int Dens() by z[i]},
assuming \texttt{z[i]} is a target variable of \texttt{Dens()}.
\begin{align*}
    \forall \sigma, \sigma', v_*, n_*, n_\mathrm{new}. \;&
    \Big((v_*,n_*) \not \in \denotation{B}(\mathcal{M},\sigma) \land
    \sigma' = \sigma[(v_*,n_*) \mapsto n_\mathrm{new}] \land
    \sigma' \vDash \mathcal{M}\Big) \Rightarrow\\
    &\Big(\denotation{\cons}(\mathcal{M},\sigma) = \denotation{\cons}(\mathcal{M},\sigma')\Big)
\end{align*}
 
\item{\bf Computable Target Variables.} The random variable set $A$ must
be computable given only the values of the conditioned variables $B$. Therefore
that any variable that a type uses to constrain the members of $A$ must be
contained in $B$. This is necessary to ensure integration within an inference
procedure is correct. For example, the if the density \texttt{Dens()} has the
target variable set  \texttt{z, x\{i in Dom1: z[i] == max(Dom2)\}} would make
Shuffle's DINT type rule invalid for the program \texttt{int Dens() by z}.
\begin{align*}
    \forall \sigma, \sigma', v_*, n_*, n_\mathrm{new}. \;&
    \Big((v_*,n_*) \not \in \denotation{B}(\mathcal{M},\sigma) \land
    \sigma' = \sigma[(v_*,n_*) \mapsto n_\mathrm{new}] \land
    \sigma' \vDash \mathcal{M}\Big) \Rightarrow\\
    &\Big(\denotation{A}(\mathcal{M},\sigma) = \denotation{A}(\mathcal{M},\sigma')\Big)
\end{align*}
 
\end{itemize}

\paragraph{\bf Density Validity}

{Another example of an invalid model is one where a density does not
integrate to $1$ over its target random variables, resulting in an invalid
probability space.} Given a definition of the form $(x,d,(\qv,\delta),A,B,\cons) \in \pi_3(\mathcal{M})$ we define several validity properties.

\begin{itemize}

\item {\bf {Self Contained.}} A density must only depend on random variables that are in either its target random variable set or conditioned random variable set.  
\begin{align*}
    \forall \sigma, \sigma', v_*, n_*, n_\mathrm{new}. \;&
    \Big((v_*,n_*) \not \in \denotation{A \comma B}(\mathcal{M},\sigma) \land
    \sigma' = \sigma[(v_*,n_*) \mapsto n_\mathrm{new}] \land
    \sigma' \vDash \mathcal{M}\Big) \Rightarrow\\
    &\Big(\denotation{d}(\mathcal{M},\sigma) = \denotation{d}(\mathcal{M},\sigma')\Big)
\end{align*}

\item {\bf Normalized.} A density must integrate to $1$ over its target variables. 
\begin{center}
    $\forall \sigma  , \Gamma_m.   \Big(\sigma \vDash \mathcal{M}, \Gamma_m[\qv \mapsto \delta] \Big) \Rightarrow
        \Big(\int_{\denotation{A}(\sigma)}
        \denotation{d}(\sigma)= 1 \Big)$
\end{center}

\item{\bf Unique.} The quantified semantics of a density defines the
density of a random variable at most once for all instantiations of its
quantifiers. Namely, given two instantiations of a density's quantifiers, the
sets of target variables co,rresponding to each instantiation do not intersect.
\begin{align*}
    \forall \sigma, \sigma', \Gamma_m. \; \Big(& \sigma \vDash \mathcal{M},\Gamma_m[\qv \mapsto \delta] \land (\exists n \in \denotation{\delta}(\mathcal{M}). \; \sigma' = \sigma[q \mapsto n]) \; \\
    & \land (\denotation{A}(\mathcal{M},\sigma') \cap \denotation{A}(\mathcal{M},\sigma) \not = \emptyset)\Big) \Rightarrow \sigma' = \sigma
\end{align*}

\item {\bf Disjoint.}
The variable sets $A$ and $B$ must be disjoint. This enforces that $\int_A
\mathcal{J}(A|B) = 1$, and is sufficient for ensuring the model is valid.
\[
    \forall \sigma, \Gamma_m. \; \Big(\sigma \vDash \mathcal{M}, \Gamma_m[\qv \mapsto \delta]\Big)
    \Rightarrow \Big(\denotation{A}(\mathcal{M}, \sigma) \cap \denotation{B}(\mathcal{M},\sigma) = \emptyset\Big)
\]

\end{itemize}

\paragraph{\bf Model Schedulability.} For a model to be valid, it must be {\em
schedulable}. This means there must be an ordering of density-integer pairs
$((\eta_0,n_0), (\eta_1,n_1), \dots, (\eta_n,n_k)) \in H \times \mathbb{N}$
such that the conditioned variables in any density definition come before the
target variables. If no such ordering exists, the model's densities can fail to
represent the conditional densities implied by their types. For example, if in
the Burglary model from Figure~\ref{fig:BurglaryModel}, the developer had
specified two densities $d_1$ and $d_2$ with type \texttt{density(earthquake)},
then $\mathcal{J}(\texttt{earthquake}) = \mathcal{J}(\texttt{earthquake} |
\emptyset)$ would be equal to $\denotation{(d_1 \opfmt{*} d_2) \opfmt{/}
\intd{d_1 \opfmt{*} d_2}{\texttt{earthquake}}}$. This is not necessarily equal
to either $d_1$ or $d_2$, which means Shuffle's type judgments -- which rely on
the fact that the type of a density corresponds to its conditional density --
would not hold. An ordering $((\eta_0,n_0), (\eta_1,n_1), \dots, (\eta_n,n_k))$
is a valid schedule if, using the notation $\eta_i = (x_i,d_i,(\qv_i,\delta_i),A_i,B_i,\cons_i)$

\begin{itemize}
\item {\bf Complete: } A schedule $S = ((\eta_0,n_0), (\eta_1,n_1), \dots,
    (\eta_n,n_k))$ is complete if all densities in the model and all values of
        a density's quantifier variable from the appropriate domain appear in
        the schedule.
        \[
            \forall \eta_i,n_i. \; \Big(\eta_i \in \pi_3(\mathcal{M}) \wedge n \in \denotation{\delta_i}(\mathcal{M})\Big) \Rightarrow (\eta_i,n_i) \in S
        \]

\item {\bf Serial:} A schedule is $((\eta_0,n_0), (\eta_1,n_1), \dots,
    (\eta_n,n_k))$ serial if the conditioned variables for a density $\eta_i$
        invoked with parameter $n_i$ are target variables in a previous density
        in the schedule. Namely, for each variable in $\eta_i$'s conditioned
        variable set, there exists $\eta_j, n_j$ such that $j < i$ and the
        variable is in $\eta_j$'s target variable set when $\eta_j$ is invoked
        with $n_j$.
        \[
            \forall (v,n), \sigma. \; \Big(\sigma \vDash \mathcal{M} (v,n) \in \denotation{A_i}(\mathcal{M},\sigma[\qv_i \mapsto n_i)\Big) \Rightarrow
            \Big(\exists j. \; j < i \land (v,n) \in \denotation{B_j}(\mathcal{M},\sigma[\qv_j \mapsto n_j])\Big)
        \]

\item {\bf Linear:} A schedule $((\eta_0,n_0), (\eta_1,n_1), \dots,
    (\eta_n,n_k))$ is linear if any random variable appears as a target
        variable in  a density at most once in the schedule. In other words,
        for every variable there is exactly one $i$ such that the variable is
        in $\eta_i$'s target variables when invoked with $n_i$. 
        \[
            \forall (v,n), . \Big(\sigma \vDash \mathcal{M} \land n \in \pi_1(\pi_2(\mathcal{M})(v)) \Big) \Rightarrow \Big(\exists! i. \; (v,n) \in \denotation{A_i}(\mathcal{M},\sigma[\qv_i \mapsto n_i)\Big)
        \]

\end{itemize}

Together these properties ensure that the model's specification does not
prescribe circular dependencies between the model's random variables.

\subsection{Type Validity}
\label{sec:typevalidity}

\paragraph{\bf Inference Variable Set Validity.} Shuffle's type rules use a
variant of variable set validity. The notation for this predicate is
$\propnolog{\mathcal{M}}{\Gamma}{\mathrm{ValidInfer}(A|B,\cons)}$ and it checks
the same conditions as the corresponding $\mathrm{Valid}(A|B,\cons)$, but under
any possible instantiation of the domains.  \[
    \propnolog{\mathcal{M}}{\Gamma}{\mathrm{ValidInfer}(A|B,\cons)} = \exists
    \Gamma_m \in (Q \rightarrow \Delta). \; \Gamma \vDash \Gamma_m \land
    \forall D.\;
    \propnolog{\mathcal{D},\pi_2(\mathcal{M}),\pi_3(\mathcal{M})}{\Gamma_m}{\mathrm{Valid}(A|B,\cons)}
\]

\paragraph{\bf Implementation.} Shuffle checks type validity by translating the above condition to constraints and solving it with Z3.

\section{Additional Type Rules}
\label{sec:otherrules}

\begin{figure*}[hp]
\small
\begin{mdframed}
 \centering
\begin{mathpar}
   \env\\
   \shfcond
\end{mathpar}
\end{mdframed}
\vspace{-.15cm}
\caption{Type rules for if statements and invocations.}
\label{fig:defrules}
\vspace{-.15cm}
\end{figure*}

\begin{figure*}
\small
\begin{mdframed}
    \centering
    \begin{mathpar}
        \defr \and \defrec\and
        \defind
     \end{mathpar}
\end{mdframed}
\vspace{-.25cm}
\caption{Type rules for definitions.}
\label{fig:defrules}
\vspace{-.25cm}
\end{figure*}

\paragraph{\bf Variable Set Choice.} We use the notation $(\phi, A, B)$ to
denote a variable set choice, which yields $A$ when $\phi$ is true and yields
$B$ when $\phi$ is false. Variable set choice extends the grammar in
Figure~\ref{fig:models} with the production $V_g \rightarrow (\cons, V_g,
V_g)$. Its denotation is given by
$\denotation{(\phi, A, B)}(\sigma) = 
     \begin{cases}
         \denotation{A}(\mathcal{M}, \sigma)  &  \hphantom{\neg} \denotation{\phi}(\mathcal{M},\sigma) \\
         \denotation{B}(\mathcal{M},\sigma)  &  \neg \denotation{\phi}(\mathcal{M},\sigma) \\
     \end{cases}
$

\paragraph{\bf Conditionals.} Figure~\ref{fig:defrules} shows Shuffle's IF rule
for handling conditionals. This rule enables the developer to construct
inference procedures whose behavior differs based on whether a constraint is
true or false. The system must check that the resulting type is valid, because
the constraint may introduce an illegal dependency in the random variables or
the constraint in the resulting type.

We denote standard capture-avoiding substitution on the free quantifier
variables of a type $t$ by the notation $t[a/\qv]$ where $a$ is a term
and $\qv$ is a quantifier variable.

\paragraph{\bf Invocation.} Shuffle checks type validity on invocations because
for two reasons: 1) A type written by the developer must be valid, and 2) the
substitution in the invocation rule (INV Figure~\ref{fig:defrules}) may yield
an invalid type if, for example, it results in $A$ depending on the value of a
random variable that is not in $B$.

\paragraph{\bf Definition.} Figure~\ref{fig:defrules} shows Shuffle's rules for
defining procedures. These give the developer the ability to encapsulate
components of the inference procedure while ensuring these components are
correctly defined and invoked. We use the notation $\qv \in \delta$ as syntactic sugar for the constraint $\mathrm{\tt min(}\delta\mathrm{\tt)} \opfmt{<=} \qv \opfmt{\&\&} \qv \opfmt{<=} \mathrm{\tt max(}\delta\mathrm{\tt)}$

\paragraph{\bf Recursion.} Shuffle's type system imposes restrictions on
recursive programs. The DEF-REC rule in Figure~\ref{fig:defrules} enforces that
procedures may only recurse on their first argument, and must pass all other
arguments through unchanged. 
In the DEF-REC rule in Figure~\ref{fig:defrules}, we use the notation
$\qv'$ to mean a fresh quantifier variable.

\paragraph{\bf Base Case {Analysis}.} 
The DEF-REC rule makes use of the base-case predicate
$\propnolog{\mathcal{M}}{\Gamma}{\mathrm{BaseCase}(\qv,\delta,A)}$ This
enforces that a recursive procedure's base case always corresponds to the set
of target random variables $A$ being empty. This justifies the default base
cases in Shuffle's semantics.
\begin{align*}
    \mathcal{M},& {\Gamma} \vDash {\mathrm{BaseCase}(\qv,\delta,A)} = 
    \exists \Gamma_m \in (Q \rightarrow \Delta). \; (\Gamma \vDash \Gamma_m) \wedge \Big(\\
        &\forall \mathcal{D}, \mathcal{M}' = (\mathcal{D},\pi_2(\mathcal{M}),\pi_3(\mathcal{M})), \sigma.\; \big(\sigma \vDash \mathcal{M}',\Gamma_m \big) \wedge \big(\denotation{\qv}(\sigma) < \denotation{\mathrm{\tt min(}\delta\mathrm{\tt)}}(\mathcal{M}')\big) \Rightarrow \denotation{A}(\mathcal{M}',\sigma) = \emptyset\Big)
\end{align*}

\begin{figure*}
\small
\begin{mdframed}
\begin{mathpar}
\coerce \and \coerceind \\
        \progcompose \and
            \envvar \and
            \envbasex \and
            \envbaseq \and
            \consprec \and
            \consneg \and
            \conscon \and
            \consdis \and
        \coerceindtype \and \and \model\\
\end{mathpar}
\end{mdframed}
\vspace{-.25cm}
\caption{{Type rules for if statements, definitions, and invocations.}}
\label{fig:defrules}
\vspace{-.25cm}
\end{figure*}

\paragraph{\bf Additional Structural Rules.} Figure~\ref{fig:defrules} shows
additional structural rules that connect different pieces of Shuffle's type
system together. These include 1) ENV rules that instantiate types from the
environment and determine whether Shuffle terms belong to a named domain 2) C,
IND, and CIND rules which apply {\em normal} and {\em independent} coercions
CONSTRAINT rules which ensure constraints have the boolean type, 4) the MODEL
rule instantiating the types in the model $\mathcal{M}$, which serve as axioms
for Shuffle's type system, and 5) the PROG-COMPOSE rule for composing programs.

\subsection{Coercions}
\label{subsec:coercions}
\begin{figure}[b]
    \begin{mdframed}
    \begin{center}
        \logtc

        \vspace{.4cm}
        \logtcindep
    \end{center}
    \end{mdframed}

    \caption{Rules for coercion side predicates}
    \label{fig:coercions}

\end{figure}

Figure~\ref{fig:coercions} shows Shuffle's rules for type coercions.

\paragraph{\bf Normal Coercions} Shuffle uses a normal coercion of the form $t_1 \rightarrow t_2$ to assert that a type judgment $t_1$ implies another type judgment $t_2$. These require additional predicates that encode logical formulae. Shuffle employs the Z3 theorem prover~\cite{z3} to verify that these predicates are true. These predicates are:
\begin{itemize}
    \item $\mathcal{M} \vDash \cons \Rightarrow A \equiv B$. This predicate states that whenever, $\cons$ is true, the variable sets $A$ and $B$ must be equivalent. Shuffle checks this by constructing, for each random variable $v$ specified by the model $\mathcal{M}$, the formulas $\cons_{vA}$ and $\cons_{vB}$ which specify the set of indices $n$ such that $(v,n) \in \denotation{A}(\sigma)$ or $(v,n) \in \denotation{A}(\sigma)$, respectively. Shuffle then checks whether $\cons \Rightarrow (\cons_{vA} \iff \cons_{vB})$.
    \item $\mathcal{M} \vDash \cons_1 \Rightarrow \cons_2$. This predicate states that the constraints $\cons_1$ imply the constraints $\cons_2$.
    \item $\mathcal{M} \vDash \cons \Rightarrow (A \cap B) = \emptyset$. This predicate determines that the variable groups $A$ and $B$ are disjoint.
\end{itemize}
The semantics of each predicate are defined as follows

\begin{alignat*}{2}
    &\mathcal{M}, \Gamma_m \vDash \cons \Rightarrow A \equiv B && = \forall \sigma. \; \sigma \vDash \mathcal{M}, \Gamma_m \land \denotation{\cons}(\mathcal{M}, \sigma) \Rightarrow (\denotation{A}(\mathcal{M},\sigma) = \denotation{B}(\mathcal{M},\sigma))\\
    &\mathcal{M},\Gamma_m \vDash \cons_1 \Rightarrow \cons_2 && = \forall \sigma. \; \sigma \vDash \mathcal{M}, \Gamma_m \land \denotation{\cons_1}(\mathcal{M},\sigma) \Rightarrow \denotation{\cons_2}(\mathcal{M},\sigma)\\
    &\mathcal{M},\Gamma_m \vDash \cons \Rightarrow (A \cap B) = \emptyset && = \forall \sigma. \; \sigma \vDash \mathcal{M}, \Gamma_m \land \denotation{\cons}(\mathcal{M},\sigma) \Rightarrow \denotation{A}(\mathcal{M},\sigma) \cap \denotation{B}(\mathcal{M},\sigma) = \emptyset\\
\end{alignat*}

\paragraph{\bf Independence Coercions} Shuffle uses a normal coercion of the
form $t_1 \rightarrow_I t_2$ to assert that a type judgment $t_1$ implies
another type judgment $t_2$. This requires the assumption log $\assumplog$ to
entail independence amongst certain variables present in $t_1$ and $t_2$.

\section{Estimators}
\label{sec:estimators}

\subsection{Syntax}
\begin{align*}
\estsyntax
\end{align*}

\estsyntaxtext

\subsection{Semantics}
\estsemantics

\subsection{Types}
\esttypefigure

\paragraph{\bf Statement.} \eststatement

\paragraph{\bf Type Rules.} \esttyperules


\section{Proofs}
\label{sec:proof}

This appendix is structured as follows: in Section~\ref{sec:proofprelim} we
establish two lemmas we require to complete the proofs; in
Section~\ref{sec:proofvalid} we establish the {\em model validity} theorem; in
Sections~\ref{sec:proofdens}, \ref{sec:proofsamp}, \ref{sec:proofkern},
\ref{sec:proofest}, and \ref{sec:proofstruct} we prove the soundness of
densities, samplers, kernels, estimators, and structural rules, respectively;
and in Section~\ref{sec:proofpres} we establish the {\em preservation theorem}.

\subsection{Preliminaries}
\label{sec:proofprelim}
\paragraph{\bf Integration by Substitution.} The following proofs make use of a property  of integrals known as the {\em substitution rule}. For measurable functions $f$ and $g$,
\[\psi(r) = \int_{x \in (-\inf, r]} g(x) \Rightarrow \int_{x \in \psi[S]} f(x) = \int_{x \in S} f(\psi(x)) * g(x)\]
where the notation $\psi[S]$ means the set obtained by mapping the function $\psi$ over $S$. This is a standard property of integration.

\paragraph{\bf Environment-Substitution Lemma.} The following proofs make use of a lemma. Let $A$ be a variable set and $\cons$ be a constraint. We have the following equivalences, for any environment $\sigma$:
\[
    \denotation{A[\argchar/\qv]}(\sigma) = \denotation{A}(\sigma[\qv \mapsto \denotation{\argchar}(\sigma))
    \denotation{\cons[\argchar/\qv]}(\sigma) = \denotation{\cons}(\sigma[\qv \mapsto \denotation{\argchar}(\sigma))
\]

\paragraph{\bf Proof.} This lemma follows from structural induction on the
syntax of constraints and variable sets/

\subsection{Model Validity}
\label{sec:proofvalid}

From the definition of $\mathcal{J}$, it must be true that
\[\mathcal{J}({A} | {B}) =
\frac{
	\int_{\mathcal{V} - (\denotation{A}(\sigma) \cup \denotation{B}(\sigma)}
	\prod_i
	\begin{cases}
		\denotation{d_i}(\sigma) & \denotation{\cons_i}(\sigma)\\
		1 & \mathrm{\bf else} \\
	\end{cases}
}{
	\int_{\mathcal{V} - \denotation{B}(\sigma)}
	\prod_i
	\begin{cases}
		\denotation{d_i}(\sigma) & \denotation{\cons_i}(\sigma)\\
		1 & \mathrm{\bf else} \\
	\end{cases}
}
\]
There exists some $i^*$ such that $d_{i^*} = d$ and $\cons_{i^*} = \cons$. 
We can partition the product in the above expression based on the schedule every valid model must have:
\[\mathcal{J}({A} | {B}) =
\frac{
	\int_{\mathcal{V} - (\denotation{A}(\sigma) \cup \denotation{B}(\sigma)}
    \prod_{i < i^*}
	\begin{cases}
		\denotation{d_i}(\sigma) & \denotation{\cons_i}(\sigma)\\
		1 & \mathrm{\bf else} \\
	\end{cases} *
    \denotation{d}(\mathcal{M},\sigma) *
    \prod_{i > i^*}
	\begin{cases}
		\denotation{d_i}(\sigma) & \denotation{\cons_i}(\sigma)\\
		1 & \mathrm{\bf else} \\
	\end{cases} 
}{
	\int_{\mathcal{V} - \denotation{B}(\sigma)}
    \prod_{i<i^*}
	\begin{cases}
		\denotation{d_i}(\sigma) & \denotation{\cons_i}(\sigma)\\
		1 & \mathrm{\bf else} \\
	\end{cases} *
    \prod_{i\ge i^*}
	\begin{cases}
		\denotation{d_i}(\sigma) & \denotation{\cons_i}(\sigma)\\
		1 & \mathrm{\bf else} \\
	\end{cases} 
}
\]
In the numerator, every element of the product over $i$ where $i > i^*$
integrates to one over its target variables, and in the denominator, every
element of the product over $i$ where $i \ge i^*$ integrates to one over its
target variables. Using the notation $\mathcal{V} < k$ to mean the subset of
$\mathcal{V} = V \times \mathbb{N}$ whose requisite target set in $H$ (the
existence of which mandated module schedule linearity), $A_i$, is such tha $i <
k$, the expression simplifies to
\[\mathcal{J}({A} | {B}) =
\frac{
    \int_{(\mathcal{V} < i^*) - \denotation{B}(\mathcal{M},\sigma)}
    \denotation{d}(\mathcal{M},\sigma) *
    \prod_{i < i^*}
	\begin{cases}
		\denotation{d_i}(\sigma) & \denotation{\cons_i}(\sigma)\\
		1 & \mathrm{\bf else} \\
	\end{cases} 
}{
    \int_{(\mathcal{V} < i^*) - \denotation{B}(\mathcal{M},\sigma)}
    \prod_{i<i^*}
	\begin{cases}
		\denotation{d_i}(\sigma) & \denotation{\cons_i}(\sigma)\\
		1 & \mathrm{\bf else} \\
	\end{cases} *
}
\]
The expression $\denotation{d}(\mathcal{M})(\sigma)$ is independent of ${(\mathcal{V} < i^*) - \denotation{B}(\mathcal{M},\sigma)}$, so we can pull it out of the integral and cancel the remaining factors.

\subsection{Density Soundness} {Recall the density soundness theorem}:
\label{sec:proofdens}

    if\dtype{\mathcal{M}}{\Gamma}{\assumplog}{d}{\qv,\delta}{A}{B}{\cons},
    then \dtype[\vDash]{\mathcal{M}}{\Gamma}{\assumplog}{d}{\qv,\delta}{A}{B}{\cons}

\paragraph{\bf Proof.} The proof follows from induction on the derivations. 
Specific rules are outlined below:

\paragraph{\bf MODEL} Because $\denotation{\qv \in \delta}(\mathcal{M},\sigma)
\Rightarrow \denotation{\qv}(\sigma) \in \denotation{\delta}(\mathcal{M})$, we can apply
Theorem~\ref{thm:validity}.

\paragraph{\bf DMUL}
The rule is sound due to the following property of $\mathcal{J}$ (we have elided the parameters $\mathcal{M}, \sigma$ for clarity)
\begin{mathpar}
\mathcal{J}(A|B,C)\mathcal{J}(B|C)= \Bigg(\frac{\int_{\mathcal{V} - (A \cup B \cup C)} \mathcal{J}}{\int_{\mathcal{V} - (B \cup C)} \mathcal{J}}\Bigg) \Bigg(\frac{\int_{\mathcal{V} - (B \cup C)} \mathcal{J}}{\int_{\mathcal{V} - C} \mathcal{J}}\Bigg)=\frac{\int_{\mathcal{V} - (A \cup B \cup C)} \mathcal{J}}{\int_{\mathcal{V} - C} \mathcal{J}}
    =\mathcal{J}(A,B|C)
\end{mathpar}
\paragraph{\bf DDIV} Applying the identity from DMUL in reverse, it must be true that
\begin{mathpar}
    \mathcal{J}(A|B,C)\mathcal{J}(B|C) = \mathcal{J}(A,B|C)
\Rightarrow\\ \mathcal{J}(A|B,C) = \frac{\mathcal{J}(A,B|C)}{\mathcal{J}(B|C)}
\end{mathpar}
which justifies the basic rule construct.
\paragraph{\bf DDIV2}

From the identity in DMUL, it must be true that
\begin{mathpar}
\mathcal{J}(A|B,C)\mathcal{J}(B|C) = \mathcal{J}(A,B|C)
\Rightarrow 
   \\ 
    \mathcal{J}(B|C) = \frac{\mathcal{J}(A,B|C)}{\mathcal{J}(A|B,C)}
\end{mathpar}
which justifies the basic rule construct.

\paragraph{\bf DINT.}
This rule relies on the following simplification of $\mathcal{J}$:
\begin{mathpar}
\int_{A} \mathcal{J}(A,B|C) = \int_{A} \frac{\int_{\mathcal{V} - (A \cup B \cup C)}\mathcal{J}}{\int_{\mathcal{V} - C}\mathcal{J}}=\frac{\int_{A} \int_{\mathcal{V} - (A \cup B \cup C)}\mathcal{J}}{\int_{\mathcal{V} - C}\mathcal{J}} =\frac{\int_{\mathcal{V} - (B \cup C)}\mathcal{J}}{\int_{\mathcal{V} - C}\mathcal{J}} 
    =\mathcal{J}(B|C)
\end{mathpar}
The second step above is justified by the fact that $A \cap C = \emptyset$, so the denominator is a constant with respect to the outer integral.

\subsection{Sampler Soundness}
\label{sec:proofsamp}

\begin{theorem}[Sampler Preservation] If \stdef[\vDash]{s}{A}{B}, \\
then  $\forall \sigma. \Big(\sigma \vDash \Gamma, \mathcal{M}, \assumplog \; \land \; \denotation{\cons}(\sigma) \Big) \Rightarrow 
    \Big(\forall \sr. \denotation{s}(\sigma,\sr) \vDash \Gamma, \mathcal{M}, \assumplog \Big)$
\end{theorem}

\paragraph{\bf Proof} Proceed by induction on the structure of type
derivations. In the case of SLIFT, note that since
$\int_{\denotation{A}(\mathcal{M},\sigma)} \mathcal{J}(A|B)(\mathcal{M},\sigma)
= 1$, so the returned value $r$ can never be larger than the maximum of the
target set of the variable being sampled. In the case of SBIND, apply the
inductive hypotheses.

Recall the sampler soundness theorem:

if\stype{\mathcal{M}}{\Gamma}{\assumplog}{d}{\qv,\delta}{A}{B}{\cons},
then\\ \stype[\vDash]{\mathcal{M}}{\Gamma}{\assumplog}{d}{\qv,\delta}{A}{B}{\cons}

\paragraph{\bf Proof.} The proof follows from structural induction on the type rules which may produce samplers. Individual cases are outlined below. 

\paragraph{\bf SLIFT.}
In the discrete case, notice that the size of the set \[\{\sr |
\denotation{\samp{\var{v}{a}}{d}}(\mathcal{M},\sigma[(v,\denotation{a}(\mathcal{M},\sigma))
\mapsto n])\}\] is exactly
$\denotation{d}(\mathcal{M},(v,\denotation{a}(\mathcal{M},\sigma)) \mapsto n)$,
and furthermore each such set is disjoint for different values of $a$.
Therefore, the integral over $\sr$ is a linear combination over these different
cases: \[\int_\sr f(\denotation{s}(\mathcal{M},\sigma)) = \sum_n
\denotation{d}(\mathcal{M},\sigma[(v,a) \mapsto n]) * f(n)\] which is exactly$\int_\denotation{A}(\sigma) f *
\mathcal{J}(A|B)(\mathcal{M},\sigma)$ in this case.

In the continuous case, the sample returned must be a real value $r$ such that
\begin{mathpar}
    \sr = \int_{x \in [-\infty, r]} \denotation{d}(\mathcal{M},\sigma[v,\denotation{a}(\sigma) \mapsto x]) = g(r) \Rightarrow \int_{\sr}f(\denotation{s}(\sigma)) = \int_\sr f(g^{-1}(\sr))
\end{mathpar}
Substituting $g$ for $\psi$ and $f \circ g^{-1}$ for $f$ in the definition for the substitution rule, it holds that
\begin{mathpar}
    \int_\sr f(\denotation{s}(\mathcal{M},\sigma,\sr)) = \int_{\denotation{A}(\mathcal{M},\sigma)} f(\sigma) * \denotation{d}(\mathcal{M},\sigma) = \int_{\denotation{A}(\mathcal{M},\sigma)} f(\sigma) * \mathcal{J}(A|B)(\mathcal{M},\sigma)
\end{mathpar}
where the above step is due to the soundness theorem for densities.

\paragraph{\bf SBIND.}
First, apply the soundness assumption for $s_1$ to find the expectation of the function $s_1 \circ f$. Then, use the assumption soundness assumption on $s_2$. This yields the equation
\begin{mathpar}
    \int_{\sr^0,\sr^1} f(\denotation{s_2}(\mathcal{M},\denotation{s_1}(\mathcal{M},\sigma,\sr^0), \sr^1)) = \int_{\denotation{A}(\mathcal{M},\sigma), \denotation{B}(\mathcal{M},\sigma)} f(\sigma) * \mathcal{J}({A} | {B \comma C})(\mathcal{M},\sigma) * \mathcal{J}({B}|{C})(\mathcal{M},\sigma)
\end{mathpar} 
Using the identity from DMUL, this simplifies to \[ \int_{\denotation{A}(\mathcal{M},\sigma), \denotation{B}(\mathcal{M},\sigma)} f(\sigma) * \mathcal{J}({A \comma B} | {C})(\mathcal{M},\sigma)\]

\subsection{Kernel Soundness}
\label{sec:proofkern}
Recall the kernel soundness theorem:

    if \ktype{\mathcal{M}}{\Gamma}{\assumplog}{k}{\hat{\qv},\hat{\delta}}{A}{B}{\cons}
    then\\ \ktype[\vDash]{\mathcal{M}}{\Gamma}{\assumplog}{k}{\hat{\qv},\hat{\delta}}{A}{B}{\cons}

\paragraph{\bf Proof.} The proof follows from structural induction on the rule derivations for kernels. We strengthen the inductive hypothesis with the proposition 
\begin{mathpar}
    \ktdef{d}{A}{B} \Rightarrow \ktdef[\vDash]{d}{A}{B}
\end{mathpar}
The specific cases are outlined below.

\paragraph{\bf KLIFT.}
This is equivalent to the statement that if $s$ is a sampler, then $\progcmd{fix} s \equiv s$. In other words, for any measurable function $f$ over the output space, the equation
\[\int_{\sr} f(\mathrm{fix}(\sigma,\sr)) = \int_{\sr,\sr'} f(\mathrm{fix}(s(\sigma,\sr'),\sr))\]
is satisfied for $\mathrm{fix} = s$. To see this, inline the definition for a sampler, which reduces the soundness property to the equation
\begin{mathpar}
    \int_{\denotation{A}(\mathcal{M},\sigma)} f(\sigma) * \mathcal{J}({A}|{B})(\mathcal{M},\sigma) = \int_{\denotation{A}(\mathcal{M},\sigma)} \mathcal{J}({A}|{B})(\mathcal{M},\sigma) \int_{\denotation{A}(\mathcal{M},\sigma)} f(\sigma) \mathcal{J}({A}|{B})(\mathcal{M},\sigma)
\end{mathpar}
which must hold because $ \int_{A} \mathcal{J}(A|B) = 1$.

\paragraph{\bf KCOMBINE.}
First, we will show that $k_1$ is invariant for the distribution $A \comma B | C$. This is true because, for a sampler $s$ such that
\[ \int_{\sr} f(s(\mathcal{M},\sigma, \sr)) = \int_{\denotation{A \comma B}(\mathcal{M},\sigma)} f(\sigma) * \mathcal{J}({A \comma B}|{C})(\mathcal{M},\sigma) \]
because of the property from DMUL, it must be true that
\begin{mathpar}
    \int_{\sr} f(s(\mathcal{M},\sigma, \sr)) = \int_{\denotation{A \comma B}(\mathcal{M},\sigma)} (f(\sigma) * \mathcal{J}({B}|{C})(\mathcal{M},\sigma)) * \mathcal{J}({A}|{B \comma C})(\mathcal{M},\sigma) 
\end{mathpar}
Substituting in $f * \mathcal{J}({B}|{C})$ for $f$ in the soundness assumption for $k_1$, it must hold that
\begin{mathpar}
    \int_{\sr^0,\sr^1} f(k_1(s(\mathcal{M},\sigma, \sr^0),\sr^1)) = 
    \int_{\denotation{A \comma B}(\mathcal{M},\sigma)} (f(\sigma) * \mathcal{J}({B}|{C})(\mathcal{M},\sigma)) * \mathcal{J}({A}|{B \comma C})(\mathcal{M},\sigma)
\end{mathpar}
Using again the identity from DMUL, this means that
\begin{mathpar} 
    \int_{\sr^0,\sr^1} f(k_1(s(\sigma, \sr^0),\sr^1)) = \int_{\denotation{A \comma B}(\sigma)} f(\sigma) * \mathcal{J}(\denotation{A \comma B}(\sigma)|\denotation{C}(\sigma))(\sigma) 
\end{mathpar}
completing the proof of the invariance property of $k_1$. By a similar logic, $k_2$ is also invariant for the distribution $A \comma B | C$. This means that $k_1 \opfmt{;} k_2$ is invariant for the distribution $A \comma B | C$. 

\subsection{Estimator Soundness}
\label{sec:proofest}
Recall the estimator soundness theorem:

    $\etdef{e}{A}{B} \Rightarrow \etdef[\vDash]{e}{A}{B}$

\paragraph{\bf Proof.} The proof follows from structural induction on the rules which may produce estimators. Individual cases are outlined below.

\paragraph{\bf ELIFT}
Since the first element of $e$ is defined to be $1$ in all cases the expression, 
\[\denotation{\cons}(\mathcal{M},\sigma) \Rightarrow \int_\sr \frac{\pi_1(\denotation{e}(\mathcal{M},\sigma,\sr)) * f(\pi_2(\denotation{e}(\mathcal{M},\sigma,\sr)))}{\int_{\sr} \pi_2(\denotation{e}(\mathcal{M},\sigma,\sr))}\]
can be simplified to
$\frac{\int_{\sr} f(\denotation{s}(\mathcal{M},\sigma,\sr))}{\int_{\sr} 1}=\int_\sr f(\denotation{s}(\mathcal{M},\sigma,\sr))$
which, according to the correctness of the sampler, must equal the expression $\int_{\denotation{A}(\mathcal{M},\sigma)} f(\sigma) * \mathcal{J}({A}|{B})(\mathcal{M},\sigma)$
as required.

\paragraph{\bf EFACT}
We use the estimator $e$ to estimate two functions, $\denotation{d} * f$ and
$\denotation{d}$, and take the ratio. This yields the equation
\begin{mathpar}
    \frac{
        \frac{
            \int_\sr \denotation{d}(\mathcal{M},\pi_2(\denotation{e}(\mathcal{M},\sigma,\sr))) *
            f(\mathcal{M},\pi_2\denotation{e}(\mathcal{M},\sigma,\sr)) *
            \pi_1(\denotation{e}(\mathcal{M},\sigma,\sr))
        }{
            \int_\sr \pi_1(\denotation{e}(\mathcal{M},\sigma,\sr))
        }
    } {
        \frac{
            \int_\sr \denotation{d}(\mathcal{M},\pi_2(\denotation{e}(\mathcal{M},\sigma,\sr))) *
            \pi_1(\denotation{e}(\mathcal{M},\sigma,\sr))
        }{
            \int_\sr \pi_1(\denotation{e}(\mathcal{M},\sigma,\sr))
        }
    } =
    
    \frac{
        \int_{\denotation{A}(\mathcal{M},\sigma)} 
        f(\sigma) *
        \mathcal{J}(C|A,B)(\mathcal{M},\sigma) * 
        \mathcal{J}(A|B)(\mathcal{M},\sigma) 
    }{
        \int_{\denotation{A}(\mathcal{M},\sigma)} 
        \mathcal{J}(C|B)(\mathcal{M},\sigma) * 
        \mathcal{J}(A|B)(\mathcal{M},\sigma) 
    }
\end{mathpar}

We notte that the left hand side of this equation simplifies to
$\denotation{\factor{e}{d}}(\mathcal{M},\sigma,\sr)$ and the right hand side
simplifies to $\mathcal{J}(A|B,C)(\mathcal{M},\sigma)$

\subsection{Structural Rule Soundness} 
\label{sec:proofstruct}

\paragraph{\bf IF} The soundness of conditionals follows from the fact that, if \(\sigma\) is such that \(\denotation{\cons_i}(\mathcal{M},\sigma)\) is true,
\begin{mathpar}
    \typegeneric[\vDash]{\mathcal{M}}{\Gamma}{\assumplog}{\prog}{\basictype{T_b}{(\cons_i,A_t,A_e)}{(\cons_i,A_t,B_t)}{(\cons_i \opfmt{\&\&} \cons_t) \opfmt{||} (\neg \cons_i \opfmt{\&\&} \cons_e)}}\\
    \iff
    \typegeneric[\vDash]{\mathcal{M}}{\Gamma}{\assumplog}{\prog}{\basictype{T_b}{A_t}{B_t}{\cons_t}}
\end{mathpar}
Otherwise, if \(\sigma\) is such that \(\denotation{\cons}(\mathcal{M},\sigma)\) is false, it must be true that
\begin{mathpar}
    \typegeneric[\vDash]{\mathcal{M}}{\Gamma}{\assumplog}{\prog}{\basictype{T_b}{(\cons_i,A_t,A_e)}{(\cons_i,A_t,B_t)}{(\cons_i \opfmt{\&\&} \cons_t) \opfmt{||} (\neg \cons_i \opfmt{\&\&} \cons_e)}}\\
    \iff
    \typegeneric[\vDash]{\mathcal{M}}{\Gamma}{\assumplog}{\prog}{\basictype{T_b}{A_e}{B_e}{\cons_e}} 
\end{mathpar}

\paragraph{\bf INV} This follows directly from the assumption that $\sigma \vDash \mathcal{M}, \Gamma, \assumplog$.

\paragraph{\bf C}

Writing the types $t_1$ and $t_2$ as $\basictype{T_b}{A_1}{B_1}{\cons_1}$ and $\basictype{T_b}{A_2}{B_2}{\cons_2}$, respectively, the added assumptions mean that, for any $\sigma$, $\denotation{A_1} = \denotation{A_2}$, $\denotation{B_1} = \denotation{B_2}$, and $\denotation{\cons_1}(\mathcal{M},\sigma) \Rightarrow \denotation{\cons_2}(\mathcal{M},\sigma)$. This means that the soundness of the judgment
\typegeneric{\mathcal{M}}{\Gamma}{\assumplog}{\prog}{t_1}
implies the soundness of the judgment \typegeneric{\mathcal{M}}{\Gamma}{\assumplog}{\prog}{t_2}.

\paragraph{\bf CIND.}

Writing the types $t_1$ and $t_2$ as $\basictype{T_b}{A}{B}{\cons}$ and $\basictype{T_b}{A}{B \comma C}{\cons}$, respectively, the added assumptions mean that, for any $\sigma$, $\mathcal{J}({C \comma A} | {B})(\mathcal{M},\sigma) = \mathcal{J}({C} | {B})(\mathcal{M},\sigma) * \mathcal{J}({A} | {B})(\mathcal{M},\sigma)$ which, applying the identity from the DMUL case, means
\(\mathcal{J}({A} | {B \comma C})(\mathcal{M},\sigma) = \mathcal{J}({A} | {B})(\mathcal{M},\sigma)\).

\subsection{Preservation}
\label{sec:proofpres}

In this section, we prove the following property, which states that definitions
{\em preserve} the satisfaction realtion for environments. Specifically, for
any type rule that produces a new type environment, if the original environment
satisfies type environment, then the environment prescribed by the semantics
satisfies the new type environment.

\begin{thm}[Preservation]
    If  $\mathcal{M}$ is valid, $\sigma \vDash \Gamma, \mathcal{M}, \assumplog$ and \typegeneric{\mathcal{M}}{\Gamma}{\assumplog}{p}{\Gamma'}, \\then $\denotation{p}(\mathcal{M},\sigma) \vDash \Gamma', \mathcal{M}, \assumplog$
\end{thm}

\paragraph{\bf Proof} Proceed by induction on the structure of derivations for types.

\paragraph{\bf DEF and DEF-IND}
Because $\denotation{\cons \opfmt{\&\&} \qv \in \delta}(\mathcal{M},\sigma) \Rightarrow \denotation{\qv}(\sigma) \in \denotation{\delta}(\mathcal{M})$
\begin{mathpar} 
    \sigma \vDash \Gamma, \mathcal{M}, \assumplog \land
    \typegeneric{\mathcal{M}}{\Gamma[\qv \mapsto \delta]}{\assumplog}{\prog}{\basictype{t_b}{A}{B}{\cons}}\\
    \Rightarrow 
    \sigma[x \mapsto ({\qv}, {\delta}, \prog)] \vDash \mathcal{M}, \Gamma[x \mapsto ({\qv}, {\delta}, \basictype{t_b}{A}{B}{\cons \opfmt{\&\&} {\qv} \in {\delta}})], \assumplog
\end{mathpar}

\paragraph{\bf DEF-REC} First, we show that $\typegeneric[\vDash]{\mathcal{M}}{\Gamma[\qv \mapsto \delta]}{\assumplog}{\prog}{\basictype{t_b}{A}{B}{\cons}}$. We proceed by strong induction on $\qv$.
\begin{itemize}
    \item {\bf Base Case.} If $\denotation{\qv}(\sigma) < \denotation{\texttt{min(}\delta\texttt{)}}(\mathcal{M})$, then because $\mathcal{J}(\emptyset|B)(\mathcal{M},\sigma) = 1$,\\ $\typegeneric[\vDash]{\mathcal{M}}{\Gamma}{\assumplog}{\prog}{\basictype{t_b}{A}{B}{\cons}}$.
    \item {\bf Inductive step.} The inductive step derives directly from the rule assumption involving $\prog$:
        \[
            \typegeneric[\vDash]{\mathcal{M}}{\Gamma[ \qv \mapsto \delta ][x \mapsto (\qv', \delta, \basictype{t_b}{A[{\qv'}/{\qv}]}{B[{\qv'}/{\qv}]}{(\cons[{\qv'}/{\qv}]) \opfmt{\&\&} {\qv'} < {\qv}})]}{\assumplog}{\prog}{\basictype{t_b}{A}{B}{\cons}}\\
        \]
\end{itemize}
The remainder of the proof follows from the same logic in the DEF case that $\denotation{\cons \opfmt{\&\&} \qv \in \delta}(\mathcal{M},\sigma) \Rightarrow \denotation{\qv}(\sigma) \in \denotation{\delta}(\mathcal{M})$

\section{Optimization Example}
\label{sec:optimization-detail}

\begin{figure}
\begin{subfigure}{0.45\linewidth}
\begin{lstlisting}[
    language=C, escapeinside={(*}{*)},
    keywordstyle=\color{blue},
    escapeinside={(*}{*)}, 
    keywordstyle=\color{blue}, 
    numbers=left,
    frame=single]
for(i = 0; i < N; i++) {
  s = 0;
  for(j = 0; j < N; j++) {
    if(i != j && x[i] == y[j])
      s += a[j];
  }
  b[i] = s;
}
\end{lstlisting}
\caption{Original}
\label{fig:opt-example-orig}
\end{subfigure}\quad\quad\quad
\begin{subfigure}{0.45\linewidth}
\begin{lstlisting}[
    language=C, escapeinside={(*}{*)},
    keywordstyle=\color{blue},
    escapeinside={(*}{*)}, 
    keywordstyle=\color{blue}, 
    numbers=left,
    frame=single]
// unroll first iteration of outer
i = 0;
s = {0 ... 0 };
for(j = 0; j < N; j++) {
  if(i != j)
    // partition s by x[j]
    s[y[j]] += a[j];
}
b[i] = s[x[i]];
for(i = 1; i < N; i++) {
  // incrementally update s
  s[y[i - 1]] += a[i - 1];
  s[y[i]] -= a[i];
  b[i] = s[x[i]];
}
\end{lstlisting}
\caption{After enabling transformation and incrementalization}
\label{fig:opt-example-after}
\end{subfigure}
\caption{Example optimization}
\label{fig:opt-example}
\end{figure}

An simplified example transformation is shown in Figure~\ref{fig:opt-example}. 
The original program in Figure~\ref{fig:opt-example-orig} has asymptotic complexity $O(N^2)$, 
while the program in Figure~\ref{fig:opt-example-after} has complexity $O(N)$.
Shuffle's optimization pass identifies loop pair patterns like the above and greedily 
performs this optimization. Therefore, an optimized inference algorithn generated by
Shuffle can have significantly better asymtotic complexity. Table~\ref{tab:resultsperfunopt}
illustrates such effect.

\section{Performance Discussion}
\label{sec:perf-discussion}

\begin{table*}
\begin{tabular}{|c|c|c|c|c|c|c|}
	\hline
	
	\multirow{2}{*}{Model} & \multicolumn{2}{c|}{Gibbs} & \multicolumn{2}{c|}{Metropolis-Hastings} & \multicolumn{2}{c|}{Likelihood Weighting} \\
	
	\cline{2-7} 
	
	& Shuffle & Venture  & Shuffle & Venture  & Shuffle & Venture  \\
	
	\hline
	GMM & \SI{3.3e-2}{\ms} & \SI{8.3e-1}{\ms}  & \SI{3.5e-1}{\ms} & \SI{8.9e-1}{\ms} & \SI{7.8e-4}{\s} & \SI{1.8}{\s} \\
	LDA & \SI{5.2e-5}{\s}  & \SI{4.1e-2}{\s} & \SI{3.5e-2}{\ms}  & \SI{5.3}{\ms} & \SI{1.2e-1}{\s} & \SI{23}{\s}  \\
	DMM & \SI{0.011}{\s} & \SI{15}{\s} & \SI{4.7e-2}{\s} & \SI{0.54}{\s} & \SI{1.1}{\s} &  \SI{13}{\s} \\
	\hline
\end{tabular}
\caption{Standard deviation for measurements in Table~\ref{tab:resultsperf}}
\label{tab:resultsperfvar}

\begin{tabular}{|c|c|c|c|c|c|c|c|c|c|}
	\hline
	
	\multirow{2}{*}{Model} & \multicolumn{3}{c|}{Gibbs} & \multicolumn{3}{c|}{Metropolis-Hastings} & \multicolumn{3}{c|}{Likelihood Weighting} \\
	
	\cline{2-10} 
	
	& Shuffle & Venture & Ratio & Shuffle & Venture & Ratio & Shuffle & Venture & Ratio \\
	
	\hline
	GMM & 39 \si{\mega\byte} & $2.1 \si{\giga\byte}$ & ${\bf 53x}$ & $39  \si{\mega\byte}$  & $2.1 \si{\giga\byte}$  & ${\bf 53x}$ & 39  \si{\mega\byte}  & 2.2  \si{\giga\byte}  & ${\bf 56x}$ \\
	LDA &118  \si{\mega\byte} & 63 \si{\giga\byte}   & ${\bf 533x}$ & 122 \si{\mega\byte} & 63 \si{\giga\byte} & ${\bf 516x}$ & 127 \si{\mega\byte} & 70 \si{\giga\byte} & ${\bf 551x}$ \\
	DMM & 88 \si{\mega\byte} & 45 \si{\giga\byte} & ${\bf 511x}$ & 88 \si{\mega\byte}  & 45 \si{\giga\byte} & ${\bf 511x}$ & 77 \si{\mega\byte} & 50 \si{\giga\byte} & ${\bf 649x}$ \\
	\hline
\end{tabular}
\caption{Peak memory consumption. Ratios are computed as Venture over Shuffle.}
\label{tab:resultsmem}
\vspace{-.5cm}
\end{table*}

The performance tradeoffs between Shuffle and Venture are primarily due to
incremental optimization opportunities. For example, in the Gaussian mixture
model, an efficient inference procedure would maintain statistics regarding the
number and sum of samples attributed to a given cluster center. When the
inference procedure modifies the cluster assignment, it incrementally updates
these statistics instead of computing them from scratch.

Venture facilitates incremental optimizations through its stochastic procedure
interface. When a developer defines a stochastic procedure, he or she must
implement methods that add or remove observations of the procedure's output.
Venture's runtime builds a computation graph of stochastic procedures, and
traverses this graph to determine which stochastic procedures require updates.
For example, in the Gaussian mixture model, each cluster forms a stochastic
procedure, and when the cluster assignment of a sample changes -- meaning a
sample is moved from one cluster to another -- the runtime calls the update
methods on the appropriate clusters, which in turn update their internal
statistics regarding the number and sum of the samples in a given cluster.  We
hypothesize Venture's computation graph contributes to its slow performance on
LDA with Gibbs and Metropolis-Hastings. LDA is a larger model than
GMM, and Gibbs and Metropolis-Hastings require more graph manipulation
operations than Likelihood Weighting. We have noticed that on these benchmarks, 
Shuffle's performance remains constant when varying the corpus size. On the other hand,
Venture manifests sublinear scaling on LDA-\{Gibbs, MH\} benchmarks, 
which causes magnitudes of performance ratio compared to Shuffle. We suspect that Venture's 
complicated data structure operations cause the run time to depend on the size of the corpus, whereas 
Shuffle's lean data structures do not have this dependence. Specifically, from profiling 
we observed that Venture \texttt{deepcopy}s its internal datastructure every sampling step 
which accounts for approximately 90 percent of the run time on these two benchmarks. 
The high memory usage of Venture  also justifies that the data structure manipulations Venture 
leverages for performing incremental updates bottlenecks the inference procedure's performance.

In contrast, Shuffle requires developers to express their inference procedures
with density arithmetic. While the resulting procedure computes the same
statistics as the corresponding Venture stochastic procedure, the natural way
to compile a Shuffle procedure is to recompute the statistics from scratch. For
example, in the Gaussian mixture model, the developer uses multiplication,
division, and integration to compute conditional probability distributions
pertaining to an individual cluster. Shuffle recognizes subexpressions of this
computation as having a closed-form solution, and replaces these with
expressions that compute counts and sums of the samples. As an additional step,
Shuffle optimizes the procedure, translating it to an equivalent form that
maintains values for subexpressions and incrementally updates them. Shuffle
currently cannot optimize all possible subexpressions in the DMM model, which
is the reason for its sporadic performance on this model. 

Lastly, Shuffle without optimization scales poorly, and suffers orders of magnitude slow down
when compared to Shuffle with optimization. This behavior re-confirms our
hypothesis on the importance of incremental optimization.

\paragraph{\bf Other Considerations.}
Another factor contributing to Shuffle's relative performance over Venture is
the fact that Venture performs dynamic checks to ensure the values of the
random variables belong to the proper domain, whereas Shuffle's type system
makes this unnecessary.

\section{More on Benchmarks}
\label{sec:morebenchmarks}

\begin{table}
\begin{subtable}[t]{.45\linewidth}
\begin{tabular}{|l|c|c|}
    \hline
    Benchmark & Shuffle & BLOG \\
    \hline
    Burglary & 19 & 13 \\
    Context-Specific Inference & 14 & 11 \\
    Healthiness & 48 & 31 \\
    Hurricane & 22 & 21 \\
    Weather & 31 & 17 \\
    
    \hline
\end{tabular}
\caption{Microbenchmark models, measuring lines of code for both Shuffle and BLOG.}
\label{tab:microbenchmodel}
\end{subtable}
\hfill
\begin{subtable}[t]{.45\linewidth}
\begin{tabular}{|l|c|c|}
    \hline
    Benchmark & Shuffle & Venture \\
    \hline
    Gaussian Conjugate & 18 & 72 \\
    Dirichlet Conjugate & 18 & 91 \\
    \hline
\end{tabular}
\caption{Microbenchmark inference proceudres, measuring lines of code for both Shuffle and Venture.}
\label{tab:microbenchinfer}
\end{subtable}
\caption{Microbenchmarks we have implemented in Shuffle.}
\label{tab:microbench}
\vspace{-0.8cm}
\end{table}

\vspace{-.1cm}
\paragraph{\bf BLOG} BLOG is a probabilistic programming language that is
designed to handle {\em identity uncertainty}, wherein the probabilistic model
admits uncertainty regarding the source of an observation.   We
compare BLOG models to Shuffle models, but do not compare BLOG's inference to
Shuffle's inference. BLOG developers write models by hand but use BLOG's
inference compiler~\citep{swift} to generate inference procedures. By contrast,
Shuffle developers write both the model and inference procedure by hand.

Shuffle is less expressive than BLOG and Venture in that Shuffle only supports
models with finite size, whereas BLOG and Venture models may be statically
unbounded. We note that even though Shuffle models must be bounded in size,
that bound need not be specified at type checking time. Type checking results
hold for all runtime instantiations of the model's domains.

\paragraph{\bf Microbenchmarks.} To demostrate Shuffle's ability to represent a
variety of statistical models and inference procedures, we ported examples from
the BLOG and Venture probabilistic programming systems~\cite{blog, venture}.

Table~\ref{tab:microbench} lists the microbenchmarks we have implemented. For
the BLOG microbenchmarks in Table~\ref{tab:microbenchmodel}, we reimplemented
each BLOG model as a Shuffle model.    For the Venture microbenchmarks in
Table~\ref{tab:microbenchinfer}, we implemented a Shuffle inference procedure
that performs the {\em sampling} method of an equivalent Venture stochastic
procedure. For the conjugate models we consider here, this corresponds to
sampling a new observation conditioned on a set of previous observations from
the model.   

\subsection{Verification Burden}

\subsubsection{Methodology}

To quantify the verification burden for the microbenchmarks, we compared the
length of Shuffle models against that of BLOG models and the length of
Shuffle's inference procedure against the length of Venture's stochastic
procedure implementation.

\subsubsection{Results}

Tables~\ref{tab:microbench} and~\ref{tab:microbenchinfer} present the size of
the microbenchmarks as implemented in each system.  Shuffle requires more lines
of code to represent a probabilistic model than BLOG due to Shuffle's
additional type annotations.  Shuffle requires fewer lines of code to implement
these inference procedures than in Venture because Venture's stochastic
procedure interface requires the developer to implement methods that are
{unnecessary for these particular inference tasks. By contrast, Shuffle
developers only have to implement the densities that are required.}

\end{document}